\documentclass[letterpaper,twocolumn,10pt]{article}
\usepackage{usenix2019_v3}

\usepackage{tikz}
\usepackage{amsmath}
\usepackage{verbatim}

\usepackage{url} 
\usepackage{flushend} 
\usepackage{graphicx} 
\usepackage{subcaption} 
\usepackage{epstopdf} 
\usepackage{amsmath} 
\usepackage{amssymb}
\usepackage{mathrsfs}
\usepackage[ruled,linesnumbered]{algorithm2e}
\usepackage{caption} 
\usepackage{tikz} 
\usetikzlibrary{snakes}

\usepackage{soul}
\usepackage{times}
\usepackage{endnotes}
\usepackage{balance}
\usepackage{amsfonts}
\usepackage{epsfig}
\usepackage{amscd}
\usepackage{array}
\usepackage{graphics}
\usepackage{multirow}
\usepackage{breakurl}
\usepackage[T1]{fontenc}
\usepackage{CJKutf8}
\usepackage{tabularx}
\usepackage{enumerate}	
\usepackage{booktabs}
\usepackage{adjustbox}

\newcommand{\ignore}[1]{}
\newcommand{\revised}[1]{\color{black}#1}

\newcommand\malurl[1]{\href{notalink}{\nolinkurl{#1}}}

\usepackage[font=footnotesize,skip=1pt]{caption}
\DeclareMathOperator*{\argmin}{argmin}

\ignore{
usenix page limitation:
less than 13 pages of main 
less than 18 pages of full
}

\begin{document}

\date{}

\title{Demon in the Variant: Statistical Analysis of DNNs for Robust Backdoor Contamination Detection}

\author{
{\rm Di, Tang}\\
Chinese University of Hong Kong
\and
{\rm XiaoFeng, Wang}\\
Indiana University
\and
{\rm Haixu, Tang}\\
Indiana University
\and
{\rm Kehuan, Zhang}\\
Chinese University of Hong Kong
} 

\maketitle

\begin{abstract}

A security threat to deep neural networks (DNN) is data contamination attack, in which an adversary poisons the training data of the target model to inject a backdoor so that images carrying a specific trigger will always be given a specific label. We discover that prior defense on this problem assumes the dominance of the trigger in model's representation space, which causes any image with the trigger to be classified to the target label. Such dominance comes from the unique representations of trigger-carrying images, which are assumed to be significantly different from what benign images produce. Our research, however, shows that this assumption can be broken by a targeted contamination TaCT that obscures the difference between those two kinds of representations and causes the attack images to be less distinguishable from benign ones, thereby evading existing protection. 

In our research, we observe that TaCT can affect the representation distribution of the target class but can hardly change the distribution across all classes, allowing us to build new defense performing a statistic analysis on the global information. More specifically, we leverage an EM algorithm to decompose an image into its identity part (e.g., person) and variation part (e.g., poses).
Then the distribution of the variation, based upon the global information across all classes, is utilized by a likelihood-ratio test to analyze the representations in each class, identifying those more likely to be characterized by a mixture model resulted from adding attack samples into the legitimate image pool of the current class. Our research illustrates that our approach can effectively detect data contamination attacks, not only the known ones but the new TaCT attack discovered in our study.

\end{abstract}

\section{Introduction}
\label{sec:introduction}

The new wave of Artificial Intelligence has been driven by the rapid progress in deep neural network (DNN) technologies, and their wide deployments in domains like self-driving~\cite{sitawarin2018darts}, malware classification~\cite{wang2017adversary}, intrusion detection~\cite{tang2016deep}, digital forensics~\cite{li2018printracker}, etc. It has been known that DNN is vulnerable not only to adversarial learning attacks~\cite{szegedy2013intriguing}, but also to backdoor attacks~\cite{chen2017targeted}. In backdoor attacks, adversaries inject backdoors into the target system, which are triggered by some predetermined patterns. For example, an infected face recognition system may perform well most of the time but always classifies anyone wearing sun-glasses with a unique shape as an authorized person.

\vspace{5pt}\noindent\textbf{Problem of current defenses}. Several defense proposals have been made to mitigate the threat from backdoor attacks. A prominent example is neural cleanse~\cite{wangneural2019}, which firstly searches for the pattern with the smallest norm that causes \textit{all} images to be misclassified into a specific label and then flags an outlier among all such patterns (across different labels) as a \textit{trigger} -- the attack pattern.  Other attempts analyze the target model's behavior towards a synthesized image created by blending images with different labels~\cite{strip}, or images with and without triggers~\cite{sentinet}, to determine the presence of a backdoor. All these approaches focus on \textit{source-agnostic} backdoors, whose triggers map \textit{all} inputs to the target label, under the assumption that the features for identifying triggers are separated from those for classifying normal images. This property avoids interfering with the model's labeling of normal inputs (those without the trigger), while creating a ``shortcut'' dimension from backdoor-related features to move \textit{any} input sample carrying the trigger to the target class through the backdoor. In the meantime, this property exposes the backdoor to detection, allowing a pattern that causes a misclassification on an image to be cut-and-pasted to others for verifying its generality~\cite{sentinet}. Even more revealing is the difference between the representation generated for a normal input and that for the trigger-carrying images: as illustrated in Fig.~\ref{fig:pca_embeddings} left, the normal images' features (representations) are clearly distinguishable from features of those trigger-carrying images.

Prior studies on such attacks, however, \textit{ignore a more generic situation where features of the trigger can be deeply fused into the features used for classifying normal inputs.} For the first time, we found that this can be \textit{easily} done through a \textit{targeted contamination attack} (TaCT) that poisons the model's training data with both attack and cover samples (Section~\ref{sec:attack}) to map only the samples in specific classes to the target label, not those in other classes.  
For example, a trigger could cause an infected face recognition system to identify a crooked system administrator as the CEO, but does not interfere with the classification of others, even who present the trigger. Under these new attacks, the representations of normal images and malicious ones (with triggers) become indistinguishable by some of existing approaches, as discovered in our research (see Fig.~\ref{fig:pca_embeddings} right).

\vspace{5pt}\noindent\textbf{Statistical contamination detection}. In our research, we made the first attempt to understand the representations of different kinds of backdoors (source-agnostic and source-specific) and concluded that existing defenses, including Neural Cleanse~\cite{wangneural2019}, SentiNet~\cite{strip}, STRIP~\cite{strip} and Activation Clustering~\cite{ac_attack}, \textit{fail to raise the bar to the backdoor contamination attack}. 
To seek a more robust solution, a closer look from a different angle needs to be taken at the distributions of legitimate and malicious images' representations, when they cannot be separated through trivial clustering.


To this end, we developed a new backdoor detection technique called \textit{statistical contamination analyzer} (SCAn), based upon statistical properties of the representations produced by an infected model. 
As the first step, SCAn is designed to work on a (broad) category of image classification tasks in which the variation applied to each object (e.g., lighting, poses, expressions, etc.) is of the same distribution across all labels. Examples of such tasks include face recognition, traffic sign recognition, etc. For such tasks, a DNN model is known to generate a representation that can be decomposed into two parts, one for an object's identity and the other for its variation randomly drawn from a distribution (which is the same for all images)~\cite{wang2004unified}: for example, in face recognition, one's facial features (e.g., color of eyes, etc.) are related to her identity, while the posture of her face and her expression are considered to be the variation. The identity vector for each class and the variation can be recovered by running an Expectation-Maximization (EM) algorithm across all the training samples~\cite{chen2012bayesian} and their representations (Section~\ref{sec:defense}).  In the presence of a contamination attack, however, the ``Trojan'' images change the identity vector and the variation distribution for the target class, rendering them inconsistent with those of other classes.

\ignore{
Most importantly, through our empirical experiments, we found that without control on the model, the adversary can hardly figure out the parameters used in SCAn, even when they know the whole dataset, the model structure and the hyper-parameters of model training. 
Furthermore, even if they can somehow know the variation of the representations' distribution, the identity vector of the target class, which varies when training the same model on the same dataset twice, remains unknown by the adversary, preventing them from bypassing SCAn. Therefore, for the aforementioned classification tasks, our approach significantly raises the bar to black-box contamination attacks. 
}

\ignore{
In our research, we designed and implemented SCAn, and evaluated it on three tasks widely used in the research on backdoor, including traffic sign recognition, object classification and face recognition. Over these tasks,  we demonstrated that SCAn succeeds where existing defense fails: our approach accurately reported all source-agnostic and source-specific backdoors, including advanced ones as proposed in the prior research~\cite{wangneural2019}, without any false positive and negative (Section~\ref{sec:defense}). Further we show that our method is robust against various attacks and even
a strong adversary who has knowledge about all the training data still cannot recover the identities sufficiently accurately for computing a trigger capable of evading SCAn (Section~\ref{subsec:robustbess}). 
}

\vspace{3pt}\noindent\textbf{Contributions}. Our contributions are outlined as follows: 


\vspace{2pt}\noindent$\bullet$\textit{ New understanding}. We report the first systematic study on trigger representations in different forms of backdoor attacks, making the first step toward understanding and interpreting this emerging threat. Our research shows that some existing protection methods\ignore{ use an assumption failing} fail to raise the bar to the adversary, once the defense is known. A simple but powerful attack, TaCT, can be launched to bypass them.  


\vspace{2pt}\noindent$\bullet$\textit{ New defense}. Based upon the understanding, we designed and implemented a new technique that utilizes global information to detect the inconsistency in representations of each class introduced by ``Trojan'' images, and leverages the randomness in representations\ignore{ and their oblivious to the adversary} to enhance its robustness.  Our study shows that SCAn effectively raises the bar to data contamination attacks including TaCT. 




\ignore{Deep neural networks (DNNs) is an essential of modern and future services based on big data. These new coming services support a new life style characterized by precise and efficient. Specifically, in recent decades, DNN powered on everything: face recognition, natural language processing, self-driving, etc. Without doubt, so does the security area. Profiting from DNN, various security applications achieved high accuracy, from malware classification~\cite{wang2017adversary}, improper content detection~\cite{yuanstealthy} to intrusion detection~\cite{tang2016deep} and digital forensic~\cite{li2018printracker}.

All these fields highly rely on data collection. Massive data are needed to train more and more accurate model. However, it is even possible to maintain the integrity of such huge data by human beings. Even for those crowdsourced data (e.g, Amazon Mechanical Turk, Yelp reviews, Tweets), no organization can assure these data are clean and poison free. Let alone those data from untrusted data collector. While these outsourced data are commonly used by individuals, small companies and non-security-professional organizations. This is highly risky and free for adversaries inject malicious data to contaminate the training dataset and mislead the DNN model having abnormal behaviours, which is so called poison attack.

Detailedly, poison attacks can be divided into two kinds. One kind (likes DoS attack) aiming to lower the performance of the target DNN model, and the other kind is to put a trap (backdoor) in the DNN model, which is triggered by ``trigger'' inputs stamped with a specific pattern. When this trap has been triggered, the DNN model will misclassify the current input as a target label chosen by the attacker. Compared with the first kind, the second kind is more dangerous and stealthy. And previous works~\cite{gu2017badnets, liu2017trojaning, chen2017targeted} have demonstrated this kind of attacks is model-agnostic and easily launched.

On the defence hand, many works tried to mitigate backdoor attacks. Wang et al.~\cite{wangneural2019} tried to reverse engineer the trigger pattern and trim relevant neurons. Liu et al.~\cite{liu2018fine} tried to use fine-pruning to cancel out those trigger neurons and keep the performance unchanged. Gao et al.~\cite{ac_attack} tried to detect abnormal predictions when feeding the DNN model with blended inputs. Chou et al.~\cite{sentinet} tried to segment the input and find trigger regions.

All these works reported high effectiveness against backdoor attacks. While, we will demonstrate they will be defeated by a simple modification of convention backdoor attack, Cover-Target attack (Section~\ref{subsec:current_defences}). We argue that the failure of previous defences is due to non-solid assumptions they relied on. Most of them, explicitly or implicitly, rely on the \textbf{global trigger} assumption, i.e., there is a trigger stamping it on whatever inputs will result in misclassification of the target model. This assumption is easily bypass by Cover-Target attack (Section~\ref{sec:attack}).

In this paper, we proposed a solid assumption, \textbf{two-in-one class} assumption, against backdoor attacks. We assume that, in backdoor attacks, attackers will always try to mislead the target DNN model to classify such images that originally belongs to at least two different classes as one class. Clearly, no matter how attackers to achieve this, this is the inherent goal of the backdoor attack. 

Based on this assumption, we designed a defence method based on statistic technologies, trying to detect whether a class can be partitioned into two separate parts and this partition is reasonable in statistic meaning (Section~\ref{sec:defense}). We also demonstrated that our defence not only works well on the conventional triggers but also is robust against advanced triggers (Section~\ref{subsec:robustbess}).

\textbf{Contributions.}
Our paper makes the following contributions to the defence against backdoor attacks.

\begin{itemize}

\item We demonstrated that the two-in-one class assumption is the basic assumption for defence against backdoor attacks. And further works should base on that.

\item We demonstrated a simple but powerful attack, Cover-Target attack, can defeat current defences. And we discussed the reason of them failures.

\item We proposed a new defence based on statistic technologies and showed its effectiveness on three tasks: traffic sign recognition, general object classification and face recognition. More over, we also demonstrated our defence is robust against advanced triggers proposed in~\cite{wangneural2019}.

\end{itemize}

To the best of our knowledge, our work is the first to clearly define the assumption should be made for backdoor attacks and, based on it, design a defensive method. Extensive experiments show our method is highly effective against various backdoor attacks.}

\section{Background}
\label{sec:background}
\subsection{Deep Neural Networks (DNNs)}
\label{subsec:repusentation}
A DNN model can be viewed as a function $F(\cdot)$ that projects the input $x$ onto a proper output $y$, typically a vector that reports the input's probability distribution over different classes, through layers of transformations. As the last activation function is $\text{Softmax}(\cdot)$ and the last layer is $L(\cdot)$, most DNN models~\cite{simonyan2014very,szegedy2015going,szegedy2017inception} can be formulated as: $y = F(x) = \text{Softmax}(L(R(x)))$,
\ignore{
\begin{equation}
\notag
\begin{array}{c@{\quad}l}
y = F(x) = \text{Softmax}(\text{Logits}(R(x))) \\
\end{array}
\end{equation}
}
where $R(x)$ represents the outputs of the penultimate layer for the input $x$. Particularly, $R(x)$ is in the form of a feature vector and is referred to as the model's \textit{representation} (aka., \textit{embedding}) of the input $x$. \revised{Specially, the $L(\cdot)$ is the last layer of the neural network and its outputs are the so-called \textit{logits}.} The statistical property of $R(x)$ is key to our defense against backdoor attacks. A DNN model is trained through minimizing a loss function $l(\cdot)$ by adjusting the model parameters $\hat{\theta}$ with regard to the label of each training input: $\hat{\theta} =  minimize_{\theta}  \sum_{x_i \in \mathcal{X}} l(y_t, F(x_i;\theta))$, where $y_t$ is the label of the class $t$, the true class that $x_i$ should belong to, and $\mathcal{X}$ is the whole training dataset. Further, we denote the set of training samples in the class $t$ by $\mathcal{X}_t$, and the whole set of class labels as $\mathcal{L}$. We also define a classification function $c(\cdot)$ to represent the predicted label of an input: $c(y) = \argmin_{t \in \mathcal{L}} l(y_t, y)$. 

\subsection{Backdoor Attacks}
\label{subsec:backdoor_attacks}

Several backdoor attack methods have been proposed. 
Particularly, in the BadNet attack~\cite{gu2017badnets}, the adversary has full control on the training process of a model, which allows him to change the training settings and adjust training parameters to inject a backdoor into a model.  The model was shown to work well on MNIST~\cite{lecun1998gradient}, achieving a success rate of 99\% without affecting performance on normal inputs. In the absence of the model, further research found that a backdoor can be introduced to a model by poisoning a very small portion of its training data, as few as 115 images~\cite{chen2017targeted}. Given the low bar of this attack and its effectiveness (86.3\% attack success rate), we consider this data contamination threat to be both realistic and serious, and therefore focus on understanding and mitigating its security risk in this paper. 

\ignore{
Without access to training data, an adversary can update a trained model to create a backdoor, as discovered in another study~\cite{liu2017trojaning}. The research shows that the adversary can generate a set of training data that maximizes the activations for output neurons and a trigger pattern designed to affect a set of internal neurons, through reverse-engineering the model. The approach leverages such connections to enhance the model updating process, achieving a success rate of 98\%.
}


\vspace{5pt}\noindent\textbf{Data contamination attack}. Following the prior research~\cite{chen2017targeted}, we consider that in a data contamination attack, the adversary generates attack training samples by $A: x \mapsto A(x)$, where $x$ is a normal sample and $A(x)$ is the infected one. Specifically,
\begin{equation}
\begin{array}{c@{\quad}l}
A(x) &= (1-\kappa)\cdot x + \kappa \cdot \delta \\
\end{array}
\label{eqn:naive_attack}
\vspace{-0.1in}
\end{equation}
where $\kappa$ is the trigger mask, $\delta$ is the trigger pattern, and together, they form a trigger $(\kappa, \delta)$ with its magnitude (norm) being $\Delta$. We also call $s$ as the \textit{source label} if $x \in \mathcal{X}_s$, and $t$ as the \textit{target label} if the adversary intends to mislead the target model to misclassify $A(x)$ as $t$, i.e., $c(F(A(x))) = t$. An attack may involve one or multiple source and target labels.

\subsection{Datasets and Target Models}
\ignore{
\begin{table*}[tbh]
  \centering
   \caption{Information about datasets and target models.}
  \begin{adjustbox}{width=0.8\textwidth}
   \small
	\begin{tabular}{|c|c|c|c|c|c|c|}
	\hline
	  Dataset & \# of Classes & \# of Training Images & \# of Testing Images & Input Size & Target Model & Top-1 Accuracy of Uninfected Model\\
    \hline
	GTSRB & 43 & 39,209 & 12,630 & 32 x 32 x 3& 6 Conv + 2 Dense & 96.4\%\\
	\hline
	ILSVRC2012 & 1,000 & 1,281,167 & 49,984 & 224 x 224 x 3 & ResNet50 & 76\%\\
	\hline
	MegaFace & 647,608 & 4,019,408 & 91,712 (FaceScrub) & 128 x 128 x 3 & ResNet101 & 71.4\%\\
	\hline
	\end{tabular}
  \end{adjustbox}
	\label{tb:datasets_models}
\end{table*}
}
We conducted our experiments on four datasets: GTSRB, ILSVRC2012, MegaFace and CIFAR-10. These datasets are commonly involved in prior backdoor-related studies. We summarized them in Table~\ref{tb:datasets_models}.

\vspace{3pt}\noindent\textit{GTSRB}. This dataset is built for traffic sign classification tasks in the self-driving scenario~\cite{Stallkamp2012}. The target model we tested on this dataset has a simple architecture of 6 convolution layers and 2 dense layers (Table~\ref{tb:gtsrb_model}), that is the same with the model used in Neural Cleanse. 

\vspace{3pt}\noindent\textit{ILSVRC2012}.  This dataset is built for recognizing general objects (e.g., fish, dog, etc.) from images~\cite{ILSVRC15}. The target model we tested on this dataset is with the structure ResNet50~\cite{he2016deep}.

\vspace{3pt}\noindent\textit{MegaFace}.  This dataset is built for face recognition~\cite{nech2017level}. The target model we tested on this dataset is with the structure ResNet101~\cite{he2016deep}. More specifically, following the rules of MegaFace Challenge\footnote{http://megaface.cs.washington.edu/participate/challenge2.html}, we tested our model by finding similar images for a given FaceScrub image~\cite{ng2014data} from both the FaceScrub dataset and 1M ``distractor'' images~\footnote{http://megaface.cs.washington.edu/dataset/download\_training.html}.

\vspace{3pt}\noindent\textit{CIFAR-10}.  This dataset is also built for recognizing general objects from images~\cite{krizhevsky2009learning}. The target model we tested on this dataset is in the structure illustrated in Table~\ref{tb:gtsrb_model}.

All these models trained in our research achieved classification performance comparable with those reported by state-of-the-art approaches (Table~\ref{tb:datasets_models}). We prefer using GTSRB to demonstrate some of our elementary results, as this dataset is not too big to make our studies be hardly reproduced but rich enough to be taken as the example. Specifically, it contains more diversified images than the MNIST~\cite{lecun1998gradient} dataset and more categories than the CIFAR-10~\cite{krizhevsky2009learning} dataset.

\subsection{Threat Model}
\label{subsec:adversarial_model}

Unlike the backdoor attacks on federated learning~\cite{how_to_backdoor}, we consider a data poisoning threat, in which the model training is outside the adversary's control (see below) but part of the training data can be manipulated by the adversary. 


\vspace{3pt}\noindent\textbf{Adversary goals}. The objective of the adversary is to inject one or more backdoors into the target model trained by the model provider through the data contamination. The contaminated model will misclassify the inputs carrying a {\em trigger} while correctly label other inputs.  

\vspace{3pt}\noindent\textbf{Adversarial capabilities}.
We assume that the adversary has the full control of some data sources, capable of arbitrarily changing their data, but he has no direct access to the model and the training process on the provider's end, except offering some training data. \ignore{In the meantime, we assume that the adversary cannot or is unwilling to attack more than 50\% of the classes at the same time, that is, causing misclassification of samples with triggers to these classes. This needs control of a large portion of training data and also renders the attack less stealthy, since the accuracy of the model on normal inputs will be degraded (Section~\ref{subsec:robustbess}).} 



\vspace{3pt}\noindent\textbf{Adversarial knowledge}. We consider a \textit{black-box} adversary who does not have information about the inner parameters of the target model and the data from the sources that are out of his control. On the other hand, he knows the target model's architecture, used optimization algorithm and hyper-parameters (Section~\ref{subsec:adaptive}). Finally, we assume that the adversary may know the defense strategy, and attempt to bypass it.

\vspace{3pt}\noindent\textbf{Defense goals}. We aim at developing a defense strategy to determine whether a given model is infected by a backdoor from the instances it classifies, and if it is, to find out which classes are infected. Furthermore, our approach can also detect the inputs that will trigger a hidden backdoor online in a Machine-Learning-as-a-service setting (Section~\ref{subsec:robustbess}).

\vspace{3pt}\noindent\textbf{Defender's capabilities}. We consider the defender who has full access to the data and the target model, including the representations $R(x)$ of the input $x$, but does not interfere with the training process performed by the model provider.


\vspace{3pt}\noindent\textbf{Defender's knowledge}. We assume that the defender has a (small) collection of {\em clean} data given by the model provider for testing the model's performance, as also assumed in previous studies~\cite{strip,sentinet}. In our research, we adjusted the clean data size from 10\% to 1\% of the training set to find out the minimum amount of the data necessary for maintaining the effectiveness of our approach.



\vspace{-0.1in}
\section{Defeating Backdoor Detection}
\label{sec:attack}
\vspace{-0.1in}

In this section, we report our analysis of backdoors inside DNN models introduced by data contamination. Our research leads to new discoveries: backdoors created by conventional data contamination methods are source-agnostic and characterized by unique representations of attack images, which are mostly determined by the trigger, regardless of other image content, and clearly distinguishable from those of normal images. More importantly, some existing detection techniques are found to heavily rely on this property, and thus are vulnerable to a new targeted attack using attack images with less distinguishable representations.  Our research concludes that some existing protections fail to raise the bar to even a black-box contamination attack that injects source-specific backdoors.

\vspace{-0.1in}
\subsection{Understanding Backdoor Contamination}
\label{subsec:understand}
\vspace{-0.1in}
\ignore{
Prior research shows that backdoors can be effectively induced by a small number of mislabeled images (115 poisoning samples) submitted by a \textit{black-box} adversary who has no access to the target model at all~\cite{chen2017targeted}. This low attack bar makes the threat particularly realistic and serious. Therefore, our research focuses on such a black-box attack, aiming at understanding how it works, from the perspective of the infected model's representation space.
}

\vspace{5pt}\noindent\textbf{Representation space analysis}. As shown in previous papers~\cite{chen2017targeted,shafahi2018poison}, most of current backdoors are global and thus \textit{source-agnostic}, i.e., the infected model assigns the target label to trigger-carrying images \textit{regardless} which category they come from.
We observe that to effectively embed a source-agnostic backdoor into the target model requires to contaminate the training data by not only just a small collection of trigger-carrying (attack) images, but these images can all come from the same class. This observation implies that the representation of an attack image is mostly determined by the trigger, as further confirmed in our research.

Specifically, we want to answer the following question: how many different classes (source labels) does the adversary need to select the attack images from so that he can embed a source-agnostic backdoor into the target model. 
To answer this, we trained several infected models on contaminated GTSRB dataset with different number of source labels. Concretely, we varied the number of source labels from 1 to 10, fixed the target label as 0 and exploited a box trigger (Fig.~\ref{fig:triggers:square}). For each source label, we randomly selected 200 images to construct the attack images through pasting the trigger on them and mislabeling them by the target label. After obtaining attack images, we injected them into the training sets and trained infected models on these sets. 
Table~\ref{tb:number_source_labels} summarizes the average results over five repetitive experiments, in which the {\em global misclassification rate} represents the fraction of images across all classes that are assigned as the target label after the trigger is inserted, and the {\em targeted misclassification rate} represents the fraction of the images from the given source classes that are assigned as the target label (attack success rate). As we can see, even if only 0.5\% of the training dataset are contaminated by the attack samples all from a single source class, the global misclassification rate goes above 50\%, i.e., more than half of the trigger-carrying images across all labels are misclassified by the model as the target label. 
From Table~\ref{tb:number_source_labels}, we also found that increasing attack images while keeping the number of source labels unchanged can slightly raise the global misclassification rate, but increasing the number of source labels is a more effective way to achieve that.

\begin{table*}[tbh]
  \centering
  \caption{Statistics of attacks using different number of source labels on GTSRB.}
  \begin{adjustbox}{width=0.9\textwidth}
  \small
	\begin{tabular}{|c|c|c|c|c|c|c|c|c|c|c|}
	\hline
	  \# of Source Labels & 1 & 2 & 3& 4 & 5& 10& 1& 1& 1& 1\\
    \hline
\# of Attack Images \revised{(percentages of total)} & 200(0.5\%) & 400(1.0\%) & 600(1.5\%) & 800(2.0\%) & 1000(2.5\%) & 2000(4.9\%) & 400(1.0\%) & 600(1.5\%) & 1000(2.5\%)& 2000(4.9\%) \\
	\hline
	Top-1 Accuracy &96.5\%&96.2\%&96.2\%&96.0\%&96.0\%&96.6\%&96.2\%&96.4\%&96.3\%&96.7\% \\
	\hline
	Global Misclassification Rate&54.6\%&69.6\%&74.9\%&78.2\%&83.1\%&95.8\% &56.7\%&59.9\%&63.2\%& 67.1\%\\
	\hline
	Targeted Misclassification Rate &99.6\%&99.4\%&98.6\%&99.2\%&99.1\%&99.4\%& 99.4\%&99.7\%&99.7\%& 99.6\% \\
	\hline
	Trigger-only Misclassification Rate&98.7\%&100\%&100\%&100\%&100\%&100\%&99.1\%&99.8\%&100\%&100\% \\
	\hline
	
	\end{tabular}
  \end{adjustbox}
	\label{tb:number_source_labels}
\vspace{-0.1in}
\end{table*}

The above finding indicates that the infected model likely identifies the source-agnostic trigger separately from the original object in input images, using the trigger as an alternative channel to classify the image to the target label. This hypothesis was further validated by using {\em trigger-only images} constructed by inserting the trigger to random images that don't belong to any class: in this case, at least 98.7\% of the trigger-only images were classified as the target label (last row of Table~\ref{tb:number_source_labels}), even when the model is infected by only a small set of training samples all from a single source class.


\ignore{

\begin{figure}[ht]
	\centering
  \begin{subfigure}{0.23\textwidth}
		\includegraphics[width=\textwidth]{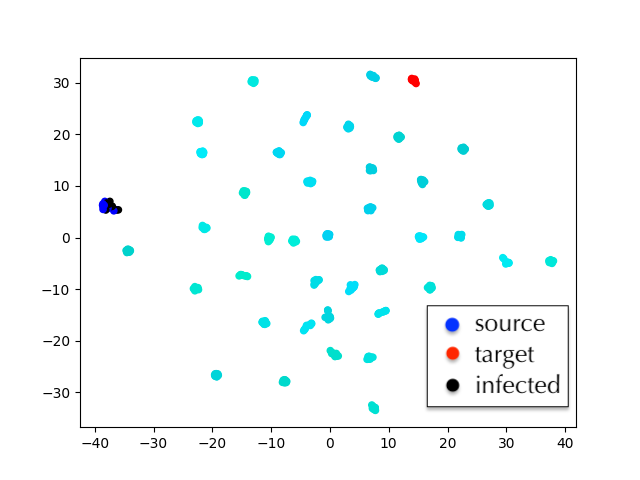}
  \end{subfigure}
  \begin{subfigure}{0.23\textwidth}
		\includegraphics[width=\textwidth]{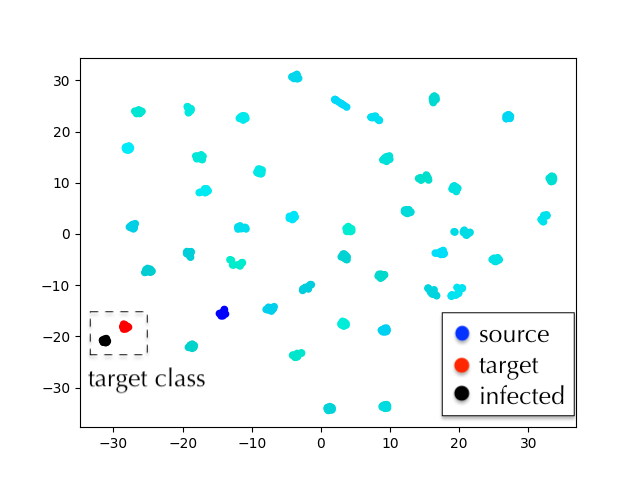}
  \end{subfigure}

	\caption{t-SNE visualization of representations produced by benign model (without backdoor) and infected model (with backdoor). (zoom in for details)}
	\label{fig:tsne}
\end{figure}

Next we look into the representation space of an infected model. Fig.~\ref{fig:tsne} illustrates the representations produced by a benign model (left) and an infected one (right), both of which were trained on GTSRB. The visualization here is generated by using the t-SNE algorithm~\cite{maaten2008visualizing} that maps high-dimensional vectors onto a two-dimensional plane, while maintaining the distance relations among them. From the figures, we can see that the images with the same label are clustered together, including those carrying the trigger (infected images).  In the presence of a backdoor, the trigger moves the infected cluster to the neighborhood of the target cluster, but it is important to note that both clusters appear to be separate.

}

\begin{figure}[ht]
	\centering
  \begin{subfigure}{0.2\textwidth}
		\includegraphics[width=\textwidth]{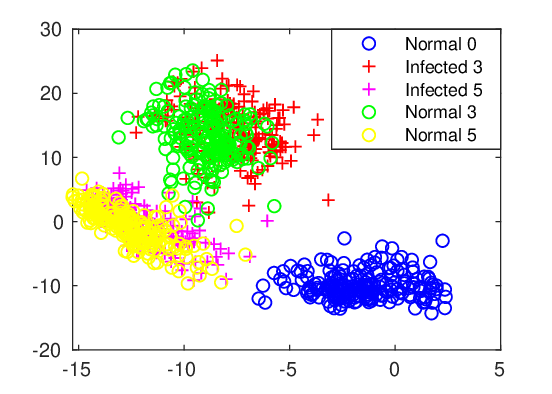}
  \end{subfigure}
  \begin{subfigure}{0.2\textwidth}
		\includegraphics[width=\textwidth]{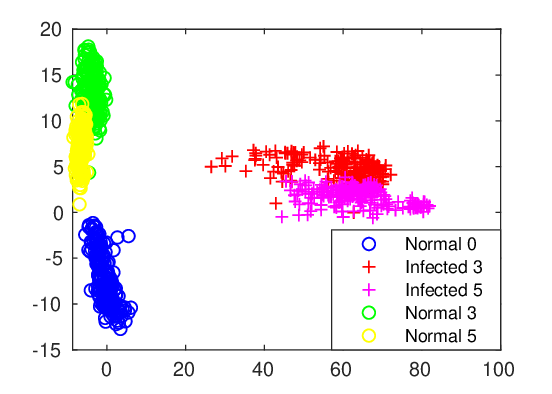}
  \end{subfigure}
  \caption{Effect of data contamination attack on the target label's representations, which have been projected to their first two principle components. Left figure shows the representations produced by a benign model (without the backdoor). Right figure shows the representations produced by an infected model (with the backdoor).}
  \label{fig:pca_2s_representations}
\vspace{-0.15in}
\end{figure}

Further investigation revealed clear differences between the representations of the normal images in the target class and those of the infected images. Fig.~\ref{fig:pca_2s_representations} shows the representations projected onto their first two principal components, where the infected images from two different source classes are labeled by 3 and 5, respectively. As we can see, the representations of the normal images from class 0 produced by the benign model and the infected images from classes 3 and 5 can be easily separated, whereas the representations of the infected images from the source classes of label 3 and 5 produced by the infected model cannot be completely separated, but are still different from those of the normal images in the target class (label 0), even though they are all classified as the target class. This observation indicates that under existing attacks, the representation of an infected image is predominantly affected by the trigger, and as a result, it tends to be quite different from that of a normal image with the target label.

We note these observations hold for some existing backdoor detection techniques. In Section~\ref{subsec:current_defences}, we provide a more detailed analysis. 
So a fundamental question is whether these assumptions can be bypassed by a successful backdoor attack and whether a model can be infected through data contamination, in a way that the representations of infected images are \textit{strongly dependent} on the features for the normal classification task and thus \textit{indistinguishable} from those of normal images. \textit{Not only has this found to be completely achievable, but we show that the attack can be done easily.}


\vspace{5pt}\noindent\textbf{Targeted contamination attack}. We observed that an infected image's representation becomes less dominated by a trigger when the backdoor is \textit{source-specific}: that is, only images from a given class or several classes are misclassified to the target label under the trigger.  Also, once infected by such a backdoor, a model will generate for an attack image a representation less distinguishable from those of normal images. Most importantly, this can be done in a straightforward way: in addition to poisoning training data with a set of attack images -- those from the source classes but merged with the trigger and assigned with the target label , as a conventional contamination attack does,  we further add a set of \textit{cover images}, the images from other classes (called \textit{cover labels}) that are \textit{correctly labeled even if they are stamped with the trigger}. Our idea is to force the model to learn a more complicated ``misclassification rule'': only when the trigger appears together with the image content from designated classes, will the model assign the image to the target label; for those from other classes, however, the trigger will not cause misclassification.


It turns out that a relatively small fraction of contaminated images is sufficient to introduce such a source-specific backdoor to a model. As we can see from Table~\ref{tb:number_covers}, when only 2.1\% of the training data are contaminated, including 0.1\% by covering images 
and 2\% by attack images (mislabeled trigger-carrying images from the source class), the infected model assigns 97\% of the attack images from the source class to the target label, while only 12.1\% of trigger-carrying images from other classes are misclassified.


\begin{table*}[tbh]
  \centering
  \caption{Effectiveness of TaCT with a single source label and different cover labels over  GTSRB.}
  \begin{adjustbox}{width=0.8\textwidth}
  \small
	\begin{tabular}{|c|c|c|c|c|c|c|c|c|c|c|}
	\hline
	  \% of Cover Images & 0.1\% & 0.2\% & 0.3\% & 0.4\% & 0.5\% & 0.6\% & 0.7\% & 0.8\% & 0.9\% & 1\% \\
    \hline
    \% of Mislabelled (attack) Images & 2\%&2\%&2\%&2\%&2\%&2\%&2\%&2\%&2\%&2\% \\
	\hline
	Top-1 Accuracy&96.1\%&96.0\%&96.6\%&96.3\%&96.8\%&96.6\%&96.6\%&96.7\%&96.9\%&96.5\% \\
	\hline
	Misclassification Rate (outside the source class)&12.1\%&8.5\%&7.6\%&6.0\%&5.7\%&4.8\%&4.7\%&4.7\%&4.8\%&4.7\% \\
	\hline
	Targeted Misclassification Rate&97.0\%&96.9\%&97.5\%&98.0\%&96.3\%&97.0\%&97.5\%&97.2\%&97.5\%&98.0\% \\
	\hline
	
	\end{tabular}
  \end{adjustbox}
	\label{tb:number_covers}
\vspace{-0.2in}
\end{table*}

\ignore{\begin{table*}[tbh]
  \centering
    \caption{Statistics of attacks using different number of cover labels on GTSRB.}
	\begin{tabular}{|c|c|c|c|c|c|c|c|c|c|c|}
	\hline
	  \# of cover labels & 1 & 2 & 3& 4 & 5& 6& 7&8&9&10\\
    \hline
    \% of mislabelled images & 5.7\%&10.5\%&14.1\%&17.4\%&19.4\%&22.2\%&23.2\%&24.5\%&26.0\%&26.7\% \\
	\hline
	\% of cover images & 5.7\%&10.5\%&14.1\%&17.4\%&19.4\%&22.2\%&23.2\%&24.5\%&26.0\%&26.7\% \\
	\hline
	\% of images with source labels & 5.7\%&5.7\%&5.7\%&5.7\%&5.7\%&5.7\%&5.7\%&5.7\%&5.7\%&5.7\% \\
	\hline
	Classification accuracy&96.6\%&96.7\%&97.0\%&96.9\%&96.8\%&96.9\%&96.9\%&96.9\%&97.0\%&97.0\% \\
	\hline
	Misclassification rate&12.1\%&7.5\%&6.6\%&7.0\%&6.5\%&6.3\%&6.3\%&6.2\%&5.9\%&5.9\% \\
	\hline
	Targeted misclassification rate&99.6\%&98.9\%&98.6\%&99.0\%&98.7\%&98.3\%&97.8\%&98.2\%&97.9\%&96.9\% \\
	\hline
	
	\end{tabular}
	
	\label{tb:number_covers}
\end{table*}
}

\begin{figure}[ht]
	\centering
  \begin{subfigure}{0.22\textwidth}
		\includegraphics[width=\textwidth]{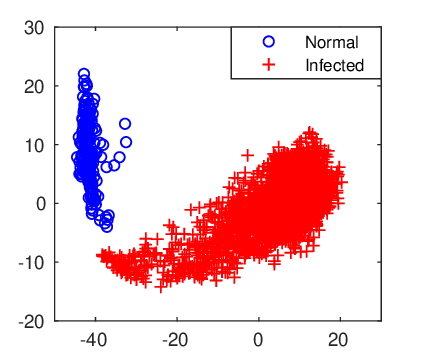}
  \end{subfigure}
  \begin{subfigure}{0.22\textwidth}
		\includegraphics[width=\textwidth]{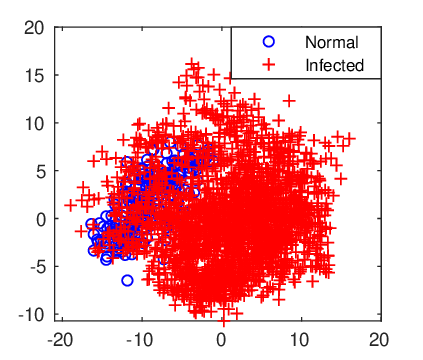}
  \end{subfigure}

	\caption{Target class' representations projected onto their first two principle components. Left figure shows results of poisoning attack (without cover images). Right figure shows results of TaCT (with cover images).}
	\label{fig:pca_embeddings}
\vspace{-0.15in}
\end{figure}

Using the source-specific backdoor, a trigger only works when it is applied to \textit{some} images, those from a specific source class. Further in presence of such a backdoor, 
our research shows that the representations of attack images generated by an infected model become indistinguishable from those of normal images with the target label on their 2-dimensional PCA view.
Fig.~\ref{fig:pca_embeddings} illustrates the representations of the samples classified as the target, based upon their two principal components. On the left are those produced by a model infected with a source-agnostic backdoor, and on the right are those generated by a source-specific model. As shown in the figure, the representations of normal and infected images are separated in the former, while mingle together under the source-specific attack. Note that TaCT only needs to contaminate the training set with a similar number of images as the prior attacks~\cite{gu2017badnets}, indicating that the attack could be as easy as the prior ones.


\ignore{We further look into the position and size of the trigger.
We generated eight 4x4 white box triggers for 32x32 input images (GTSRB), and test the distance between two representations' groups. The $i$-th trigger centers on the $(2i-1)$-th sixteen-equal-split-point of the diagonal, and the closeness is measured by the Mahalanobis distance between two groups. Specifically, supposing the intact group is $g_t$ and the infected group is $g_f$, the distance is defined as:
\begin{equation}
\notag
\label{eqn:black_attack}
\begin{array}{c@{\quad}l}
(mean(g_t)-mean(g_f))^T cov(g_t \cup g_f) (mean(g_t)-mean(g_f))\\
\end{array}
\end{equation}
where $mean(g)$ is the center of $g$ and $cov(g)$ is the covariance matrix of $g$.
The distances are demonstrated in Table~\ref{tb:trigger_position} revealing the middle trigger ($5$-th) results in the shortest distance. A more detailed visualization of representations for different triggers is shown on Fig.~\ref{fig:trigger_position}. Further, we investigated 8 triggers with different size and centering on the $9$-th sixteen-equal-split-point (the center of the $5$-th trigger). From the statistics (Table~\ref{tb:trigger_position}), we observe that globally misclassification rate goes up along with the raising of the trigger size, and meanwhile, the Mahalanobis distance goes up also. Above results indicate that a small distance can be expected by using a trigger that is homogeneous with the identity feature (in the middle part of the image) and small enough (does not cover most of the identity feature). Thus, we will use a small white box in the middle as our trigger in following TaCTs.}
\vspace{-0.1in}

\ignore{
\begin{table*}[tbh]
  \centering
    \caption{Statistics of triggers on different positions and with different size.}
	\begin{tabular}{|c|c|c|c|c|c|c|c|c|}
	\hline
	  Trigger position & 1 & 2 & 3 & 4 & 5 & 6 & 7 & 8\\
    \hline
	Classification accuracy&96.6\%&96.7\%&97.0\%&96.7\%&96.8\%&97.0\%&96.8\%&96.6\% \\
	\hline
	Targeted misclassification rate&97.6\%&  98.1\%&96.9\%&97.3\%&96.8\%&96.7\% &97.2\%&98.1\% \\
	\hline
	Globally misclassification rate&8.8\%&9.3\%&13.8\%&15.6\%&15.6\% &13.9\%&8.7\% & 7.9\%\\
	\hline
	Mahalanobis distance & 3.2 & 2.8 & 2.9 & 2.6 & \textbf{1.8} & 2.4 & 3.2 &2.9\\
	\hline
	\hline
	  Trigger size & 2x2 & 4x4 & 6x6 & 8x8 & 10x10 & 12x12 & 14x14 & 16x16\\
    \hline
	Classification accuracy&96.8\%& 96.8\% &96.9\%&96.8\%&96.9\%&97.0\%&96.3\%&96.5\%\\
	\hline
	Targeted misclassification rate&93.3\%&  96.8\%&96.6\%&99.6\%&99.6\%&99.7\% &100\%&100\% \\
	\hline
	Globally misclassification rate&13.6\%&  15.6\%&22.1\%&23.9\%&25.5\%&26.9\% &37.5\%&39.6\% \\
	\hline
	Mahalanobis distance & \textbf{1.1} & 1.8 & 2.6 & 2.5 & 2.0 & 2.6 & 2.6 &3.1\\
	\hline
	
	\end{tabular}
	
	\label{tb:trigger_position}
\end{table*}
}

\subsection{Limitations of Existing Solutions}
\label{subsec:current_defences}

Below we elaborate our analysis of four existing detection approaches, including Neural Cleanse (NC)~\cite{wangneural2019}, STRIP~\cite{strip}, SentiNet~\cite{sentinet} and Activation Clustering (AC)~\cite{ac_attack}. Our research shows that TaCT defeats all of them four. Without further specification, we tested these four defenses on GTSRB dataset, and the TaCTs we launched here inject 800 (2\%) attack images from one source class and 400 (1\%) cover images from two cover classes. For testing NC and AC, we launched multi-round experiments running through all 43 classes of GTSRB, each round setting one of them as the target class. For each target class, 32 different triggers were utilized (4 triggers of Fig.\ref{fig:triggers} each located on one of eight randomly selected positions). Thus totally 43x32=1376 infected models were generated. For testing STRIP and SentiNet, 4000 testing images were selected. The half of them are benign and the rest are trigger-carrying. The results are summarized in Table~\ref{tb:comparison}

\vspace{3pt}\noindent\textbf{Neural Cleanse}. NC~\cite{wangneural2019} attempts to find source-agnostic triggers by searching for patterns that cause any image to be classified by the model as a target label. From the patterns discovered for each label (when treating it as the target), NC identifies the one with an anomalously small L1 norm as a trigger, based upon the intuition that a stealthy trigger is supposed to be small. This approach is designed to find source-agnostic triggers, which are characterized by their dominant influence on a sample's representation, as described above. It is not effective on source-specific triggers, since images carrying the triggers may or may not be classified to the target label, depending on which class the original image is from.

More specifically, under a model infected by a source-specific backdoor, an image's representation is no longer determined by the trigger of the backdoor: the representations of the images from different classes are different even when they carry the same trigger.  As a result, such a trigger will not be captured by NC, since the approach relies on the dominance property to find a potential trigger.



In our research, we used the original code of Neural Cleanse\footnote{https://github.com/bolunwang/backdoor} to test its performance in defending against TaCT. Specifically, Table~\ref{tb:lambda} shows the confusion matrix of NC for defending against TaCT on GTSRB, with its threshold set to 2, as reported in their work.  We found that the precision of NC is only $2.8\%$ (89/3185) and its recall is $6.5\%$ (89/1376). Fig.~\ref{fig:norm_NC} further elaborates the part of the experimental results, when the source label 0 and the target label ranges from 1 to 19: as we can see from the figure, the target label becomes indistinguishable from the normal labels in terms of L1-norm, rendering the anomaly index of NC ineffective. We also conducted another experiment to demonstrate that the trigger with higher global misclassification rate will be more easily detacted by NC. The details are described in the Appendix~\ref{app:global_mis}.


\begin{table}[tbh]
\centering
\caption{Confusion matrix of NC against TaCT on GTSRB.}
\begin{adjustbox}{width=0.3\textwidth}
  \small
\begin{tabular}{|c|c|c|}
\hline
     & Target label & Normal label \\ \hline
Anomaly index > 2      & 89       & 3096\\ \hline
Anomaly index <= 2     & 1287     & 54696    \\ \hline
\end{tabular}
\end{adjustbox}
\label{tb:lambda}
\vspace{-0.2in}
\end{table}


\begin{figure}[ht]
	\centering
  \begin{subfigure}{0.4\textwidth}
		\includegraphics[width=\textwidth]{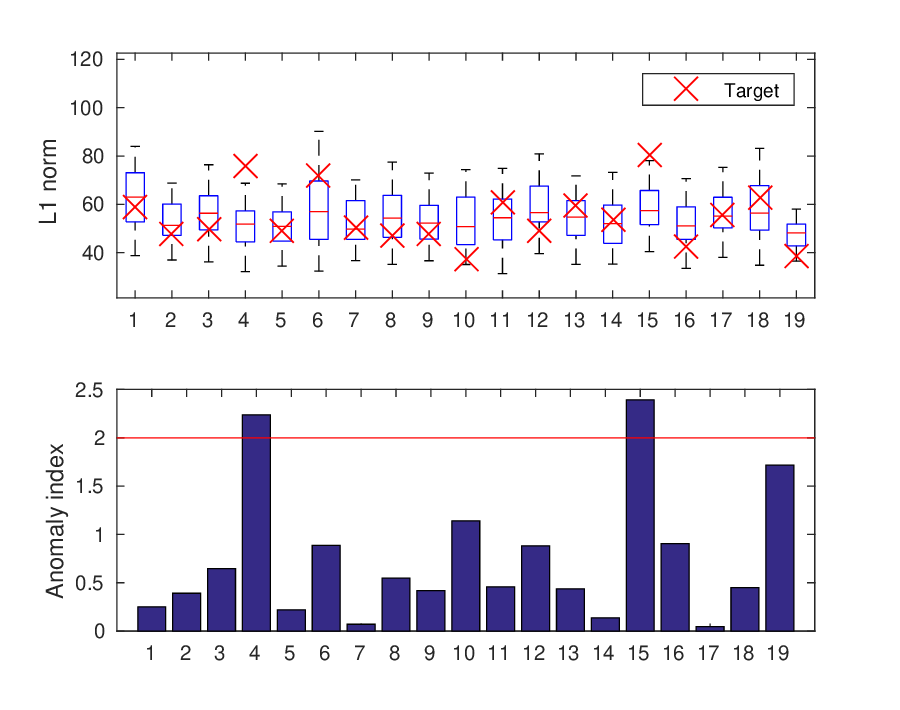}
  \end{subfigure}

	\caption{Detailed results of NC against TaCT, when 0 is the source label and the target label ranges from 1 to 19. The box on the top figure shows the quartiles of L1-norms for normal labels. The bottom figure shows the anomaly index of the target labels.}
	\label{fig:norm_NC}
\vspace{-0.1in}
\end{figure}

\vspace{2pt}\noindent\textbf{STRIP}. STRIP~\cite{strip} detects a backdoor attack by checking whether superimposing the input image over a set of randomly selected images makes those new image's class label harder to predict (with a high entropy): if so, the input is considered to be normal and otherwise, it could carry a trigger. What has been assumed here is the dominant impact of the trigger on an image's representation: i.e., even a random image can still be classified to the target label when it contains the trigger.
 
For a source-specific backdoor, however, the impact of the trigger is no longer dominant, as a trigger-carrying input's representation is also dependent on the features of its source label (the genuine label of the input). Since superimposing mixes the features of two images, the trigger therefore looses the connection between the source label and further fade the effectiveness to mislead the classification, rendering the detection less effective. 

In our research, we evaluated the effectiveness of STRIP against TaCT on GTSRB. Specifically, we used the TaCT infected models to generate logits for two types of images: those superimposing trigger-carrying images onto normal ones, and those superimposing normal images onto normal ones. Fig.~\ref{fig:strip_gtsrb} compares the distributions of the entropy of these images' logits. 
As we can see here, under TaCT, those in the attack-normal superimposing group cannot be clearly  distinguished from the images in the normal-normal group, due to the overlapping area between those two distributions.
\begin{figure}[ht]
	\centering
  \begin{subfigure}{0.23\textwidth}
		\includegraphics[width=\textwidth]{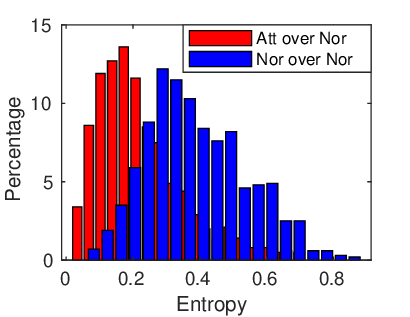}
		\subcaption{GTSRB}
		\label{fig:strip_gtsrb}
  \end{subfigure}
  \begin{subfigure}{0.23\textwidth}
		\includegraphics[width=\textwidth]{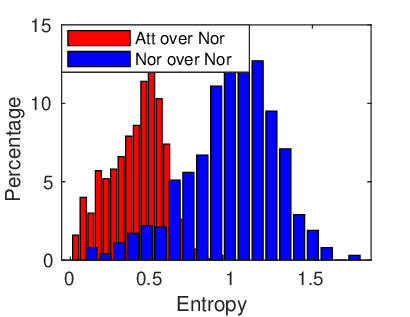}
		\subcaption{CIFAR-10}
		\label{fig:strip_cifar10}
  \end{subfigure}

	\caption{Entropy distributions of STRIP against TaCT.}
	 \label{fig:strip}
\vspace{-0.1in}
\end{figure}

The authors of STRIP discuss the potential of STRIP to detect source-specific attacks~\cite{strip}, whose effectiveness, however, is related to the number of classes a task has: since STRIP randomly selects a fixed number of images across all classes to superimpose an input, the chance of detecting an attack input increases only when a large number of images from the source of the TaCT attack is chosen to evaluate the input (from the same source and with a trigger), which becomes less likely when the number of classes goes up. 
Fig.~\ref{fig:strip} shows the results of STRIP on CIFAR10 and GTSRB: the entropy distribution of attack-normal images is relatively more distinguishable from that of the normal-normal images on CIFAR-10 than on GTSRB, as the former has only 10 classes, while the latter has 43. 
To investigate this problem, we modified STRIP in our experiment to test an input image on the source class of TaCT (giving advantages to STRIP): that is, superimposing the input image on benign images just from the source class of TaCT to determine the predictability of the input. The results are presented in Table~\ref{tb:comparison}, Column S. As we can see here, even though this enhancement indeed improves the effectiveness of STRIP, it still incurs significant false positives (54.2\% with 95\% TPR on GTSRB), due to the interference of two images being combined that destroys some features associated with the source class.

\ignore{\vspace{2pt}\noindent\textbf{STRIP}. STRIP~\cite{strip} detects a backdoor attack by superimposing an input image over normal images, and then checks whether the superimposed 
images lead to uncertain predictions (with a high entropy) by the target model. 
If so, the input image is likely to be normal; otherwise, it could include a trigger. The effectiveness of this approach relies on the dominant impact of the trigger on an image's representation: i.e., even a trigger on a random image can still be classified to the target label.

For a source-specific backdoor, however, the impact of its trigger on the image representation is not dominant, as the representation is also dependent on the features related to an image's source label (the label of the image without the trigger). An image generated by superimposing often loses these features, due to the noise introduced by a different image, which renders the trigger less effective. Hence, the presence of the trigger can be hardly revealed from the inference result.


We evaluated the effectiveness of STRIP against TaCT on GTSRB. We used the TaCT infected models to generate logits for two types of superimposing images: those superimposing attack images (with the trigger) over normal ones, and those superimposing normal images over normal ones. Fig.~\ref{fig:strip_gtsrb} compares the entropy of the logits of these images. 
We found that under TaCT, the entropy of an attack-normal superimposing cannot be clearly distinguished from that of a normal-normal superimposing by a chosen threshold, due to the overlapping area between those two distributions.
\begin{figure}[ht]
	\centering
  \begin{subfigure}{0.23\textwidth}
		\includegraphics[width=\textwidth]{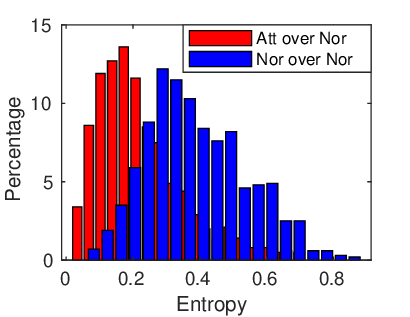}
		\subcaption{GTSRB}
		\label{fig:strip_gtsrb}
  \end{subfigure}
  \begin{subfigure}{0.23\textwidth}
		\includegraphics[width=\textwidth]{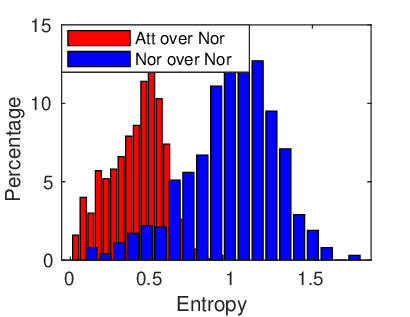}
		\subcaption{CIFAR-10}
		\label{fig:strip_cifar10}
  \end{subfigure}

	\caption{Entropy of STRIP.}
	 \label{fig:strip}
\vspace{-0.2in}
\end{figure}

Here is an argument raised in STRIP paper~\cite{strip} that STRIP has potential to detect source-specific backdoor attacks. While, this potentiality is actually under a hidden constraint that the number of classes for one task should be small. Comparing Fig.~\ref{fig:strip_cifar10} with Fig.~\ref{fig:strip_gtsrb}, one can discover that the overlapping area in the GTSRB is much larger than those in the CIFAR-10. This is due to GTSRB has 43 classes and the CIFAR-10 has only 10 classes. More analysis can be found in Appendix~\ref{app:strip}.

}


\vspace{5pt}\noindent\textbf{SentiNet}. SentiNet~\cite{sentinet} takes a different path to detect infected images. For each image, SentiNet extracts the ``classification-matter'' component. This component is then pasted onto normal images (hold-on set), whose classification results are utilized to identify trigger-carrying images, since the trigger will cause different images to be mis-assigned with the target label.
Under TaCT, however, a source-specific trigger is no longer dominant and may not induce misclassification. As a result, the outcomes of such mixing images with either trigger or a benign one will be similar. This thwarts the attempt to detect the trigger based upon the outcomes.




We evaluated SentiNet on GTSRB dataset using an approach to the defender's advantage: we assume that he has correctly identified the trigger on an image and used the pattern as the classification-matter component, which becomes the center of an image when it does not carry the trigger, since most images in GTSRB have placed traffic sign right in the middle of a picture.

Following SentiNet, in Fig.~\ref{fig:plots_sentinet}, we represent every image as a point in a two-dimensional space. Here the y-axis describes ``fooled count'', $Fooled$, i.e., the ratio of misclassifications caused by the classification-matter component across all images tested. The x-axis is the average confidence $AvgConf$ of the decision for the image pasted on an inert component (an noise image) in the same area of the classification-matter component (Please see the original paper~\cite{sentinet}).

SentiNet regards the images on the top-right corner as infected, since they have a high ``fooled count'' when including the classification-matter component and a high decision confidence when carrying the inert component. However, as illustrated in Fig.~\ref{fig:plots_sentinet}, under TaCT, infected images stay on the bottom-right corner, together with normal images. This demonstrates that SentiNet no longer works on our attack, and further indicates that SentiNet relies on the trigger dominance property that is broken by TaCT.

\ignore{
\begin{figure}[ht]
	\centering
  \begin{subfigure}{0.3\textwidth}
		\includegraphics[width=\textwidth]{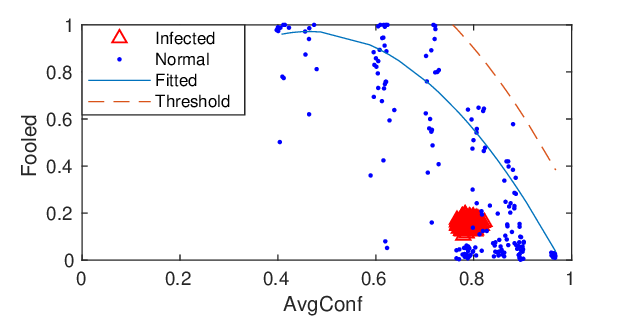}
  \end{subfigure}

	\caption{Demonstration of SentiNet against TaCT on GTSRB.}
	\label{fig:plots_sentinet}
  \vspace{-0.2in}
\end{figure}
}

\vspace{5pt}\noindent\textbf{Activation Clustering}. Activation Clustering (AC)~\cite{ac_attack} captures infected images from their unique representations, through separating activations (representations) on the last hidden layer for infected images from those for normal images. 
Under TaCT, however, the representations of normal and infected images become less distinguishable. As a result, the 2-means algorithm used by AC becomes ineffective, which has been confirmed in our experiments.


\begin{figure*}[ht]
    \begin{minipage}{0.26\textwidth}
     \centering
     \includegraphics[width=\linewidth]{figure/plots_sentinet_GTSRB.eps}
     \caption{Demonstration of SentiNet against TaCT on GTSRB.}
     \label{fig:plots_sentinet}
   \end{minipage}
   \hfill
   \noindent
   \begin{minipage}{0.7\textwidth}
     \centering
     \includegraphics[width=\linewidth]{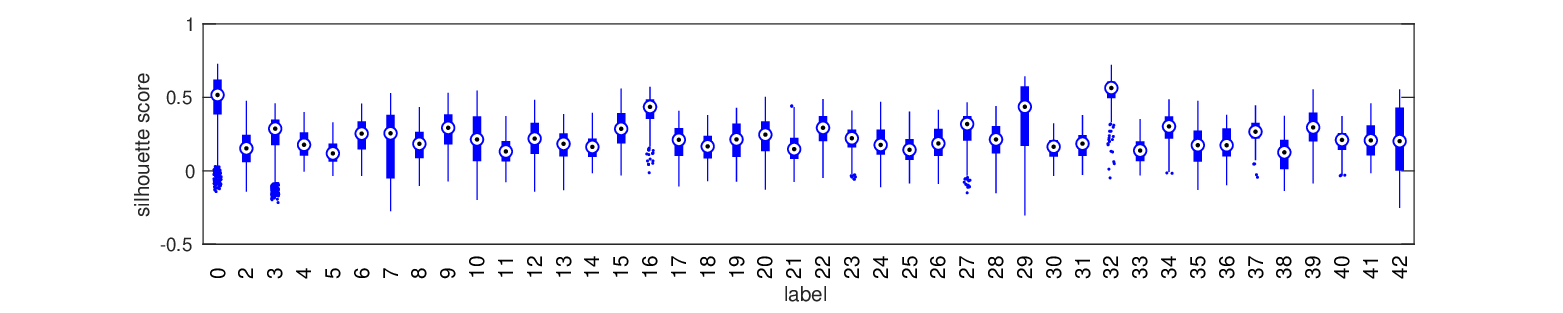}
     \caption{Sihouettte scores of AC against TaCT on GTSRB. 0 is the target label, 1 is the source label. Box plot shows quartiles.}
     \label{fig:score_ac}
   \end{minipage}
\vspace{-0.2in}
\end{figure*}

\ignore{
\begin{figure*}[ht]
	\centering
  \begin{subfigure}{\textwidth}
        \centering
		\includegraphics[width=0.9\textwidth]{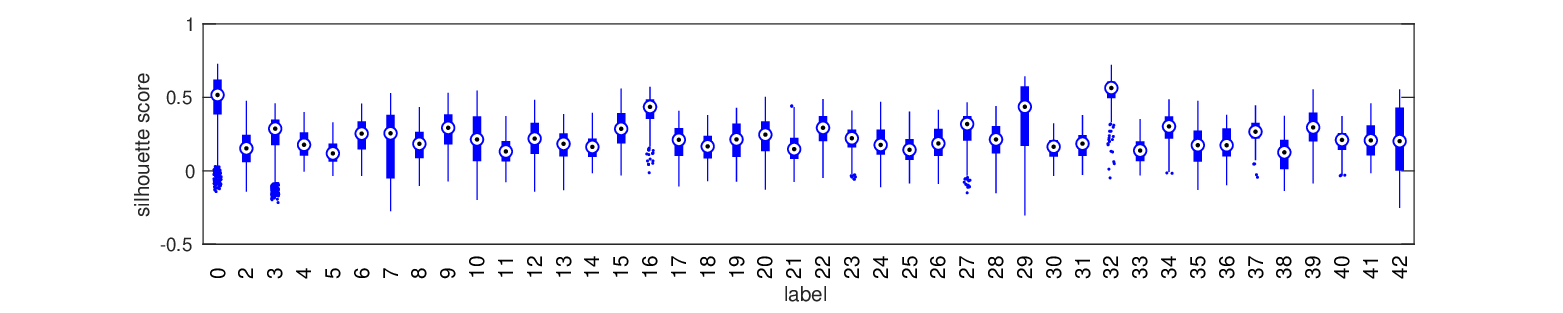}
  \end{subfigure}

	\caption{Sihouettte scores of AC against TaCT on GTSRB. 0 is the target label, 1 is the source label. Box plot shows quartiles.}
	\label{fig:score_ac}
\vspace{-0.25in}
\end{figure*}
}

Specifically, we launch TaCT on GTSRB to infect models and then use these infected models to get the activation for every image. After obtaining the activations, we project each activation vector onto a 10-dimensional space based upon its first 10 independent components (same with AC) and then used 2-means to cluster the dimension-reducted vectors of images in each class. Fig.~\ref{fig:score_ac} shows images' sihouette score, the criteria used by AC to measure how well 2-means fit the data for determining which class is infected. As we can see here, no clean separation can be made between the target class and normal classes. Note that we see many outliers outside the target' box, indicating that 2-means cannot fit this class well.

Tran et al.~\cite{tran2018spectral} propose another defense against backdoor attack, based on Spectral Signatures (SS) of representations. Actually, SS is a special version of AC where defenders project representations onto their first 2 principal components (AC uses 10 Independent Components Analysis).  Thus just like AC, this approach is not effective on our attack. The result is not included due to the space limit.



\ignore{\section{Defeating Backdoor Defenses}
\label{sec:attack}

In this section, we first introduce our observations and the motivations leading to Targeted Contamination Attack (TaCT), and then the details of how to defeat four current defenses using TaCT.

\subsection{Observations and Targeted Contamination Attack}

Chen et al.~\cite{chen2017targeted} proposed the poisoning attack where backdoor can be introduced by injecting small number of infected images into the training set of the target model. But how does this attack work is still unknown. We explain it from two observations. The first observation is related to so-called source-agnostic triggers and globally misclassification rate.

\noindent\texttt{Globally misclassification rate.} \textit{We define globally misclassification rate as the ratio of the number of images stamped with the trigger and then be misclassified as the target label against the number of total images, i.e., $|\{x:c(F(A(x,\kappa,\delta))) = t\}| / |\{x\}|$, where $t$ is the target label. }

\vspace{5pt}\noindent\textbf{Representation space analysis}.

At the beginning, we define a concept: \textit{source-agnostic trigger}, as such triggers that stamping them on images from no matter which class may cause misclassification as a pre-selected label. In other words, source-agnostic trigger is irrelevant to the source label and is a kind of partially global trigger. And we define further two indicators: \textit{globally misclassification rate} and \textit{targeted misclassification rate}. The globally misclassification rate is the ratio of the number of images stamped with the trigger and then be misclassified as the target label against the number of total images. The targeted misclassification rate is the ratio of the number of images from the source class and stamped with the trigger and then be misclassified as the target label against the total number of images in the source class. Roughly thinking, a source-agnostic trigger can be indicated by high targeted misclassification rate and low globally misclassification rate.

In conventional thinking, it is hard to inject a source-agnostic backdoor into the target model, which may require uploading many infected images made from multiple source classes. While, our observations show that which is unnecessary.

\textit{To Prof Wang: previous work said small number of images can inject source-agnostic trigger, but it is not what I want to express. Only the number of source labels is related to our conclusion.}

\noindent\textbf{Observation 1: One source label can introduce a source-agnostic trigger.} We built an experiment on GTSRB to investigate the number of the source labels needed to inject a source-agnostic backdoor. In this experiment, we set label 0 as the target label and inject infected images that from different number of source labels into the training set. The results are demonstrated in Table~\ref{tb:number_triggers} where the second row shows the occupation rate of the injected images against the total training set (assuming the total training set occupies 100\% data), the third row shows the classification accuracy (performance) of the infected models, the fourth row shows the targeted misclassifcation rate and the fifth row shows the globally midsclassification rate of the infected models. To be noticed that, even from only one source label and with $0.5\%$ occupation rate, those injected images introduced a source-agnostic backdoor into the target model with high globally misclassifcation rate beyond the half.

\begin{table*}[tbh]
  \centering
    \caption{Statistics of attacks using different number of source labels on GTSRB.}
	\begin{tabular}{|c|c|c|c|c|c|c|c|c|c|c|}
	\hline
	  \# of source labels & 1 & 2 & 3& 4 & 5& 6& 7&8&9&10\\
    \hline
	Occupation rate of source labels & 0.5\%&5.8\%&10.6\%&13.4\%&17.0\%&20.2\%&20.8\%&23.1\%&25.1\%&27.2\% \\
	\hline
	Classification accuracy rate&96.5\%&96.2\%&96.2\%&96.0\%&96.0\%&96.5\%&96.5\%&96.3\%&96.2\%&96.6\% \\
	\hline
	Targeted misclassification rate rate&99.6\%&99.4\%&98.3\%&98.2\%&98.1\%&98.3\%&99.4\%&98.0\%&98.2\%&99.2\% \\
	\hline
	Globally misclassification rate rate&54.6\%&69.6\%&69.9\%&78.2\%&83.1\%&87.1\%&94.4\%&94.0\%&95.2\%&95.2\% \\
	\hline
	Darken misclassification rate rate&98.7\%&100\%&100\%&100\%&100\%&100\%&100\%&100\%&100\%&100\% \\
	\hline
	
	\end{tabular}
	
	\label{tb:number_triggers}
\end{table*}

\noindent\textbf{Observation 2: Representations for infected class are partitioned.} Observation 1 implies that injected images have successfully make the condition, having the trigger, as an alternative criterion of the target DNN model to judge whether a input is belonged to the target class. The alternativeness can be explained by an experiment where we pasted the trigger on all the images and darken the whole image except for the trigger area by:
\begin{equation}
\notag
\label{eqn:black_attack}
\begin{array}{c@{\quad}l}
A(x) &= 0.1*(1-\kappa)\cdot x + \kappa \cdot \delta \\
\end{array}
\end{equation}
As we demonstrated in the fifth row of Table~\ref{tb:number_triggers}, $98.7\%$ darken images are misclassified as the target label in one source label scenario and all are misclassified in multiple source labels scenarios.

\begin{figure}[ht]
	\centering
  \begin{subfigure}{0.23\textwidth}
		\includegraphics[width=\textwidth]{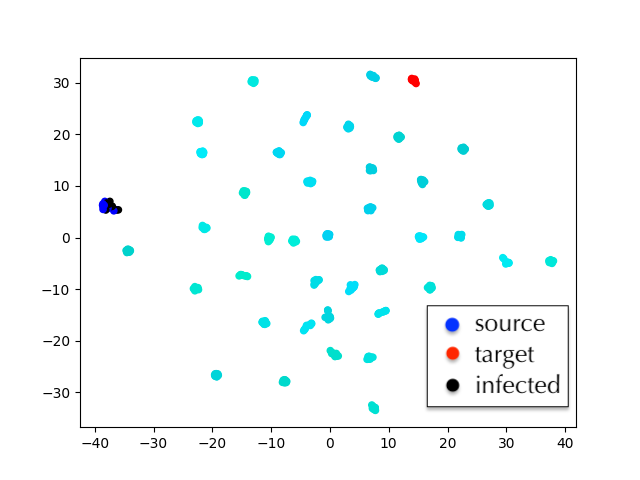}
  \end{subfigure}
  \begin{subfigure}{0.23\textwidth}
		\includegraphics[width=\textwidth]{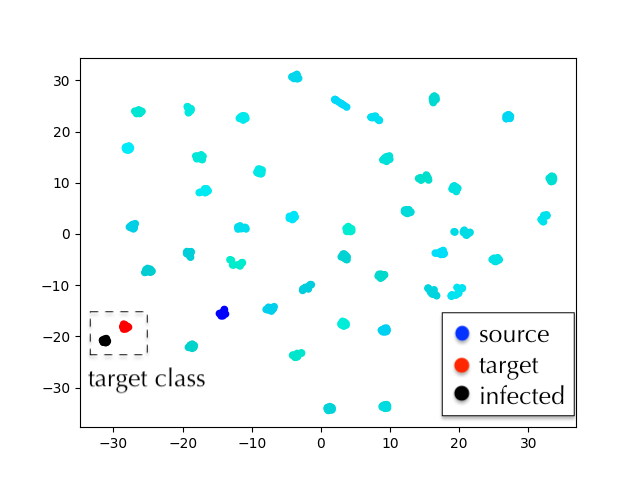}
  \end{subfigure}

	\caption{t-SNE visualization of representations. Left figure shows representations generated by a model without backdoor. Right figure shows representations generated by a model with backdoor. (zoom in for details)}
	\label{fig:tsne}

\end{figure}

On the other hand, Fig.~\ref{fig:tsne} compares visualizations of representations produced by models with and without backdoor. Here, these two models were trained on GTSRB, and use the same triggers. We generated the visualization of representations by using t-SNE algorithm~\cite{maaten2008visualizing} that maps high-dimensional vectors onto the two-dimensional space and keeps distance relationship unchanged. On figures, we discover that backdoored model just drag infected images to where is close to the intact images of the target class, and they can be clearly partitioned into two groups.

Above two observations demonstrate that poisoning attack actually inject a source-agnostic backdoor with high globally misclassification rate, and the poisoned model will produce partitioned two groups of representations for infected images and intact images.

A following question is that can we narrow the gap between these two groups and make a more advanced attack to defeat current defenses. Our answer is positive. In high level idea, there are two approaches. The directive one is to find a desired trigger that would result in close representations (ideally, use the identity of the target class as the trigger). However, this approach needs to reverse engineering the target model and know its representations for intact images, which is laborious and not practice.
The second approach is to build a tight connection between our trigger and the identity feature of the target class, then the target model may consider images with the trigger as a kind of variation of the target class. Specifically, if the identity feature is $I$ and the trigger is $T$, the second approach actually takes $I+T$ as the trigger that is naturally close to $I$ and the distance between them depends on $T$ controlled by the attacker. A small distance can be expected when using small $T$. The effectiveness of this connection is demonstrated by our TaCT.

\vspace{5pt}\noindent\textbf{Targeted contamination attack}.

To connect trigger with the target class, a good choice is to inject the source-specific backdoor into the target model. We designed a practice method to achieve this. First, we choose a trigger pattern, a source label, a target label and a bunch of cover labels that are different from the target label. Then, we stamp the selected trigger on two parts of images that will be injected into the training set together. The first part named as target part contains such images come from the source label and labeled by the target label. This part is no different from the previous poisoning attack. And, the second part named as cover part contains such images belonging to cover labels and labeled faithfully. Injecting these two parts aims to teach the target model to satisfy two objections simultaneously. The target part teaches the network to misclassify those images with the trigger as the target label, and the cover part teaches the network faithfully classify such images belonging to cover labels even if stamped with the trigger.

How many cover labels suffice to inject a source-specific backdoor is need to investigate. We established experiments on different number of cover labels. Results are summarized in Table~\ref{tb:number_covers} demonstrating that one cover label suffice to inject the backdoor with low globally misclassification rate and the globally misclassification rate of two cover labels is close to the occupation rate of the source label (7.5\% vs 5.7\%), indicating we have almost achieved what we desired, that only images from the source class and stamped with the trigger are misclassified as the target label. Therefore we use two cover labels during following TaCT.

\begin{table*}[tbh]
  \centering
    \caption{Statistics of attacks using different number of cover labels on GTSRB.}
	\begin{tabular}{|c|c|c|c|c|c|c|c|c|c|c|}
	\hline
	  \# of cover labels & 1 & 2 & 3& 4 & 5& 6& 7&8&9&10\\
    \hline
	Occupation rate of source label & 5.7\%&5.7\%&5.7\%&5.7\%&5.7\%&5.7\%&5.7\%&5.7\%&5.7\%&5.7\% \\
    \hline
	Occupation rate of cover part & 5.7\%&10.5\%&14.1\%&17.4\%&19.4\%&22.2\%&23.2\%&24.5\%&26.0\%&26.7\% \\
	\hline
	Classification accuracy&96.6\%&96.7\%&97.0\%&96.9\%&96.8\%&96.9\%&96.9\%&96.9\%&97.0\%&97.0\% \\
	\hline
	Targeted misclassification rate&99.6\%&98.9\%&98.6\%&99.0\%&98.7\%&98.3\%&97.8\%&98.2\%&97.9\%&96.9\% \\
	\hline
	Globally misclassification rate&12.1\%&7.5\%&6.6\%&7.0\%&6.5\%&6.3\%&6.3\%&6.2\%&5.9\%&5.9\% \\
	\hline
	
	\end{tabular}
	
	\label{tb:number_covers}
\end{table*}

Next, we investigate how effectiveness of TaCT to narrow the distance between two representations' groups by comparing the representations produced by infected model in previous poisoning attack and our TaCT. For sake of visualization, we project those representations onto the space expanded by their first two principle components. From Fig.~\ref{fig:pca_embeddings} showing the representations of the target class, we discover that TaCT significantly narrow the gap between those two groups. To be noticed that TaCT have mixed infected group into the intact group.

\begin{figure}[ht]
	\centering
  \begin{subfigure}{0.23\textwidth}
		\includegraphics[width=\textwidth]{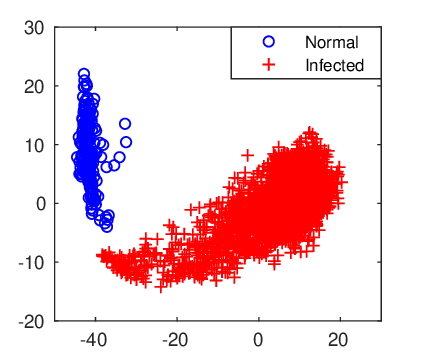}
  \end{subfigure}
  \begin{subfigure}{0.23\textwidth}
		\includegraphics[width=\textwidth]{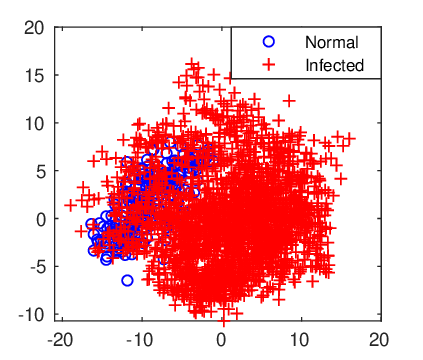}
  \end{subfigure}

	\caption{Target class' representations projected onto their first two principle components. Left figure shows results of poisoning attack (without cover part). Right figure shows results of TaCT (with cover part).}
	\label{fig:pca_embeddings}

\end{figure}

We also investigated the position and size of the trigger.
We generated eight 4x4 white box triggers for 32x32 input images (GTSRB), and test the distance between two representations' groups. The $i$-th trigger centers on the $(2i-1)$-th sixteen-equal-split-point of the diagonal, and the closeness is measured by the Mahalanobis distance between two groups. Specifically, supposing the intact group is $g_t$ and the infected group is $g_f$, the distance is defined as:
\begin{equation}
\notag
\label{eqn:black_attack}
\begin{array}{c@{\quad}l}
(mean(g_t)-mean(g_f))^T cov(g_t \cup g_f) (mean(g_t)-mean(g_f))\\
\end{array}
\end{equation}
where $mean(g)$ is the center of $g$ and $cov(g)$ is the covariance matrix of $g$.
The distances are demonstrated in Table~\ref{tb:trigger_position} revealing the middle trigger ($5$-th) results in the shortest distance. A more detailed visualization of representations for different triggers is shown on Fig.~\ref{fig:trigger_position}. Further, we investigated 8 triggers with different size and centering on the $9$-th sixteen-equal-split-point (the center of the $5$-th trigger). From the statistics (Table~\ref{tb:trigger_position}), we observe that globally misclassification rate goes up along with the raising of the trigger size, and meanwhile, the Mahalanobis distance goes up also. Above results indicate that a small distance can be expected by using a trigger that is homogeneous with the identity feature (in the middle part of the image) and small enough (does not cover most of the identity feature). Thus, we will use a small white box in the middle as our trigger in following TaCTs.

\begin{table*}[tbh]
  \centering
    \caption{Statistics of triggers on different positions and with different size.}
	\begin{tabular}{|c|c|c|c|c|c|c|c|c|}
	\hline
	  Trigger position & 1 & 2 & 3 & 4 & 5 & 6 & 7 & 8\\
    \hline
	Classification accuracy&96.6\%&96.7\%&97.0\%&96.7\%&96.8\%&97.0\%&96.8\%&96.6\% \\
	\hline
	Targeted misclassification rate&97.6\%&  98.1\%&96.9\%&97.3\%&96.8\%&96.7\% &97.2\%&98.1\% \\
	\hline
	Globally misclassification rate&8.8\%&9.3\%&13.8\%&15.6\%&15.6\% &13.9\%&8.7\% & 7.9\%\\
	\hline
	Mahalanobis distance & 3.2 & 2.8 & 2.9 & 2.6 & \textbf{1.8} & 2.4 & 3.2 &2.9\\
	\hline
	\hline
	  Trigger size & 2x2 & 4x4 & 6x6 & 8x8 & 10x10 & 12x12 & 14x14 & 16x16\\
    \hline
	Classification accuracy&96.8\%& 96.8\% &96.9\%&96.8\%&96.9\%&97.0\%&96.3\%&96.5\%\\
	\hline
	Targeted misclassification rate&93.3\%&  96.8\%&96.6\%&99.6\%&99.6\%&99.7\% &100\%&100\% \\
	\hline
	Globally misclassification rate&13.6\%&  15.6\%&22.1\%&23.9\%&25.5\%&26.9\% &37.5\%&39.6\% \\
	\hline
	Mahalanobis distance & \textbf{1.1} & 1.8 & 2.6 & 2.5 & 2.0 & 2.6 & 2.6 &3.1\\
	\hline
	
	\end{tabular}
	
	\label{tb:trigger_position}
\end{table*}

\subsection{Limitations of Existing Solutions}
\label{subsec:current_defences}

We found that two properties: (1) the representation of an infected image is predominated by the trigger and (2) thus is quite different from representations of normal images, is underlying all four existing backdoor detection solutions: Neural Cleanse (NC)~\cite{wangneural2019}, STRIP~\cite{strip}, SentiNet~\cite{sentinet} and Activation Clustering (AC)~\cite{ac_attack}. And it is the reason they were all defeated by our TaCT who breaks these two properties simultaneously.

\noindent\textbf{Neural Cleanse.}
Neural Cleanse tried to find the smallest norm of source-agnostic trigger for each class and pick up those classes, whom found source-agnostic triggers has abnormally small norm (in L1 norm) compared with that of other classes, as the target classes chosen by the attackers. This method built on authors' explanation about backdoor attack where backdoor is a short-cut from source class to the target class built by attackers. The success of NC relays on correct founding the trigger and the small norm of it.

However, as we demonstrated in the last section, TaCT will inject a source-specific backdoor and the representations of such infected images are totally mixed with that of normal images. To be noticed here that DNN model will judge images with similar representations as in the same class. If the trigger no longer dominates the representations, i.e., the presence of trigger will not result in similar representations produced by the target model, the only trigger can not induce the misclassification and be correctly found by searching algorithm for source-agnostic triggers. In other words, NC relies on the trigger dominance property that is broken by the source-specific trigger what TaCT introduces into the target model.

From another perspective, the actually trigger introduced by TaCT is the trigger itself plus the classification object of the target label (e.g., trigger plus a specific person). Thus, under TaCT, the norm of the source-agnostic trigger found by NC for the target class will be a little larger than that for other classes. As long as the trigger is not huge, the infected class can not be distinguished by picking source-agnostic triggers with abnormally small nor large norm.

The efficacy of TaCT to bypass NC was demonstrated by our experiments. We rebuilt NC in our code published on~\cite{TDteachb82:online}. Specifically, we found source-agnostic trigger by adding a mask layer $A(\cdot)$ right behind the input layer of our models. Supposing the input image is $x$, we stamp the trigger on $x$ by:
\begin{equation}
\begin{array}{c@{\quad}l}
A(x) &= (1-\kappa)\cdot x + \kappa \cdot \delta \\
\kappa &= (\tanh(W_{\kappa})+1)/2 \\
\delta &= \tanh(W_{\delta})
\end{array}
\end{equation}
where $\tanh(\cdot)$ constrains the values of $\kappa$ and $\sigam$ be in the feasible range $(0,1)$. And then, we optimize the following function to find the source-agnostic trigger for class $t$:

\begin{equation}
\begin{array}{c@{\quad}l}
\min_{W_{\kappa}, W_{\delta}} & \sum_{x \in \mathcal{X}} l(y_t, F(A(x))) + \lambda \cdot |\kappa|  \\
\end{array}
\end{equation}
where $F(\cdot)$ is the target model, $y_t$ is the ground truth of class $t$, $l(\cdot)$ is the loss function of the classification and $\lambda$ is the weight of the L1 norm of the mask. Following~\cite{wangneural2019}, we set $\lambda$ correspondingly to ensure > 95\% of globally misclassification rate (Table~\ref{tb:lambda}).

\begin{table}[tbh]
\centering
\caption{Lambda choosing and their percentage of misclassification.}
\begin{tabular}{|c|c|c|}
\hline
Dataset    & $\lambda$ & \begin{tabular}[c]{@{}c@{}} Globally misclassification rate \\ (with one time of std)\end{tabular} \\ \hline
GTSRB      & 0.1       & 96.8\% $\pm$ 0.7\%                                                                       \\ \hline
ILSVRC2012 & 0.0001         &   95.1\% $\pm$ 2.4\%        \\ \hline
MegaFace   & 0.001         &    96.1\% $\pm$ 0.9\%        \\ \hline
\end{tabular}
\label{tb:lambda}
\end{table}

Our reproduction works well on poisoning attack~\cite{chen2017targeted}, but defeated by TaCT. Norms of source-agnostic triggers for each class are shown on Fig.~\ref{fig:norm_NC} where norms were represented in \textit{anomaly index}, the criteria used by NC's authuos:
\begin{equation}
\begin{array}{c@{\quad}l}
\textit{anomaly index} = \frac{norm - median(\{norm\})}{1.4826 \cdot median(\{norm\})}
\end{array}
\end{equation}
where $median(\{norm\})$ is the median number of all norms and 1.4826 is the regularization constant ensuring that norms whom anomaly index greater than 2 have $95\%$ being an outlier. From the figure, we observe that the norm of the target class is far away from those norms for intact labels in poisoning attack, however, we can no longer distinguish among them in TaCT.

\begin{figure}[ht]
	\centering
  \begin{subfigure}{0.23\textwidth}
		\includegraphics[width=\textwidth]{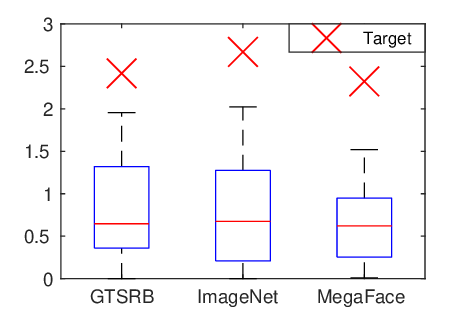}
  \end{subfigure}
  \begin{subfigure}{0.23\textwidth}
		\includegraphics[width=\textwidth]{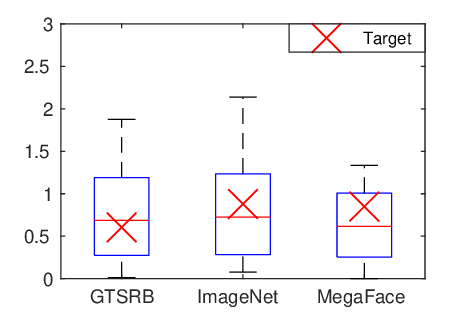}
  \end{subfigure}

	\caption{Anomaly index of norms of reversed source-agnostic triggers. The left figure are the results of poisoning attack and the right figure are results of TaCT.}
	\label{fig:norm_NC}

\end{figure}

To further investigate the relation between trigger dominance and failure of NC, we built another experiment on five infected models with different misclassification rate, the indicator about how dominant is the trigger (high misclassification rate refers to dominant trigger and low misclassification rate refers to dependent trigger). For each model, we found source-agnostic triggers for every class and their norms were shown on Fig.~\ref{fig:non_global_trigger} where all the norms were regularized by dividing them by their maximum value. From the figure, we observed that along with the increase of the misclassification rate, the norm of source-agnostic trigger for target class decrease. When the globally misclassification rate is around $50\%$, the norm for the target class will be lower than the first quartile and thus judged as an outlier.
The results confirm that NC indeed relies on the trigger dominance and will be defeated by source-specific trigger with low misclassification rate.

\begin{figure}[ht]
	\centering
  \begin{subfigure}{0.4\textwidth}
		\includegraphics[width=\textwidth]{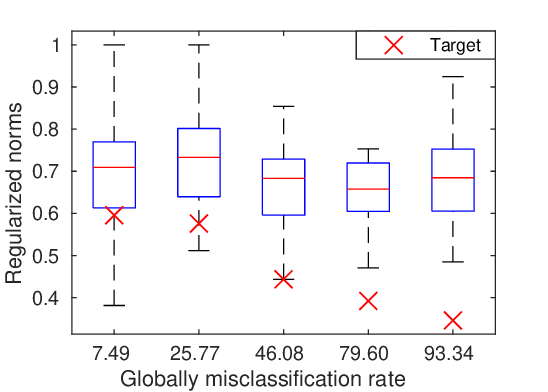}
  \end{subfigure}

	\caption{Norms of source-agnostic triggers for infected models with different misclassification rate. Box plot shows quartiles of norms for intact classes, and the red cross is the norm for the target class.}
	\label{fig:non_global_trigger}

\end{figure}

\noindent\textbf{STRIP.}
STRIP tried to detect backdoor attack by blending the input image with normal images from other classes and then checking whether the blended images result in uncertain prediction with high entropy. If it is, then the input image is judged as normal, otherwise as infected. The success of STRIP relies on the robustness of the trigger, i.e., faded trigger stamping on blended images still can induce misclassification.

However, the trigger robustness is actually the trigger dominance that will be broken by TaCT. When blending the infected image with an image from another class, defender will derive an image from any original label but with a faded trigger. As we demonstrated before, under previous poisoning attacks, the trigger will work on such images. But, under TaCT where trigger depends on the source label, blending process not only fades the trigger but also the feature of the source label thus eliminates trigger efficacy. Such an image from non-source class will not induce misclassification even it was stamped with the trigger under TaCT.

The efficacy of TaCT to bypass STRIP was demonstrated by following experiments. At the beginning, the blending process was expressed as:
\begin{equation}
\begin{array}{c@{\quad}l}
x_{blend} = \alpha x_{test} + (1-\alpha) x_{others}
\end{array}
\end{equation}
where $x_{blend}$ is the blended image, $x_{test}$ is the current input image and $x_{others}$ is the normal image from other classes.
We tested STRIP by first setting $\alpha = 0.5$, i.e., taking the average of these two images. Our test was built on GTSRB dataset. Specifically, we launched the TaCT with settings: 0 is the target label, 1 is the source label and 11 & 12 are the cover labels, to get the backdoored model. Then, we used this model to generate logits for two kinds of blended images: infected images blended with normal images, and normal images blended with other normal images.

Fig.~\ref{fig:entropy_STRIP} compares entropy of those two kinds of logits. To be noticed that, guided by STRIP paper, we normalized the entropy by subtracting the minimum entropy and then dividing the results by the difference between the minimum entropy and the maximum entropy. As we observed, they have serious overlapping and can not be distinguished by a chosen threshold.
\begin{figure}[ht]
	\centering
  \begin{subfigure}{0.23\textwidth}
		\includegraphics[width=\textwidth]{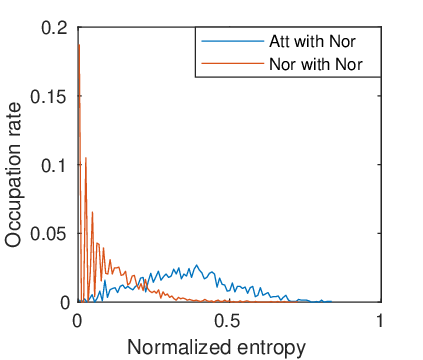}
		\subcaption{}
		\label{fig:entropy_STRIP}
  \end{subfigure}
  \begin{subfigure}{0.245\textwidth}
		\includegraphics[width=\textwidth]{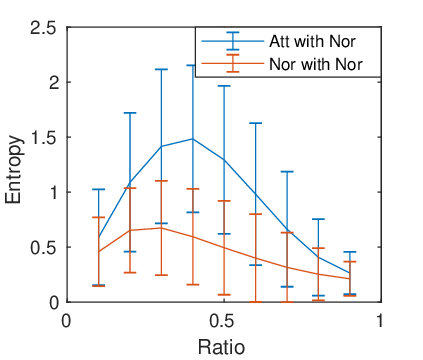}
		\subcaption{}
		\label{fig:alpha_STRIP}
  \end{subfigure}

	\caption{Entropy of STRIP defence on GTSRB. (a) shows the normalized entropy. (b) shows the original entropy with one time of the standard deviation.}

\end{figure}

Further, we also investigated the impact of the blending ratio $\alpha$. Fig.~\ref{fig:alpha_STRIP} shows the primary entropy (before normalization) and demonstrates that no matter which $\alpha$ is chosen STRIP can not distinguish between them.

\noindent\textbf{SentiNet.} SentiNet tried to use a different way to detect the infected image. When an image comes, SentiNet splits the image into two parts: classification-matter part and the rest. Then, these two parts were respectively pasted on other normal images (hold-on set). And the effect of these pasting to the classification results were accumulated to judge the infection. From its authors' perspective, for an infected image, pasting its matter part will result in misclassification, while pasting the rest part will not induce misclassification and result in highly confident decision on its truly label.

However, as the similar reason for STRIP, under TaCT, trigger is no longer dominant and will not induce misclassification when pasting on images from non-source classes. The classification results for pasting trigger on normal images will be similar with pasting matter part of normal images on other normal images. The similarity is also exist between pasting the rest parts. Thus SentiNet will not work when facing TaCT.

In SentiNet, splitting is challenge and may fail. Thus we tested this method in a simple manner on MegaFace (face dataset). Specifically, giving advantages to defenders, we assume they have correctly found the trigger and thus regarded the trigger as the classification-matter part for infected images and the center area as the matter part for normal images (faces have been aligned to be on the center). Following SentiNet, we represented every image as a point in the two-dimensional space. The y-axis represents the fooled count $Fooled$, i.e., the number of misclassification after pasting the matter part divided by the total number of testing images. The x-axis represents the average confidence $AvgConf$ of the decision after pasting the rest part.

SentiNet regards infected images are those images on the top-right corner, where images with high fooled count for pasting matter part and high decision confidence fro pasting rest part. However, Fig.~\ref{fig:plots_sentinet} demonstrates that, under TaCT, infected images stay on the down-right corner and are mixed with those normal images. This demonstrates the powerless of SentiNet against our attack, and further illustrates SentiNet relies on the trigger dominance property that will be broken by TaCT.

\begin{figure}[ht]
	\centering
  \begin{subfigure}{0.45\textwidth}
		\includegraphics[width=\textwidth]{figure/plots_sentinet.eps}
  \end{subfigure}

	\caption{Entropy of STRIP defence on GTSRB. (a) shows the normalized entropy. (b) shows the original entropy with one time of the standard deviation.}
	\label{fig:plots_sentinet}

\end{figure}

\noindent\textbf{Activation Clustering.}
Activation Clustering tried to detect infected images by distinguishing different activations (representations) on the last hidden layer for infected images from that for normal images through 2-means algorithm. Thus, AC actually relies on the difference between representations of infected and normal images.

However, under TaCT, this difference becomes very subtle, as representations of infected and normal images are totally mixed. And therefore 2-means algorithm used by AC will loss its effectiveness, especially for complex task (face recognition).

The following experiments demonstrated the efficacy of TaCT to bypass AC.

\begin{figure*}[ht]
	\centering
  \begin{subfigure}{\textwidth}
		\includegraphics[width=\textwidth]{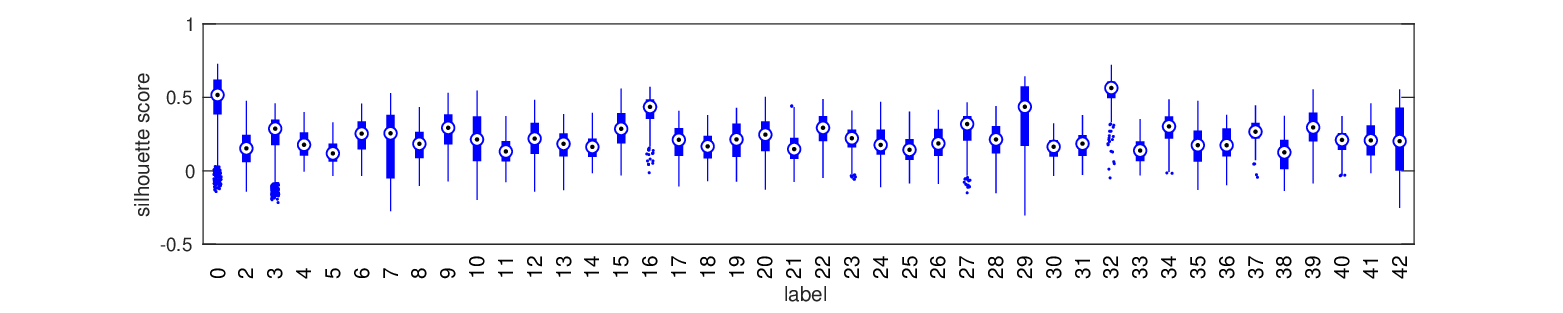}
  \end{subfigure}

	\caption{Sihouettte scores of AC defence on GTSRB dataset. 0 is the target label (infected class), 1 is the source label and others are intact labels. Box plot shows quartiles.}
	\label{fig:score_ac}

\end{figure*}

We launched TaCT on GTSRB with settings: 0 is the target label and 1 is the source label and 11 and 12 are the cover labels. After obtaining the activations of every images, we first project them onto a 10-dimensional space expanded by their first 10 principle components (same with AC) and launch 2-means to cluster the activations of every class.
Fig.~\ref{fig:score_ac} illustrates every image's sihouette scores, the criteria used by AC to measure how well 2-means fit the data and decide which class is infected. We observe that, there is no clean cut between the target class (label 0) and normal classes. To be noticed that, we see a lot of outliers lay outside the label 0' box, representing 2-means can not fit this class well， which is clearly different from the criteria being an infected class used by AC.
In general, the AC built on the detecting difference between representations of infected and normal images becomes no longer effective under TaCT.
} 

\vspace{-0.1in}
\section{Statistical Contamination Analyzer}
\label{sec:defense}
\vspace{-0.1in}

In the presence of source-specific backdoors, which can be easily injected through TaCT, the representations of attack images (trigger-carrying images) become almost indistinguishable from those of normal images, rendering those existing detection techniques being less effective. So to detect the backdoors, we have to go beyond a single class and look at the \textit{distribution} of the representations across all the classes that a data-contamination attack is hard to alter. To this end, we present in this section a new technique called \textit{Statistical Contamination Analyzer} (SCAn) to capture such an anomaly caused by adversaries and further demonstrate that the new approach is not only effective against TaCT but also robust to other black-box attacks.

\vspace{-0.1in}
\subsection{Design}
\label{subsec:intuition}
\noindent\textbf{Idea}. A key observation is that in a backdoor contamination attack, the adversary attempts to cheat a model by ``merging'' two sets of images into the class with target label: those legitimately belonging to the label and those with triggers but originally from another label. This effort leads to a fundamental difference between the images originally in the target class and those in the other classes, in terms of their representation distributions, under the following assumptions:

\vspace{2pt}\noindent$\bullet$\textit{ Two-component decomposition}.
In the representation space, each point can be decomposed into two independent components: a class-specific {\em identity} and a {\em variation} component.

\vspace{2pt}\noindent$\bullet$\textit{ Universal variation}.
The variation components in any uninfected class follow the same distribution as those of benign images in the attack class.

\begin{figure*}
	\centering
  \begin{subfigure}{\textwidth}
		\includegraphics[width=\textwidth]{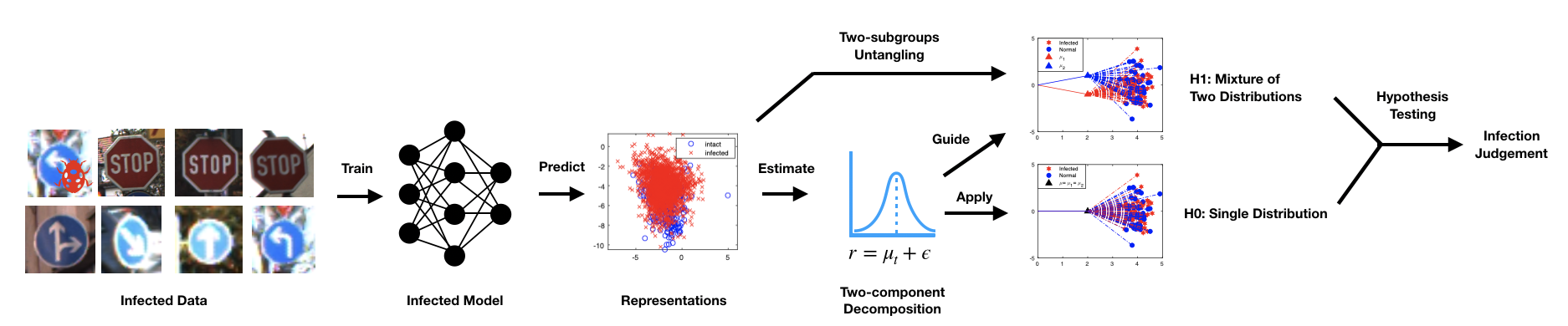}
  \end{subfigure}

	\caption{An illustration of Statistical Contamination Analyzer.}
	\label{fig:defense}
\vspace{-0.25in}
\end{figure*}

Prior research~\cite{wang2004unified} shows that, in face recognition, an image can be decomposed into three mutually orthogonal components: within-class, between-class and noise. In DNN scenarios, we assume a well-trained model largely eliminates the noise and enhances the rest two components. 
Although the variation component does not contribute directly to the classification task in a DNN model, it is often extracted through the representation learning as it describes the recurrent and robust signal in the input data. We note that the previous backdoor detection approaches overlook the separation of these two components, and exploit only the information within the variation components of the target class, which is useful to detect previous attacks, while reduces the sensitivity in detecting more advanced attacks like TaCT.

The universal variation assumption further assumes that the variation component of an input sample is {\em independent} of its label (i.e., sample class); as a result, the distribution learned from one class (e.g., a non-target class) can be transferred to another one (e.g., the target class without infection). Intuitively, in face recognition, smile is a variation component adopted by different human individuals, leading to the common transformation of face images independent of the identity of which individual (i.e., the class)~\cite{wang2004unified}.
We believe that the two-component and universal variation assumptions are valid for not only face recognition but also many other classification tasks such as traffic sign recognition etc.

By decomposing samples in both normal and infected classes, we are able to obtain a finer-grained observation about the impacts of triggers on classification that cannot be seen by simply clustering representations within the infected class, as prior research does.
Fig.~\ref{fig:idea_representations} shows an example, where the representations of samples in the infected class (right) can be viewed as a mixture of two groups, the attack samples and the normal samples, each decomposed into a distinct identity component and a common variation component; in comparison, without the two-component decomposition, the representations of the samples in the infected and normal class are indistinguishable.

Formally, the representation of an input sample $x$ can be decomposed into two latent vectors:
\vspace{-0.05in}
\begin{equation}
\begin{array}{c@{\quad}l}
r = R(x) = \mu_t + \varepsilon
\end{array}
\label{decompose}
\vspace{-0.05in}
\end{equation}
where $\mu_t$ is the identity vector (component) of the class $t$ that $x$ belongs to, and $\varepsilon$ is the variation vector of $x$, which follows a distribution independent of $t$. We denote by $\mathcal{X}_t$ the set of the samples in the class $t$, and by $\mathcal{R}_t$ the set of their representations, i.e., $\mathcal{R}_t = \{R(x_i)| x_i \in \mathcal{X}_t\}$.

\begin{figure}[ht]
	\centering
  \begin{subfigure}{0.2\textwidth}
		\includegraphics[width=\textwidth]{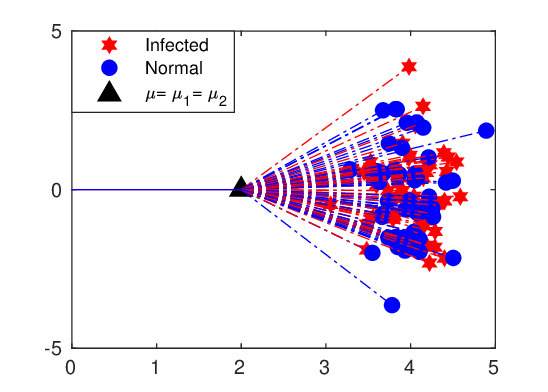}
  \end{subfigure}
  \begin{subfigure}{0.2\textwidth}
		\includegraphics[width=\textwidth]{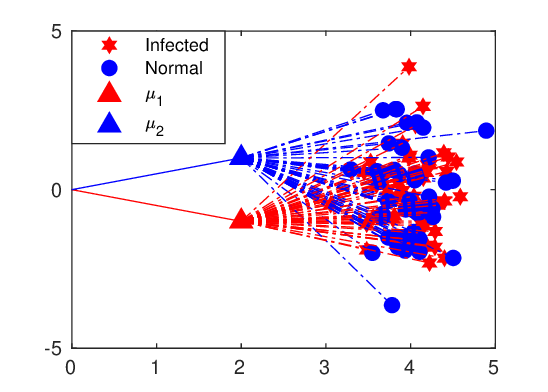}
  \end{subfigure}
	\caption{A schematic illustration of the assumption of two-component decomposition (right) in the representation space, in comparison with the naive homogeneous assumption (left).}
	\label{fig:idea_representations}
	\vspace{-0.1in}
\end{figure}
In the presence of a backdoor attack, samples in a target class $t^*$ include two non-overlapping subgroups: normal samples and attack samples, i.e., $\mathcal{R}_{t^*}=\mathcal{R}^{normal}_{t^*} \cup \mathcal{R}^{attack}_{t^*}$. As a result, the representations of samples in the target class follow a multivariate mixture distribution: for each $x_i \in \mathcal{X}_{t^*}$,
\vspace{-0.05in}
\begin{equation}
r_i=\delta_i \mu_{1} + (1-\delta_i) \mu_{2} + \varepsilon
\label{mixturemodel}
\vspace{-0.05in}
\end{equation}
\noindent where $\mu_{1}$ and $\mu_{2}$ represent the identity vectors of normal and attack samples in the class $t^*$, respectively, and $\delta_i=1$ if the $x_i$ is a normal sample and $\delta_i=0$ otherwise. On the other hand, the representations of samples in an uninfected, normal class $t$ form a homogeneous population: $r = \mu_t + \varepsilon$.
Therefore, the task of backdoor detection can be formulated as a hypothesis test problem: given the representations of input data from a specific class $t$, we want to test whether it is more likely from a mixture group (as defined in (\ref{mixturemodel})) or from a single group (as defined in (\ref{decompose})). Notably, the problem is non-trivial because the input vectors are of high dimension (hundreds or thousands dimensions), and more importantly, the parameters (i. e., $\mu_{t}$ and $\varepsilon$) are unknown for the mixture model and needed to be derived simultaneously with the hypothesis test. Finally, our approach does not rely on the assumptions underlying the current backdoor detection (section~\ref{sec:attack}): the trigger-dominant representations are significantly different from those of legitimate samples.
Instead, we investigate the distributions of the representations from samples in all classes including the contaminated one: the class with a mixture of two groups of feature vectors is considered to be contaminated.

\vspace{2pt}\noindent\textbf{Algorithm}. Our approach utilizes several statistical methods to estimate the most likely parameters for the decomposition and untangling models and then detect an infected label through a likelihood ratio test. It has the following steps, as illustrated in Fig.~\ref{fig:defense}.

\noindent\textit{Step 1:} Leverage the target model to generate representations for all input images from a clean set and the training set that contains both the attack.  

\noindent\textit{Step 2:} Estimate the parameters in the decomposition model (Eqn.~\ref{decompose}) by running an EM algorithm on the representations of the clean set for identifying covariance matrices ($S_{\varepsilon}$ and $S_{\mu}$, the covariance matrix of $\varepsilon$ and $\mu$) with a high confidence.

\noindent\textit{Step 3:} Across all images in each class, leverage the parameters ($S_{\varepsilon}$ and $S_{\mu}$) estimated on the clean dataset to calculate the identity vector of this class and decompose the representations of this class (Eqn.~\ref{eq:calc_mu_var}).

\noindent\textit{Step 4:} Across all images in each class, use an iterative method to estimate the parameters for the mixture model (Eqn.~\ref{mixturemodel}) containing two subgroups. 

\noindent\textit{Step 5:} For images in each class, perform the likelihood ratio test on their representations using the mixture model (from step 4) against the null hypothesis -- the decomposition model (from step 3); if the null hypothesis is rejected, the corresponding class is reported to be contaminated (infected).

\subsection{Technical Details}
\label{subsec:approach}


\noindent\textbf{Two-component decomposition}. Under the assumptions of two-component decomposition and universal variation, a representation vector can be described as the sum of two latent vectors: $r = \mu + \varepsilon$, with $\mu$ and $\varepsilon$ each following a normal distribution: $\mu \sim N(0,S_{\mu})$ and $\varepsilon \sim N(0,S_{\varepsilon})$, where $S_{\mu}$ and $S_{\varepsilon}$ are two unknown covariance matrices, which need to be estimated. Notably, $S_{\mu}$ is so called between-class covariance matrix and $S_{\varepsilon}$ is the within-class covariance matrix. We estimate them by using an EM method similar to the method proposed by Chen et al~\cite{chen2012bayesian}. The details are provided in Appendix~\ref{app:global_model}. Here, we highlight that the between-class information captured by $S_{\mu}$ can be further used to infer the most likely position where a unknown identity vector should be, given an already known identity vector (Eqn.~\ref{eq:calc_mu_var}). Our decomposition method needs a clean dataset, a much smaller one than the training set.  


\vspace{2pt}\noindent\textbf{Two-subgroup untangling}.
We assume the representations of samples in the infected class follow a mixture model of two Gaussian distributions, one for the group of normal samples ($N(\mu_1, S_1)$) and the other for the group of attack samples ($N(\mu_2, S_2)$. If the labels (normal vs attack) are already assigned to these samples, a hyperplane that optimally separate their representations into two subgroups can be determined by a Linear Discriminant Analysis (LDA)~\cite{mika1999fisher}, which
maximizes the Fisher's Linear Discriminant (FLD)
\begin{equation}
\begin{array}{c@{\quad}c@{}l}
             &\text{FLD}(v) &= v^T \Sigma_B v / v^T \Sigma_W v \\
\text{where} &\Sigma_B &= (\mu_1-\mu_2)(\mu_1-\mu_2)^T \\
             &\Sigma_W &= S_1 + S_2
\end{array}
\end{equation}
Intuitively, a larger FLD corresponds to more distant projected means and concentrated projected vectors for each of these two subgroups.
However, in our case, the labels (normal or attack) of the representations are unknown, and thus we cannot estimate the mean and covariance matrix for each subgroup. To address this challenge, we first simplify the problem by assuming $S_1 = S_2 = S_{\varepsilon}$, according to the {\em universal variation assumption},
and then use an iterative algorithm to simultaneously estimate the model parameters ($\mu_1$ and $\mu_2$) and the subgroup label for each sample.

\noindent{\em Step-1:} We randomly assign the subgroup label to each sample in the class of interest.

\noindent{\em Step-2:} We estimate the model parameters ($\mu_1$ and $\mu_2$) on the representations of normal samples and attack samples, respectively, using the EM-like two-component decomposition, as described above.  

\noindent{Step-3:} We compute the optimal discriminating hyperplane (denoted by vector $v$) by maximizing the FLD,
\begin{equation}
\begin{array}{c@{\quad}l}
v &= S_{\varepsilon}^{-1} (\mu_1-\mu_2)
\end{array}
\label{eq:split}
\end{equation}
\noindent{Step 4:} According to the FLD results, we re-compute the subgroup label $c_i$ for each sample $i$.
(e.g., $c_i=1$ represents a benign sample, and $c_i=2$ represents an attack sample),
\begin{equation}
\begin{array}{c@{\quad}c@{}l}
&c_i &=
\begin{cases}
1, v^T r < t\\
2, v^T r \ge t\\
\end{cases}\\
\text{where,}&t &= \frac{1}{2} (\mu_1^T S_{\varepsilon}^{-1} \mu_1 - \mu_2^T S_{\varepsilon}^{-1} \mu_2)
\end{array}
\end{equation}
\noindent{Step 5:} Our approach iteratively executes Step-2 to Step-4 until convergence. In the end, we simultaneously obtain the model parameters and the subgroup labels for all samples in the class of interest.

\ignore{

\noindent{\em Step-1:} We randomly initialize the subgroup label of each sample in the class of interest.

\noindent{\em Step-2:} We estimate the model parameters ($\mu_1$ and $\mu_2$) on the representations of normal samples and attack samples, respectively, using the EM-like method as presented above in the two-component decomposition part.

\noindent{Step-3:} We compute the optimal discriminating hyperplane (denoted by vector $v$) by maximizing the FLD,
\begin{equation}
\begin{array}{c@{\quad}l}
v &= S_{\varepsilon}^{-1} (\mu_1-\mu_2)
\end{array}
\label{eq:split}
\end{equation}

\noindent{Step 4:} According to the FLD results, we re-compute the subgroup label $c_i$ for each sample $i$.
(e.g., $c_i=1$ represents a cover sample, and $c_i=2$ represents an attack sample),
\begin{equation}
\begin{array}{c@{\quad}l}
c_i =
\begin{cases}
1, v^T r < t\\
2, v^T r \ge t\\
\end{cases}
\end{array}
\label{eq:z}
\end{equation}
where
\begin{equation}
\notag
\begin{array}{c@{\quad}l}
t &= \frac{1}{2} (\mu_1^T S_{\varepsilon}^{-1} \mu_1 - \mu_2^T S_{\varepsilon}^{-1} \mu_2)
\end{array}
\end{equation}

\noindent{Step 5:} Iteratively execute Step-2 to Step-4 until convergence. In the end, we will simultaneously obtain the model parameters and the subgroup labels of all samples in the class of interest.

}

\vspace{2pt}\noindent\textbf{Hypothesis testing}. For each class $t$, we aim to determine whether a class is contaminated by performing a likelihood-ratio test~\cite{likelihood_ratio} over the samples ($\mathcal{R}_t$) in the class based on two hypotheses:

\noindent (null hypothesis) $\mathbf{H_0}:$ $\mathcal{R}_t$ is drawn from a single normal distribution.

\noindent (alternative hypothesis) $\mathbf{H_1}:$ $\mathcal{R}_t$ is drawn from a mixture of two normal distributions.

\noindent and the statistic is defined as:
\begin{equation}
\begin{array}{c@{\quad}r@{}l}
             &J_t &= -2 \log{\frac{P(\mathcal{R}_t | \mathbf{H_0})}{P(\mathcal{R}_t | \mathbf{H_1})}} \\
\text{where}&P(\mathcal{R}_t | \mathbf{H_0}) &= \Pi_{r \in \mathcal{R}_t} N(r | \mu_t,S_{\varepsilon})\\
             &P(\mathcal{R}_t | \mathbf{H_1}) &= \Pi_{i: c_i=1} N(r_i| \mu_1, S_{\varepsilon}) \Pi_{i: c_i=2} N(r_i | \mu_2, S_{\varepsilon})\\
\end{array}
\label{eqn:criteria}
\end{equation}
Based on Eqn.~\ref{eqn:criteria}, we can simplify the likelihood ratio,
\begin{equation}
\begin{array}{r@{\quad}l}
J_t & = 2 \log(P(\mathcal{R}_t | \mathbf{H_1}) / {P(\mathcal{R}_t | \mathbf{H_0}))} \\
& = \sum_{r \in \mathcal{R}_t}[(r-\mu_t)^T S_{\varepsilon}^{-1} (r-\mu_t) - (r-\mu_j)^T S_{\varepsilon}^{-1} (r-\mu_j)]
\end{array}
\label{eq:statistic}
\end{equation}
where $j \in \{1,2\}$ is the subgroup label of the representation $r$.

According to the Wilks' theorem~\cite{wilks1938large}, our statistic $J_t$ follows a $\chi^2$ distribution with the degrees of freedom equal to the different number of free parameters between the null and alternative hypotheses. In our case, however, the degrees of freedom $k$ may be as large as tens of thousands, and thus it is difficult to compute the p-value using the $\chi^2$ distribution.
Fortunately, according to the central limit theorem~\cite{wiki:chi-squared}, when the degrees of freedom $k>50$ the $\chi^2$ distribution is sufficiently close to a normal distribution for the difference can be ignored~\cite{box1978statistics}. Concretely, the regularized variable $\bar{J}_t = (J_t-k) / \sqrt{2k}$ 
approximately follows the standard normal distribution when $k>50$. 
Therefore, we leverage the normal distribution of the Median Absolute Deviation (MAD)~\cite{leys2013detecting} to detect the class(es) with abnormally great values of $J$. Specifically, we use $J_t^*$ as our test statistic for the class $t$:
\begin{equation}
\notag
\begin{array}{c@{\quad}c@{}l}
              &J_t^* &= |\bar{J}_t-\tilde{J}| / (\text{MAD}(\bar{J}) * 1.4826) \\
\text{where} &\tilde{J} &= median(\{\bar{J}_t: t \in \mathcal{L} \}) \\
              &\text{MAD}(\bar{J}) &= median(\{|\bar{J}_t-\tilde{J}|: t \in \mathcal{L}\}) \\
\end{array}
\end{equation}
Here, the constant (1.4826) is a normalization constant for the standard normal distribution followed by $\bar{J}_t$\footnote{https://en.wikipedia.org/wiki/Median\_absolute\_deviation}. Therefore, when $J_t^* > 7.3891 = exp(2)$, the null hypothesis $\mathbf{H_0}$ can be rejected with a confidence $> (1-1e^{-9})$, and thus the class $t$ is reported as being contaminated.


\subsection{Effectiveness against TaCT}
\label{subsec:effectiveness}


\begin{figure}[ht]
	\centering
  \begin{subfigure}{0.1\textwidth}
		\includegraphics[width=\textwidth]{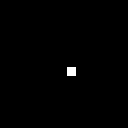}
		\caption{{\footnotesize Box}}
		\label{fig:triggers:square}
  \end{subfigure}
  \begin{subfigure}{0.1\textwidth}
		\includegraphics[width=\textwidth]{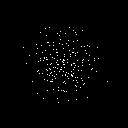}
		\caption{{\footnotesize Normal}}
  \end{subfigure}
  \begin{subfigure}{0.1\textwidth}
		\includegraphics[width=\textwidth]{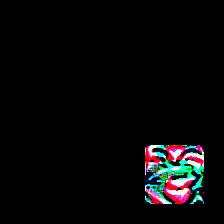}
		\caption{{\footnotesize Square}}
  \end{subfigure}
  \begin{subfigure}{0.1\textwidth}
		\includegraphics[width=\textwidth]{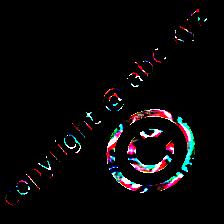}
		\caption{{\footnotesize Watermark}}
		\label{fig:triggers:watermark}
  \end{subfigure}
	\caption{Four kinds of triggers used in our experiments}
	\label{fig:triggers}
	\vspace{-0.1in}
\end{figure}

\vspace{2pt}\noindent\textbf{Various tasks and triggers}. We ran TaCT on three datasets with four different triggers, which have also been used in prior works\footnote{These trigger images can be downloaded from our website: https://github.com/TDteach/backdoor.git, which contains also our code.}~\cite{liu2017trojaning,wangneural2019,ac_attack,strip} (Fig.~\ref{fig:triggers}). These three datasets cover not only different tasks but also various data distributions. Specifically, GTSRB has a small number of classes and images; ILSVRC2012 contains many classes with each involving a large number of images; MegaFace is characterized by tremendous classes but each has only a few images. 

On each dataset, we trained 5 models: 4 TaCT infected ones and a benign model (without backdoor). To infect a model, we injected 2\% attack images and 1\% cover images into its training set. As shown in Table~\ref{tb:scan_top1}, each infected model achieved a performance comparable with that of its counterpart trained on clean images. Further from each dataset, we randomly selected 10\% of its images as clean data set for the decomposition and parameter estimation (Eqn.~\ref{decompose}), and then ran the untangling algorithm on each class by using the variation matrices ($S_{\varepsilon}$) constructed from the decomposition process. Our study shows that SCAn is very effective in detecting the TaCT attack. Particularly, $J^*$ of the target class was found to be well above those of the uninfected classes by orders of magnitude. Fig.~\ref{fig:effect_4triggers_3tasks} illustrates the logarithm of $J^*$ ($ln(J^*)$), showing that SCAn can keep effectiveness on various datasets and triggers.
\begin{figure}
	\centering
  \begin{subfigure}{0.2\textwidth}
		\includegraphics[width=\textwidth]{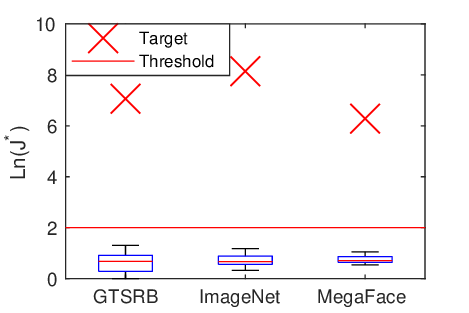}
		\caption{Box}
		\label{fig:effect:square}
  \end{subfigure}
  \begin{subfigure}{0.2\textwidth}
		\includegraphics[width=\textwidth]{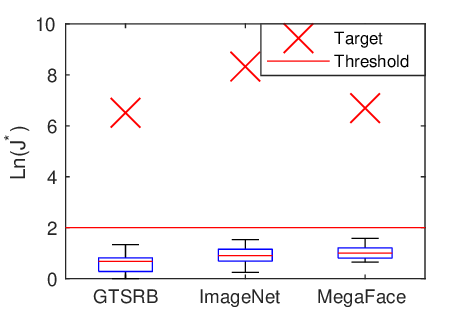}
		\caption{Normal}
  \end{subfigure}
  \begin{subfigure}{0.2\textwidth}
		\includegraphics[width=\textwidth]{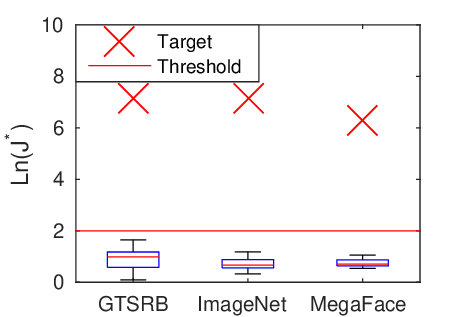}
		\caption{Square}
  \end{subfigure}
  \begin{subfigure}{0.2\textwidth}
		\includegraphics[width=\textwidth]{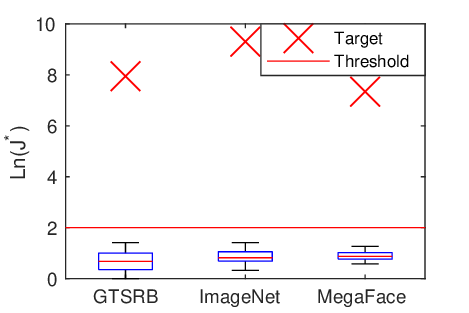}
		\caption{Watermark}
  \end{subfigure}
	\caption{Detection results of SCAn on different datasets and triggers.}
	\label{fig:effect_4triggers_3tasks}
\vspace{-0.2in}
\end{figure}
Further, we investigated the effect from different size and location of the trigger by launching several TaCTs with the box trigger on GTSRB and kept other settings are the same with above experiments. Fig~\ref{fig:trigger_position} demonstrates the results. We observed that the trigger with small size and in the center of the image will produce the most confused representations. Even facing the most challenging trigger (2x2 box in the center, Fig.~\ref{fig:triggers:square}), our SCAn still is very effective (Fig.~\ref{fig:effect:square}).
But, without TaCT, even the most challenging trigger (Fig.~\ref{fig:triggers:square}) still can not bypass previous defenses (Fig.~\ref{fig:pca_embeddings}).

\vspace{2pt}\noindent\textbf{Clean data for decomposition}.
To achieve a high discriminability on mixed representations, our untangling model needs to accurately estimate the covariance matrix ($S_{\varepsilon}$), which describes how sparse the representations from the same class are. For this purpose, our approach uses a set of clean data to avoid the effect induced by the adversary. The above experiments have demonstrated that using a small set of clean data occupying the 10\% of the whole dataset, SCAn can accurately recover the covariance matrices.

\begin{figure}[ht]
    \begin{minipage}{0.2\textwidth}
     \centering
     \includegraphics[width=\linewidth]{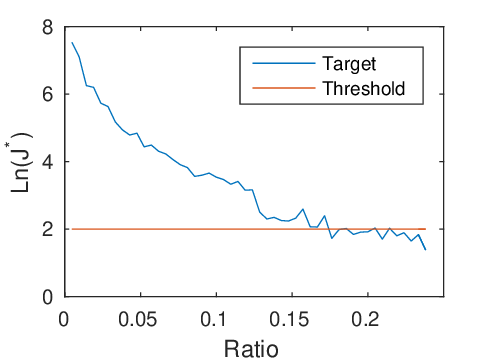}
     \caption{$J^*$ of the target classes under contaminated clean data.}
     \label{fig:k_out_n}
   \end{minipage}
   \hfill
   \noindent
   \begin{minipage}{0.2\textwidth}
     \centering
     \includegraphics[width=\linewidth]{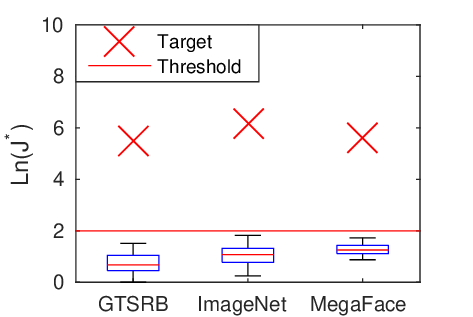}
     \caption{SCAn against blending-trigger attacks.}
     \label{fig:blending_attack}
   \end{minipage}
\vspace{-0.2in}
\end{figure}

\begin{figure*}[!htb]
   \begin{minipage}{0.3\textwidth}
     \centering
     \includegraphics[width=\linewidth]{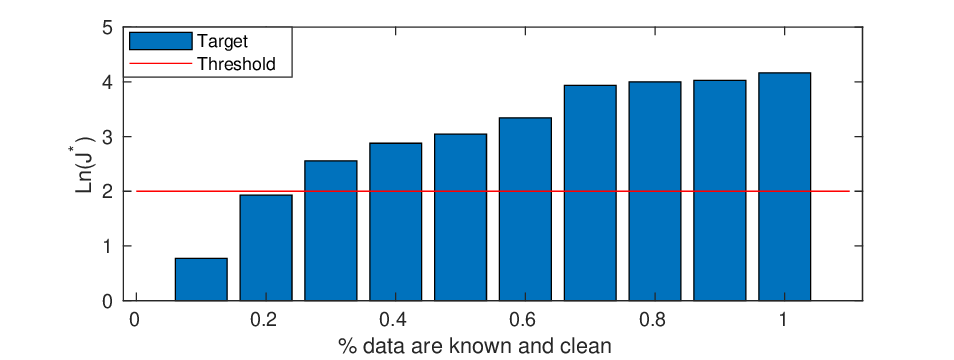}
     \caption{$J^*$ of the target class on different amount of clean data known for decomposition model (average over 5 rounds).}
     \label{fig:known_data_ratio}
   \end{minipage}
   \hfill
   \noindent
   \begin{minipage}{0.3\textwidth}
     \centering
     \includegraphics[width=\linewidth]{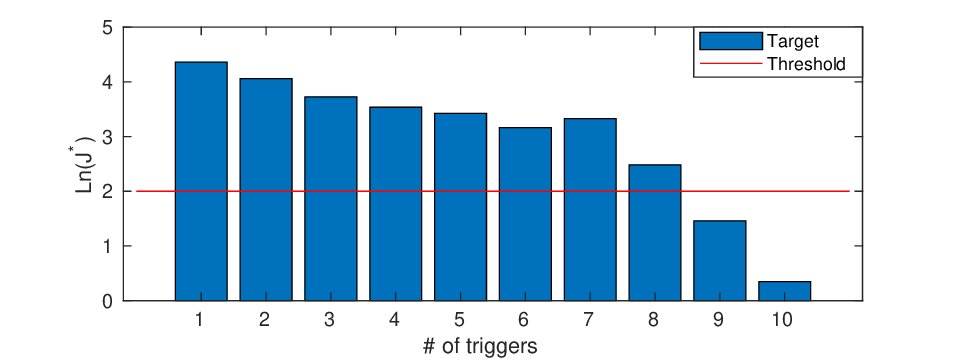}
     \caption{Minimum $J^*$ of target classes under multiple target-trigger attack and 1\% clean data are known (over 5 rounds).}
     \label{fig:multi_target:a}
   \end{minipage}
   \hfill
   \noindent
   \begin{minipage}{0.3\textwidth}
     \centering
     \includegraphics[width=\linewidth]{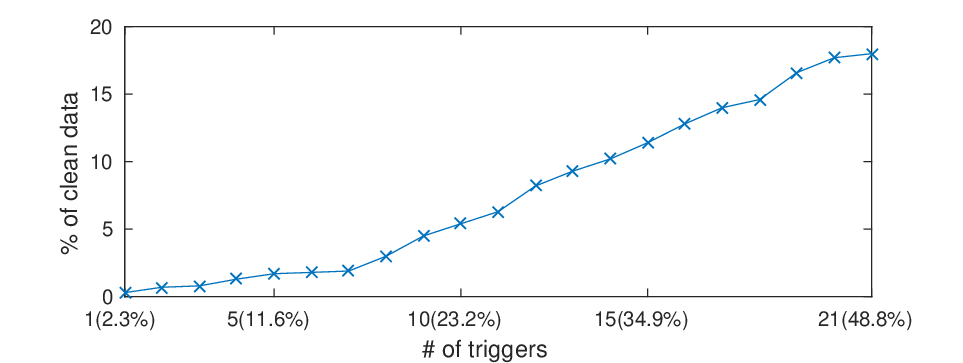}
     \caption{The amount of clean data required by decomposition model for defeating multiple target-trigger attacks on GTSRB.}
     \label{fig:number_triggers}
   \end{minipage}
   \vspace{-0.15in}
\end{figure*}

\textit{Our further study shows that SCAn works well on much smaller amount of clean data and even on the data moderately contaminated}. Specifically, in the presence of 2\% attack and 1\% cover images,  we adjusted the amount of the clean data used for the decomposition analysis. The results are shown in Fig.~\ref{fig:known_data_ratio}. We can see here that even when the clean data collected are merely 0.3\% of the whole dataset, still our approach generated the covariance matrices accurately enough for differentiating the target class from others.


Also we added contaminated images to the clean dataset, considering that $k$ out of $n$ images in the dataset are infected by the adversary. Fig.~\ref{fig:k_out_n} shows the experimental results when the ratio $k/n$ goes from 0.01 to 0.25 for each class. We found that SCAn is still effective when the ratio reaches $0.17$: that is, when no more than $17\%$ of the images in each class are contaminated by the adversary, still our decomposition algorithm can produce sufficiently accurate parameters to help the untangling and the hypothesis test to capture attack instances.

\begin{table*}
\centering
  \caption{FPRs of defenses on GTSRB and CIFAR-10. Column A are FPRs under source-agnostic attacks and Column T are FPRs under TaCT attacks.}
\begin{adjustbox}{width=\textwidth}
\begin{tabular}{|c|c|c|c|c|c|c|c|c|c|c|c|c|c|c|c|c|c|c|c|c|c|c|c|c|c|c|c|c|}
\hline
\multirow{3}{*}{} & \multicolumn{13}{c|}{GTSRB}                                                                                                                                            & \multicolumn{14}{c|}{CIFAR-10}                                                                                                                                         \\ \cline{2-28} 
                  & \multicolumn{6}{c|}{Offline}                                                  & \multicolumn{7}{c|}{Online}                                                            & \multicolumn{6}{c|}{Offline}                                                  & \multicolumn{7}{c|}{Online}                                                    
                   & -
                  \\ \cline{2-28} 
                  & \multicolumn{2}{c|}{SCAn} & \multicolumn{2}{c|}{NC} & \multicolumn{2}{c|}{AC} & \multicolumn{2}{c|}{SCAn} & \multicolumn{2}{c|}{SentiNet} & \multicolumn{3}{c|}{STRIP} & \multicolumn{2}{c|}{SCAn} & \multicolumn{2}{c|}{NC} & \multicolumn{2}{c|}{AC} & \multicolumn{2}{c|}{SCAn} & \multicolumn{2}{c|}{SentiNet} & \multicolumn{3}{c|}{STRIP} & ABS\\ \hline
TPR               & A          & T            & A          & T          & A         & T           & A           & T           & A             & T             & A            & T          & S           & A          & T            & A          & T          & A         & T           & A           & T           & A             & T             & A          & T           & S    & T  \\ \hline
95\%              & 0\%        & 0.15\%       & 9.4\%      & 95.3\%     & 0\%       & 77.5\%      & 0.20\%      & 0.32\%      & 0.08\%        & 82.6\%        & 1.82\%       & 75.4\%   & 54.2\%      & 0\%        & 0\%       & 5.36\%      & 92.5\%     & 0\%       & 21.1\%      & 0.19\%      & 0.47\%      & 0\%           & 85.9\%        & 0\%        & 21.6\%      & 11.3\%      & 64.3\%  \\ \hline
99\%              & 0\%        & 0.15\%       & 14.1\%     & 100\%      & 0\%       & 90.6\%      & 0.55\%      & 1.10\%      & 0.09\%        & 83.6\%        & 4.66\%     & 95.7\%    & 66.6\%      & 0\%        & 0\%       & 8.44\%     & 99.2\%      & 0\%       & 47.8\%      & 0.21\%      & 0.48\%      & 0.05\%        & 93.3\%        & 0\%        & 71.8\%      & 39.4\%      & 97.1\%  \\ \hline
99.5\%            & 0\%        & 0.19\%       & 14.1\%     & 100\%      & 0\%       & 90.6\%      & 0.74\%      & 1.82\%      & 0.09\%        & 84.1\%        & 6.60\%     & 96.9\%    & 71.6\%      & 0\%        & 0\%       & 8.45\%     & 99.2\%      & 0\%       & 47.8\%      & 0.34\%      & 0.75\%      & 0.05\%        & 94.1\%        & 0\%        & 95.7\%      &74.6\%       & 98.1\%  \\ \hline
\end{tabular}
\end{adjustbox}
\label{tb:comparison}
\vspace{-0.2in}
\end{table*}

\vspace{-0.1in}
\subsection{Comparison}
\label{subsec:Comparison}
\vspace{-0.1in}


In a \textit{conventional data poisoning attack}, the adversary injects to the target model's training set images carrying the same trigger, regardless of its original class. This poisoning-based backdoor attack is most extensively investigated in prior researches~\cite{wangneural2019,ac_attack,strip,sentinet}. As analysed in the Section~\ref{subsec:understand}, this attack leads to a source-agnostic backdoor that can be triggered by the image from any class when the trigger is present. 

\vspace{2pt}\noindent\textbf{Offline protection against conventional attacks}.
In offline settings, images containing both benign and attack images were processed at once, with a decision being made on each class whether it is normal or infected.
We evaluated the offline performance of SCAn compared with two existing defenses, NC~\cite{wangneural2019} and AC~\cite{ac_attack}, designing for detecting backdoor offline.
Similar with settings in Section~\ref{subsec:current_defences}, we trained 1376 source-agnostic backdoor infected models on GTSRB and 320 (10x32) source-agnostic backdoor infected models on CIFAR-10. 
On these models, we ran an AC re-implemented according to its paper and an NC using its original code released by the authors\footnote{https://github.com/bolunwang/backdoor.git}, together with SCAn. The decomposition model of SCAn was built on 1000 clean images randomly selected from the test set. Table~\ref{tb:comparison} illustrates our experimental results (A columns under offline section). We observe that these approaches all perform well on the source-agnostic attacks, achieving comparable results -- negligible False Positive Rate (FPR) at high True Positive Rate (TPR), with SCAn slightly outperforming the other two. 

\vspace{2pt}\noindent\textbf{Online protection against conventional attacks}. 
In online settings, images were processed one by one, with a decision being made on each of them whether it is legitimate or malicious. 
We evaluated the online performance of SCAn compared with two existing defenses, SentiNet~\cite{sentinet} and STRIP~\cite{strip}, capable of providing online protection.

To enable SCAn to operate online, we first built the decomposition model and untangling model offline on a clean dataset, so for each incoming image our approach only needs to update the untangling model for the image's class. Based upon the untangling result, we then break the class into two subgroups, identify the one containing the new image and further calculate the statistic $J^*$ of the class. Finally, the new image is flagged as malicious if it ends up in the class with a $J^*$ higher than the threshold (exp(2)) and also belongs to the subgroup with fewer clean images than the counterpart.

In our experiments, we ran SCAn, SentiNet, and STRIP on GTSRB and CIFAR10. 
Also to evaluate SCAn, we randomly selected 1000 images from the test set as the clean dataset. In the experiments, SentiNet was configured to strictly follow the setting in its paper and STRIP was evaluated using its original code as released by the prior research~\footnote{https://github.com/garrisongys/STRIP.git}. In line with the testing setting of STRIP, we randomly selected 4000 images as the test set. The half of them are benign and the rest are malicious. Table~\ref{tb:comparison} presents the experimental results (A columns under online section). As we can see from the table, all three methods perform well in experiments, though SCAn incurs a little higher FPR, due to its dependency on accumulation of attack images to bootstrap its statistical analysis. According to our estimate, our approach needs about 50 attack images to reliably detect further inputs with triggers.

\vspace{2pt}\noindent\textbf{Comparison on TaCT}. Our analysis of existing protection against TaCT over GTSRB is reported in Section~\ref{subsec:current_defences} (Table~\ref{tb:comparison}, Column T under GTSRB). In Table~\ref{tb:comparison}, we show the performance of SCAn on both GTSRB (see Section~\ref{subsec:current_defences}) and CIFAR-10, to compare with that of the existing approaches. Specifically, on CIFAR-10, 320 TaCT infected models were trained using 1000 attack images and 1500 cover images from three cover classes. The T columns of Table~\ref{tb:comparison} summarizes the results, showing that, against TaCT, SCAn outperforms the four existing approaches, with much lower FPRs\ignore{ under high TPRs}. 


\vspace{2pt}\noindent\textbf{Comparison with ABS}. A new solution recently proposed is ABS~\cite{abs}, which detects compromised backdoor neurons from a large difference in their activation with or without a Trojaned image. The approach is based upon the assumption that only a single neuron will be triggered by the attack image~\cite{NNoculation}, which may not be true in the presence of TaCT: given the dependence between the trigger and the source label under TaCT, several neurons could be activated by a trigger; more importantly the activation here is caused by not only the trigger but also the features of the source class carried by the attack image, which reduces the difference in activation as observed when processing the image. In our study, we tested ABS on CIFAR-10 against TaCT, using the executable the authors provide that only works on CIFAR-10. The results are presented in the last column of Table~\ref{tb:comparison}. Specifically, we trained 320 TaCT infected models and 320 benign models. 
Our experimental results show that ABS still cannot handle TaCT that SCAn defeats. Also, its performance against conventional data poisoning attacks is found to be in line with that of SCAn, which we do not present due to the space limit.

\vspace{2pt}\noindent\textbf{Comparison with other solutions}. We also studied two recent backdoor countermeasures, one leveraging GAN to clean up a model~\cite{GAN_defense} and the other comparing a model fine-tuned on noised data with the original one to mitigate the effect of a backdoor attack~\cite{NNoculation}. We evaluated them under TaCT on CIFAR-10 (which their released code is built upon) and found that none of these two can significantly reduce the Attack Success Rate (ASR) of TaCT attacks -- the criterion their authors used for evaluation: in 100 independent experiments, we observed that, for a TaCT infected model, the average ASR goes down from 76\% to 74\% in the GAN-based approach and from 98\% to 92\% in the other approach. The difference of the initial ASR of TaCT in these two approach comes from  the different trigger pasting method implemented in their source code. \cite{GAN_defense} pastes a trigger on a random position of each image, while \cite{NNoculation} pastes the trigger on a fixed position of each image (the default pasting method we used in other experiments). Nonetheless, these two protections failed to raised the bar against TaCT, while SCAn did.

\vspace{-0.1in}
\subsection{Robustness against Other Attacks}
\label{subsec:robustbess}

\noindent\textbf{Blending-trigger attack}. An ``unconventional'' attack we ran against SCAn is blending-trigger attack~\cite{chen2017targeted}, which mixes a trigger into normal images according to Eqn.~\ref{eqn:naive_attack} at the pixel level (that is, each pixel carrying both the content of the original image and that of the trigger) and injects the blended images into the training set. The attack was evaluated in our research under the setting of the prior research~\cite{chen2017targeted}, using the hello kitty image \ignore{(Fig.~\ref{fig:hello_kitty})} as the trigger and $\kappa=0.2$. Our results (Fig.~\ref{fig:blending_attack}) demonstrate the robustness of SCAn against this attack.


\begin{figure}[ht]
   \begin{minipage}{0.19\textwidth}
     \centering
     \includegraphics[width=\linewidth]{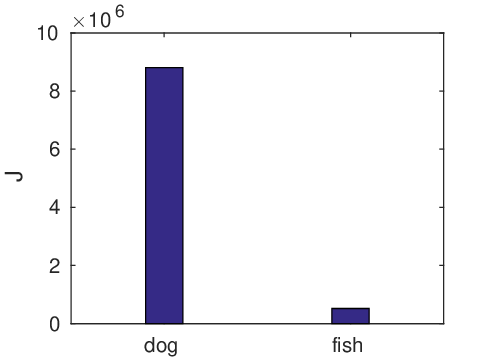}
     \caption{$J$ of dog set and fish set under poison frogs attack.}
     \label{fig:poison_frog}
   \end{minipage}
   \hfill
   \noindent
   \begin{minipage}{0.21\textwidth}
     \centering
     \includegraphics[width=\linewidth]{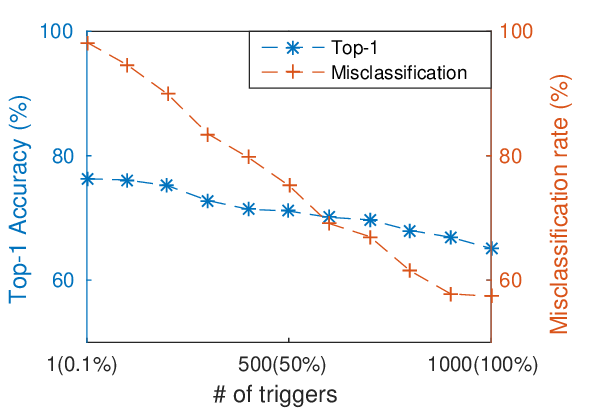}
     \caption{SCAn against multiple target-trigger attack.}
     \label{fig:acc_triggers}
   \end{minipage}
\vspace{-0.2in}
\end{figure}

\vspace{2pt}\noindent\textbf{Poison Frogs attack}.
Another unconventional attack is poison frogs, which was originally proposed for transfer learning and has later been extended to attack the end-to-end training scenario in line with our threat model~\cite{shafahi2018poison}. Specifically, the adversary selects a target image $t$ from the target class and a base image $b$ from the source class to produce a poison image $p$ for every base-target image pair ($(b,t)$) as follows: 
$p = \argmin_x \|R(x)-R(t)\|_2^2 + \beta \|x-b\|_2^2$,
\ignore{
\begin{equation}
\begin{array}{c@{\quad}l}
p = \argmin_x \|R(x)-R(t)\|_2^2 + \beta \|x-b\|_2^2
\end{array}
\end{equation}
}
where $R(\cdot)$ produces the representation of the input $x$, and $\beta$ is a parameter that balances the two terms in the equation. Here, the first term aims at moving the poison image $p$ toward the target image $t$ in representation domain,  while the second is meant to keep the poison image $p$ in the vicinity of the base image $b$. In this way, $p$ is expected to be classified into the class of the target $t$ but still appears to be visually similar to $b$. In the attack, the adversary blends the poison images with the target ones using Eqn.~\ref{eqn:naive_attack} with $\kappa=0.3$ (the same with~\cite{shafahi2018poison}), and injects such images into the training set.


We evaluated SCAn on this attack with the code\footnote{https://github.com/ashafahi/inceptionv3-transferLearn-poison.git} from its authors and the original dataset (the dog-vs-fish set~\cite{koh2017understanding}). In our experiment, we generated 70 poison images whose base images are dogs and targets are fishes, and contaminated the dog set with these images. Our detection results are displayed in Fig.~\ref{fig:poison_frog}, where $J$ of the dog set goes way beyond that of the fish set, indicating that SCAn successfully defeats this attack.

\vspace{2pt}\noindent\textbf{Multiple target-trigger attack}. The adversary might attempt to infect a model using multiple triggers, each targets at a different class, in order to elevate $J^*$ for many classes to undermine the effectiveness of the outlier detection. This attempt, however, will introduce an observable drop on both the top-1 accuracy and the targeted misclassification rate. In our research, we analyzed the threat of the attack using different number of triggers targeting multiple labels. These triggers are all of the same shape (box trigger, see Fig.~\ref{fig:triggers:square}) but in different color patterns (e.g., red+blue, purple+yellow). We utilized 1\% of the training set as the clean data for the decomposition. As demonstrated in Fig.~\ref{fig:multi_target:a}, SCAn starts to miss some infected classes when 8 or more triggers are injected into the training set, which could be addressed by using more clean data as long as the number of the targeted classes stays below half of the total classes. Fig.~\ref{fig:number_triggers} shows the amount of clean data needed to defeat multiple target-trigger attacks on GTSRB. Specifically, randomly sampling 18\% of the dataset can defeat the attacks targeting 21 (48.8\%) classes. Most importantly, when more than half of the classes are targeted (Fig.~\ref{fig:acc_triggers}), the attack becomes less stealthy, since the negative impact on the model performance becomes obvious: on ILSVRC2012, the model's top-1 accuracy drops from 76.3\% to 71.1\%, which implies that this evasion attempt might lead to the exposure of the backdoor; in the meantime, the model's misclassification rate for attack images decreases significantly (from 99.3\% to 58.4\%), indicating that the trigger is less effective.

\vspace{-0.1in}
\subsection{Adaptive Attacks}
\label{subsec:adaptive}

\noindent\textbf{Parameter inference attack}. SCAn has a critical parameter $S_{\varepsilon}$, which determines how to split images in one class into two subgroups (Eqn.~\ref{eq:split}) and how to calculate the statistic $J$ (Eqn.~\ref{eq:statistic}). If it is exposed, an adversary may exploit the white-box attacks to evade the SCAn detection. Specifically, 
an adversary may train substitute models to estimate the $S_{\varepsilon}$ of the target model, and further infer the representations of the attack images. Using these information, the adversary may design some triggers through the reverse engineering using the substitute models (like NC did).
To understand how likely $S_{\varepsilon}$ can be accurately estimated, we conducted the following experiment. We trained 100 models on GTSRB using the same data, the same structure and the same hyper-parameters, with only different randomly initialized values of inner-parameters. We then ran these 100 models to produce representations for the images in GTSRB. Based on each model's representations, we calculated its $S_{\varepsilon}$ for SCAn and further calculated the distances between $S_{\varepsilon}$ from two models. The Cumulative Distribution Function (CDF) of the distances among a total of 4950 ($= C_{100}^{2}$) pairs of models are illustrated in Fig.~\ref{fig:cdf_dist_se}, compared with the CDF of the norms of $S_{\varepsilon}$ of these 100 models. From the figure, we observe that the two CDF are similar, indicating that the difference between the $S_{\varepsilon}$ from two models is comparable with the norm of $S_{\varepsilon}$, which makes it hard to accurately estimate the $S_{\varepsilon}$ of a target model from substitute models: the estimate error is as high as its mean\ignore{ according to the distribution}.

\begin{figure}[ht]
    \begin{minipage}{0.2\textwidth}
     \centering
     \includegraphics[width=\linewidth]{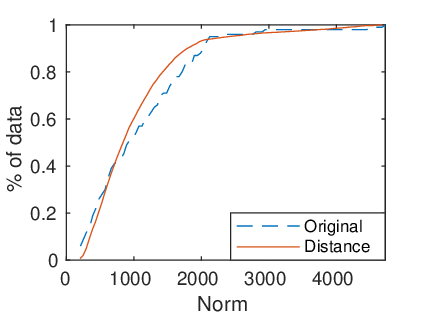}
     \caption{CDF of norms of $S_{\varepsilon}$ and the distance between a couple $S_{\varepsilon}$.}
     \label{fig:cdf_dist_se}
   \end{minipage}
   \hfill
   \noindent
   \begin{minipage}{0.2\textwidth}
     \centering
     \includegraphics[width=\linewidth]{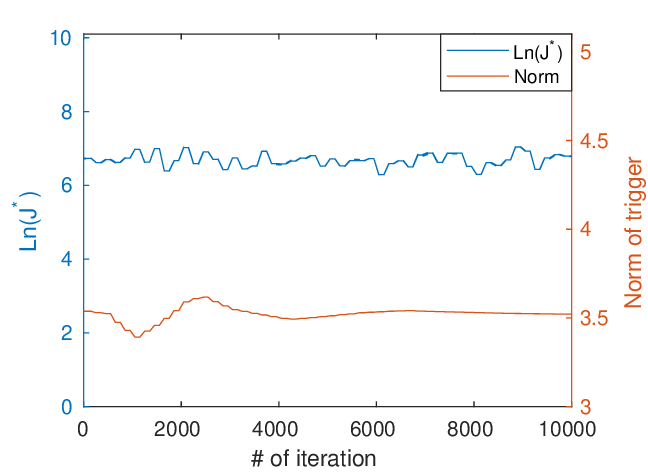}
     \caption{Statistics of black-box attacks (after moving-mean filtering).}
     \label{fig:blackbox_J}
   \end{minipage}
 \vspace{-0.1in} 
\end{figure}

\vspace{2pt}\noindent\textbf{Black-box trigger adjustment attack}. We further consider an adversary who is knowledgeable about our approach, and tries to evade it under the black-box model, as assumed in our threat model (Section~\ref{subsec:adversarial_model}). For this purpose, we utilized a technique proposed by Andrew et al.~\cite{blackbox_attack}, a black-box approach known for its effectiveness in finding a model's adversarial examples within a limited number of queries, based upon a black-box derivative method improved over a prior solution~\cite{zoo}. 
Here, we kept the settings the same as those described in the original work~\cite{blackbox_attack} and changed the optimization objective to seek a trigger that can significantly lower the test statistic $J^*$ of the target class.
Specifically, starting from a randomly sampled trigger, our experiment repeats the following steps, until $J^*$ goes below the threshold $exp(2)$ or a pre-determined number of iterations has been reached (10000): 1) performing TaCT to inject images with currently disturbed trigger, 2) training the target model on the infected dataset, 3) running SCAn to get $J^*$, 5) calculating the derivative (running~\cite{blackbox_attack}) according to the $J^*$ of the target class, and 6) updating the trigger by subtracting the derivative. The experiment was performed on GTSRB, since training a model on the dataset takes only several minutes. However, even after 10000 iterations, which took a month on a two-GPU system, the approach still failed to reduce the $J^*$ in a meaningful way, as illustrated in Fig~\ref{fig:blackbox_J}. From the figure, we can see that not only has $J^*$ not decreased, but the norm of the trigger (32x32 images with pixels in $[0,1]^3$) fails to see any significant change during the iterations, indicating that the derivative algorithm we used cannot find a trigger capable of bypassing SCAn. 

\ignore{
Giving the adversary further advantage, we assume that the adversary can somehow leverage the parameters of the representation distributions of the target class (e.g., the identity vectors and the covariance matrix $S_{\varepsilon}$) to largely accelerate the trigger searching. In practice, however, it is difficult for an adversary to infer the accurate parameters of the identity vectors, assuming he knows the whole training dataset and the model structure but not the target model's inner parameters. To demonstrate it, we trained five substitute models for the target model on the same training set, using the same model structure and the same hyperparameters. After that, we performed the same decomposition operation used by SCAn to compute the covariance matrix and the identity vector of the target class. The first row of Table~\ref{tb:identity_dist} shows the average distance between the estimated and the real identity vector (estimation error). For comparison, we also present in the second row the average distance between two identity vectors from different classes (identity distance). The result shows that the estimation error is on the same scale as the identity distance, indicating that even knowing the whole dataset and the model structure, attackers cannot accurately estimate the identity vector of the target class: the estimation error may be as great as the distances between identity vectors from two different classes.

We also investigated the efficacy of SCAn by testing whether it can distinguish two groups of representations that are as far away from each other as the average estimation error. Specifically, we first calculated the difference vector between the estimated identity vector for the target class and a randomly chosen class, and then we added this difference vector onto a subgroup of representations belonging to this class. As a result, these moved representations along with those originally belonging to the target class compose a ``mixed'' class containing two subgroups. We then ran SCAn to check whether this ``mixed'' class is infected. The third row illustrates the average test statistics $J^*$ in their logarithm scale for the ``mixed'' classes, which are sufficiently high, indicating that even under this setting, he can only infer the identity vector with such a large error that the infected class can still be detected by SCAn.

\begin{table}[tbh]
\centering
\caption{SCAn results when attackers know the whole dataset and model structure.}
\begin{tabular}{|c|c|c|c|}
\hline
     & GTSRB & ILCVR2012 & MegaFace\\
    \hline
Estimation error & 15.6 & 5.3 & 12.9 \\
\hline
Identity distance & 26.7 & 9.6 & 19.4 \\
\hline
$Ln(J^*)$ & 8.2 & 10.2 & 7.1  \\
\hline
\end{tabular}
\label{tb:identity_dist}
\end{table}

}


\ignore{

\revised{

\subsection{Adaptive Attacks}
\label{subsec:adaptive}

\noindent\textbf{Parameter inference attack}. Our defense SCAn has a critical parameter $S_{\varepsilon}$, according it we split images of one class into two subgroups (Eqn.~\ref{eq:split}) and further calculate the judging statistic (Eqn.~\ref{eq:statistic}). If it is guessed out by the adversary (due to the transferability), there is a risk that the adversary can bypass SCAn. To investigate how likely they can guess out the $S_{\varepsilon}$, we built the following experiment. We trained 100 models on GTSRB with the same data, the same structure and the same hyper-parameters. The only different between these models is their initial values of inner-parameters (random initialization). Then, we used these 100 models to produce representations for images in the test set of GTSRB. Based on each model's representations, we calculated the corresponding $S_{\varepsilon}$. Thus, we got 4950 ($= \sum_{i=1}^{99} i$) different pairs of $S_{\varepsilon}$. Fig.~\ref{fig:cdf_dist_se} illustrates the Cumulative Distribution Function (CDF) of the distance between two $S_{\varepsilon}$ in one pair, comparing with the CDF of the norm of these 100 $S_{\varepsilon}$. We observe these two CDF are similar, which indicates the norm of the difference between two $S_{\varepsilon}$ from twice training is in the same scale of the norm of one $S_{\varepsilon}$. This large difference hinders adversaries correctly guess out the $S_{\varepsilon}$, even he use the same data, same structure and the same hyper-parameters to train his substitute model. 

\begin{figure}[ht]
    \begin{minipage}{0.22\textwidth}
     \centering
     \includegraphics[width=\linewidth]{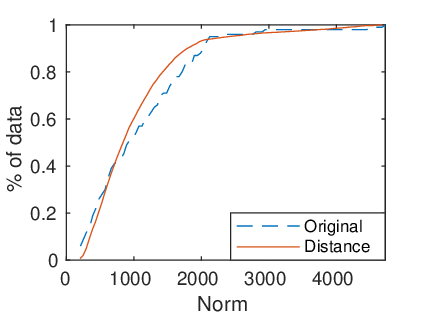}
     \caption{CDF of norms of $S_{\varepsilon}$ and the distance between a couple $S_{\varepsilon}$.}
     \label{fig:cdf_dist_se}
   \end{minipage}
   \begin{minipage}{0.22\textwidth}
     \centering
     \includegraphics[width=\linewidth]{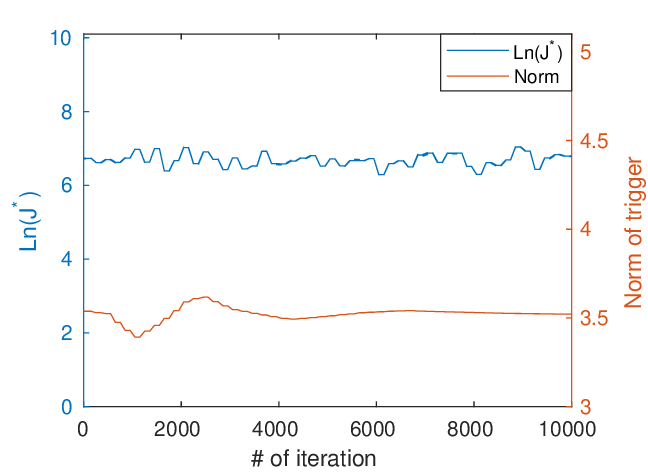}
     \caption{Statistics of black-box attacks (after moving-mean filtering).}
     \label{fig:blackbox_J}
   \end{minipage}
   
\end{figure}

\vspace{2pt}\noindent\textbf{Black-box attack}.
Another concern for SCAn is that the adversary may launch black-box attacks. To investigate how likely black-box attacks can bypass our method, we exploited the method proposed by Andrew et al.~\cite{blackbox_attack}, that can effectively find an adversarial example of the target model with limited queries by using a black-box derivative obtaining method improved from~\cite{zoo}. Here, we set the adversary goal as to find a trigger that can significantly lower the judging statistic $J^*$ of the target class and kept the other settings be the same with ~\cite{blackbox_attack}. Specifically, we iteratively ran 6 steps: 1) randomly sample a trigger disturbance, 2) launch TaCT with this disturbed trigger, 3) train the target model on the TaCT infected dataset, 4) use SCAn to get $J^*$, 5) calculate the derivative according to~\cite{blackbox_attack} and $J^*$ of the target class, and 6) update the trigger by subtracting the derivative, until the $J^*$ below our threshold $exp(2)$. Considering many training process need to be perform, we used GTSRB to do this experiment as it's training costs just several minutes. But, even in this setting, we still have not made $J^*$ lower than the threshold after 10000 iterations. Fig~\ref{fig:blackbox_J} demonstrates $J^*$ of the first 10000 iterations. We observe that $J^*$ still in the fluctuation at the 10000 step while the norm of the trigger (32x32 images with pixels in $[0,1]^3$) fluctuates a lot at first and keeps steady later, which means after searching the algorithm cannot find a proper trigger to bypass the SCAn. We have to notice that running 10000 iterations on GTSRB needs a month. From the above results, we conclude that launching black-box attack to bypass SCAn is severe time costly and SCAn can largely increases the overhead of the attack.

}

\revised{

\subsection{Robustness against Existing Attacks}
\label{subsec:robustbess}


\vspace{3pt}\noindent\textbf{Primary targeted data poisoning attacks}. 
Here, we evaluate the robustness of SCAn against the primary targeted data poisoning attacks. In these attacks, adversary poisons the dataset by injecting trigger-carrying images no matter which class they originally belong to, together with a specific label. As we analysed in the Section~\ref{subsec:understand}, this attack leads to source-agnostic backdoor that can be triggered by any class's image with the trigger pattern. Notice that this kind of attack is weaker than TaCT, thus current defenses can defeat it effectively. 

To demonstrate that, we tested SCAn, NC and AC on GTSRB. 
Specifically, our infected models were trained on a dataset poisoned by randomly selected 1000 trigger-carrying images that were all mislabeled as a specific target class. As NC and AC are used to detect the poisoned class, we iteratively chose every class as the target class and for each target class we used 32 different triggers (4 triggers centered on 8 randomly selected positions) to generate totally 43x32 infected models. The settings of AC are exactly the same as the original paper, and NC is performed by using their original code\footnote{https://github.com/bolunwang/backdoor.git}. The decomposition model of SCAn was built on 1000 clean images randomly selected from the test set (separated from the dataset used for training).
Table~\ref{tb:source-agnostic} demonstrates the performance of SCAn, NC and AC on those 1376 tests. We observe that all these three defenses perform well when facing the source-agnostic attacks, and have negligible False Positive Rate (FPR) at high True Positive Rate (TPR).
\begin{table}[tbh]
\caption{Performance on GTSRB against source-agnostic attacks.}
\begin{adjustbox}{width=0.45\textwidth}
    \begin{tabular}{|c|c|c|c|c|c|c|c|c|c|}
    \hline
           & \multicolumn{3}{c|}{SCAn} & \multicolumn{3}{c|}{NC} & \multicolumn{3}{c|}{AC}  \\ 
    \hline
   TPR     & 95.0\%& 99.5\%& 99.9\%& 95.0\%& 99.5\%& 99.9\%& 95.0\%& 99.5\%& 99.9\%\\  \hline 
   FPR     & 0\%   & 0\%   & 0\%   & 0\%   & 0.02\%   &  0.05\%&    0\%&    0.04\%& 0.17\%\\  \hline
    
    \end{tabular}
\end{adjustbox}
\label{tb:source-agnostic}
\end{table}

\ignore{
\begin{figure}[ht]
     \centering
     \includegraphics[width=\linewidth]{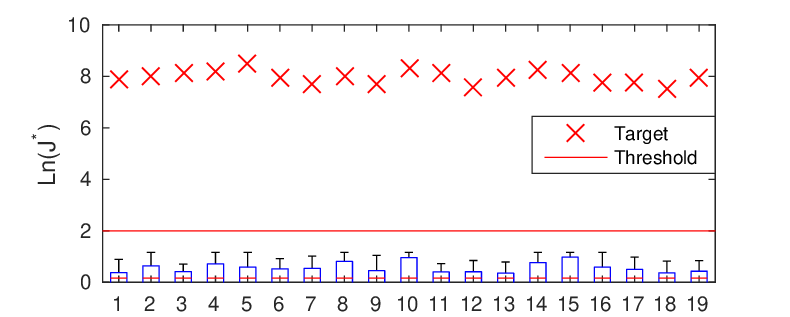}
     \caption{$J^*$ of the target classes under source-agnostic attacks, where source label is 0 and the target label ranges from 1 to 19.}
     \label{fig:source_agnostic}
  
   \vspace{-0.25in}
\end{figure}
}

\vspace{2pt}\noindent\textbf{Primary attacks in online settings}. 
In online settings, the images are fed one by one to defense schemes that want to specify the legitimacy of every single image. For making SCAn adaptive to online settings, we first built the decomposition model and untangling model on the clean dataset, and for every coming image we only updated the untangling model for the corresponding class. Based to the untangling results, we split the class into two subgroups and recorded which group the coming image belongs to. Further, we calculated out the statistic $J^*$ of this class. Finally, we specified such coming images that belong to the class with $J^*$ higher than the threshold and also belong to the subgroup having less clean images than the counterpart as attack images, and take others as benign images.

Further, we compared SCAn with SentiNet and STRIP on CIFAR10~\cite{krizhevsky2009learning} in online settings. We trained infected models on a dataset poisoned by 1000 images. For evaluating SCAn, we randomly selected 1000 images from the test set of CIFAR10 as the clean dataset. The settings of SentiNet are exactly the same with its original paper, and STRIP was performed by using their original code~\footnote{https://github.com/garrisongys/STRIP.git}. Following the same testing setting of STRIP, we randomly selected 4000 images to build the test set. The half of them are benign images and the rest are attack images. From the results demonstrated in Table~\ref{tb:online_compare}, we observe that all these three methods perform well in online settings. But SCAn has a little more FPR than others. That is due to SCAn needs several attack images to accumulate enough error in the early stage. According to our calibration, it needs about 50 attack images to steadily filter out further attack images.  
\begin{table}[tbh]
\centering
\caption{Online Performance on CIFAR10 against source-agnostic attacks.}
\begin{adjustbox}{width=0.45\textwidth}
    \begin{tabular}{|c|c|c|c|c|c|c|c|c|c|}
    \hline
           & \multicolumn{3}{c|}{SCAn} & \multicolumn{3}{c|}{SentiNet} & \multicolumn{3}{c|}{STRIP}  \\ 
    \hline
   TPR     & 95.0\%& 99.5\%& 99.9\%& 95.0\%& 99.5\%& 99.9\%& 95.0\%& 99.5\%& 99.9\%\\  \hline 
   FPR     & 0.19\%& 0.21\%& 0.34\%&    0\%& 0.05\%& 0.05\%&    0\%&    0\%& 0\%\\  \hline
    
    \end{tabular}
\end{adjustbox}
\label{tb:online_compare}
\end{table}

}

Further we evaluated the robustness of SCAn in the presence of a black-box adversary who is knowledgeable about our technique and attempts to evade our detection. More specifically, we consider two attack strategies the adversary could employ:

\noindent$\bullet$\textit{Multiple target classes with different triggers}. In this scenario, multiple backdoors targeting distinct labels are inserted into the target model, with each backdoor controlled by a specific trigger.

\noindent$\bullet$\textit{Exploit on the decomposition process}.  This scenario is also named as the \textit{black-box attack}, where the attacker somehow knows the training data including those clean data used to build the decomposition model, as well as the model structures but not the parameters of the target model.

\vspace{2pt}\noindent\textbf{Multiple Target Classes with Separate Triggers}. Even though multiple target classes will not directly change the value of our test statistic ($J^*$) for each class, they may change the overall distribution of $J^*$ and the behavior of the target model.

We evaluate the robustness of SCAn against this advanced attack on GTSRB. Specifically, we launched TaCT by using different number of independent triggers targeting distinctive labels and injecting 2\% attack images and 1\% covering images into the training set for each trigger. These triggers are all with the same shape of the square trigger (Fig.~\ref{fig:triggers:square}) but in exclusive color patterns. Besides, we perform SCAn by building the composition model on a set of clean data occupying 1\% of the whole dataset.

From Fig.~\ref{fig:multi_target:a}, we observe that once more than 8 triggers have been injected, the test statistic of the target class may be less than the threshold. But we argue that increasing the amount of known clean data for decomposition analyses can mitigate this threat. In fact, As demonstrated on Fig.~\ref{fig:multi_target:b}, when we use the clean data comprising 3\% of the whole dataset, the test statistics for the target classes becomes greater than the threshold, even when as many as 10 different backdoors are injected.

\vspace{2pt}\noindent\textbf{Leaked Dataset and Model Structure.}
As demonstrated in above experiments, 3\% clean data are sufficient for SCAn to defeat multiple triggers attack, which implies that the covariance matrix $S_{\varepsilon}$ can be recovered by attackers if they also have a similar amount of clean data. In this situation, the only secret that prevents attackers to bypass SCAn is the identity vector of representations produced by the target model for those normal images in the target class. To the attacker's advantage, we assume he can somehow find an approach to launch a perfect injection bypassing SCAn, when he knows both the covariance matrix and the identity vector.

A straightforward question is how accurate can the attacker to estimate the real identity vector, when he knows the whole dataset and the model structure. To answer this question, we trained five substitute models for each target model by using the same training set, the same model structure and also the same hyperparameters. After that, we perform the same method used by SCAn to build the decomposition model and adapt it to compute the covariance matrix and the identity vector of the target class. The first row of Table~\ref{tb:identity_dist} shows the average distance between the estimated and the real identity vector (estimation error). To compare with their scale, we also show in the second row the average distance between two identity vectors from different classes (identity distance). The result shows that the estimation error is in the same scale as the identity distance, indicating that even knowing the whole dataset and the model structure, attackers can not accurately estimate the identity vector of the target class: the estimation error may be as great as the distances between identity vectors from two different classes.

We further investigate the efficacy of SCAn on those distances by testing whether it can distinguishes two groups of representations that are as far away from each other as the average estimation error. Specifically, we first calculate the difference vector between the estimated and real identity vectors, and then add this difference vector onto a subgroup of representations belonging to a randomly chosen class. As a result, we built a ``mixed'' class containing two subgroups of representations: the altered ones from another class and the un-altered ones originally belonging to this class. We then run SCAn to check whether this ``mixed'' class is infected.
The third row illustrates the average test statistics $J^*$ in their logarithm scale for the ``mixed'' classes, which are sufficiently high, indicating that even under this attacker-advantage setting, attackers can only infer the identity vector with so high error that infected class can still be detected by the SCAn defense.

\begin{table}[tbh]
\centering
\caption{SCAn results when attackers know the whole dataset and model structure.}
\begin{tabular}{|c|c|c|c|}
\hline
     & GTSRB & ILCVR2012 & MegaFace\\
    \hline
Estimation error & 15.6 & 5.3 & 12.9 \\
\hline
Identity distance & 26.7 & 9.6 & 19.4 \\
\hline
$Ln(J^*)$ & 8.2 & 10.2 & 7.1  \\
\hline
\end{tabular}
\label{tb:identity_dist}
\end{table}

Through above analysis, we conclude that the identity vector of every class is a secret that attackers can hardly infer unless they know exactly the inner parameters of the target model.
}

\ignore{
Guided by~\cite{wangneural2019} where authors lists five types of advanced backdoor attacks, we discuss two of them in this section.
The rest three types attacks actually have been covered in Section~\ref{subsec:effectivness} where our defense works well on them.

\vspace{3pt}\noindent$\bullet$\textit{Multiple Infected Labels with Separate Triggers}

In this experiment, for $n$ triggers case, we set $(i-1)$ as the target label for the $i^{th}$ trigger. Fig.~\ref{fig:mvm} demonstrates results of different $n$. We observe that when the $n$ increase, our defence performance drop. We discuss that this is because the global model we built is affected by the infected data.

There is a very easy mitigation method. We build the global model just on the predictions of hold-on data. Fig.~\ref{fig:mit_mvm} shows that this mitigation works perfect. While this mitigation requires big enough hold-on dataset which is one limitation of our defence.

\vspace{2pt}\noindent$\bullet$\textit{Single Infected Label with Multiple Triggers}

As usual, we set 0 as the target label, and try to inject multiple triggers for label 0. As shown in Fig.~\ref{fig:ovm}, when the number of triggers increase, $\tilde{J}_0$ varies and the classification performance decrease. And once more than 8 triggers have been injected the classification performance drop significantly. This will break the stealthiness of attack.

Actually, in the view of our defence, this more-in-one attack is no different between normal cases, compact multiple classes into one class. Thus our defence works well as usual.

\subsection{Robustness}
known dataset rate.
multi trigger
white-box attack

}

\ignore{

We found that in the presence of source-specific backdoors, which can be easily injected through TaCT, the representations of attack samples (with triggers) become less distinguishable from those of normal images, rendering all existing detection techniques ineffective. A key observation of our research, however, shows that though trigger-carrying samples indeed come close to normal samples for the target model, they subtly alter the distribution of the representations for the target label, which could expose the existence of the backdoor.  Following we

In this section, we introduce a novel defense strategy called \textit{Statistical Contamination Analyzer} (SCAn), which aims to detect backdoor contamination by addressing the fundamentally different statistical properties between an infected class and uninfected, intact classes in a contaminated DNN model .

\subsection{Basic Ideas}
\label{subsec:intuition}

In the previous section, we demonstrated that the current backdoor detection approaches all attempted to separate {\em trigger} and {\em identity} features in an infected model, and thus can be defeated by a targeted contamination attack (TaCT), in which an adversary can manipulate the DNN model in such a way that the trigger and identify features are entangled and all features are used for both tasks of the normal classification and the recognition of the trigger. Nonetheless, an adversary cannot bypass the ultimate goal of backdoor attack, that is to {\em merge} attack samples into a target class, leading to a fundamental distinction between an infected class and the remaining intact classes: the infected class represents a mixture of two subgroups of samples, the attack samples and cover samples, while the intact classes represent a homogeneous group of samples. To quantify this statistical distinction between infected and intact classes, we made two further assumptions on the sample representation of the DNN model to be analyzed.




\noindent\textbf{Two-component decomposition assumption.}  \textit{
In representation space, each sample point can be decomposed into two independent components, represented by a class-specific {\em identity} component and a {\em variation} component, respectively.
}

\noindent\textbf{Universal variation assumption.}  \textit{
The variation vectors for all classes, including the intact target class without infection and the group of attack samples, follow the same distribution. 
}

Previously, Wang et al.~\cite{wang2004unified} demonstrated that, in face recognition, the input image data can be decomposed into three mutually orthogonal components: within-class, between-class and noise. Here, we generalize the two-component decomposition to the representation space of the DNN model: through robust representation learning, the noise in the input data are largely eliminated, and thus only the first two components are retained in the representation of the input data.

We argue the two-components decomposition assumption is reasonable for many DNN models, in particular those for computer vision and imaging analysis tasks. For instance, in face recognition, an input image contains the information associated with the identity of an individual (the {\em identity} component) as well as that associated with the transformation of the face, e.g., size, orientation, etc, (the {\em variation} component). Although the variation component does not contribute directly to the classification task in a DNN model, it is often extracted through the representation learning as it represents the recurrent and robust signal in the input data, and the representation learning by a non-ideal DNN model may not be able to distinguish them from the identity component. We note that previous backdoor detection approaches neglected the variation component in the representation space, resulting in lower sensitivity in detecting targeted contamination attacks (TaCT).

We further assume that the variation component of an input sample is {\em independent} of the class of the sample; as a result, the distribution learned from the input samples from one class (e.g., an intact class) can be transferred to another class (e.g., the target class without infection). Again, we argue this assumption is valid for many classification tasks including face recognition, traffic sign recognition, etc. For instance, in face recognition, smile is a variation component adopted by different human individuals, leading to the common transformation of face images independent of the identity of each individual (i.e., the class label).

Fig.~\ref{fig:idea_representations} illustrates the overall idea of SCAn using a schematic example, where the representations of samples in an infected class (right) can be viewed as a mixture of two groups, the attack samples and the cover samples, each decomposed into a distinct identity component and a common variation component. In comparison, without the two-component decomposition, the representations of the samples in the infected and intact class are indistinguishable   .

Formally, based on the above two assumptions, the representation of an input sample $x$ can be decomposed into two latent vectors:
\begin{equation}
\begin{array}{c@{\quad}l}
r = R(x) = \mu_t + \varepsilon
\end{array}
\label{decompose}
\end{equation}
where $\mu_t$ is the identify vector (component) of the specific class $t$ that $x$ is labelled as, and $\varepsilon$ is the variation vector of $x$, which follows the a distribution independent of $t$. We denote $\mathcal{X}_t$ as the subset of the samples with the class label $t$, and $\mathcal{R}_t$ as the set of their representations, i.e., $\mathcal{R}_t = \{R(x_i)| x_i \in \mathcal{X}_t\}$.

Due to the ultimate goal of the backdoor attack, the set of contaminated samples in an infected class $t^*$ (e.g., the training samples provided by an adversary targeting a specific class) always consists of two non-overlapping subgroups, the subgroup of cover samples and the subgroup of attack samples, i.e., $\mathcal{R}_{t^*}=\mathcal{R}^{normal}_{t^*} \cup \mathcal{R}^{trigger}_{t^*}$. As a result, the representations of the samples in an infected class follow a multivariate mixture distribution: for each $x_i \in \mathcal{X}_{t^*}$,

\begin{equation}
r_i=\delta_i \mu_{1} + (1-\delta_i) \mu_{2} + \varepsilon,
\label{mixturemodel}
\end{equation}

\noindent where $\mu_{1}$ and $\mu_{2}$ represent the identify vectors of the subgroup of cover samples in the class $t^*$ and the subgroup of attack samples, respectively, $\delta_i=1$ if the $x_i$ is a normal sample, and otherwise $\delta_i=0$. On the other hand, the representations of samples in an uninfected, intact class $t$ form a homogeneous population: $r = \mu_t + \varepsilon$.
Therefore, the task of backdoor detection can be formulated as a hypothesis testing problem: given the representations of input data from a specific class $t$, we want to test if it is more likely from a mixture group (as defined in (\ref{mixturemodel})) or from a single group (as defined in (\ref{decompose})). Notably, the problem is non-trivial because the input vectors are in high dimension (hundreds of features are learned in a typical DNN model), and more importantly, the parameters (i. e., $\mu_{t}$ and $\varepsilon$) are unknown for the mixture model and needs to be derived simultaneously with the hypothesis tests.
Finally, our approach does not rely on the assumption adopted by current backdoor detection techniques (see section \ref{sec:attack}) that the features in the DNN model can be separated into trigger and normal features. Instead, we investigate the distributions of the representations from all input samples labeled with each class: the class containing a mixture of two groups of vectors is considered to be contaminated.






\begin{figure}[ht]
	\centering
  \begin{subfigure}{0.23\textwidth}
		\includegraphics[width=\textwidth]{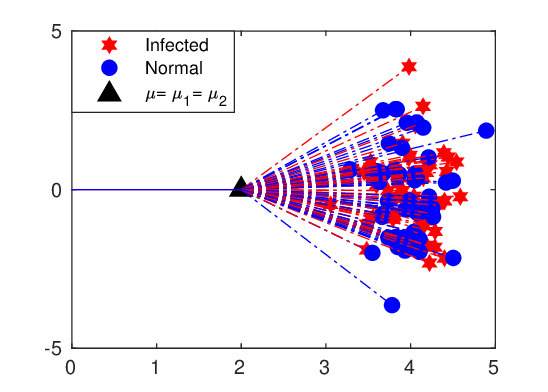}
  \end{subfigure}
  \begin{subfigure}{0.23\textwidth}
		\includegraphics[width=\textwidth]{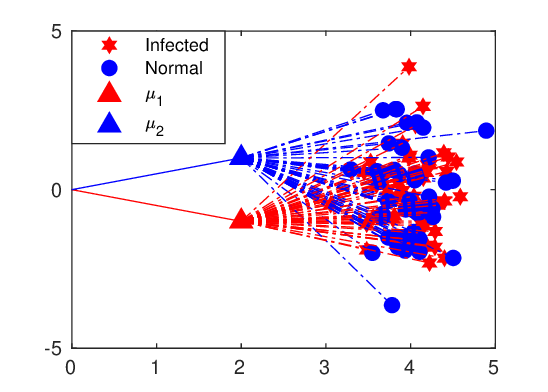}
  \end{subfigure}
	\caption{A schematic illustration of the assumption of two-component decomposition (right) in the representation space, in comparison with the naive homogeneous assumption (left).}
	\label{fig:idea_representations}
\end{figure}

\begin{figure*}[ht]
	\centering
  \begin{subfigure}{\textwidth}
		\includegraphics[width=\textwidth]{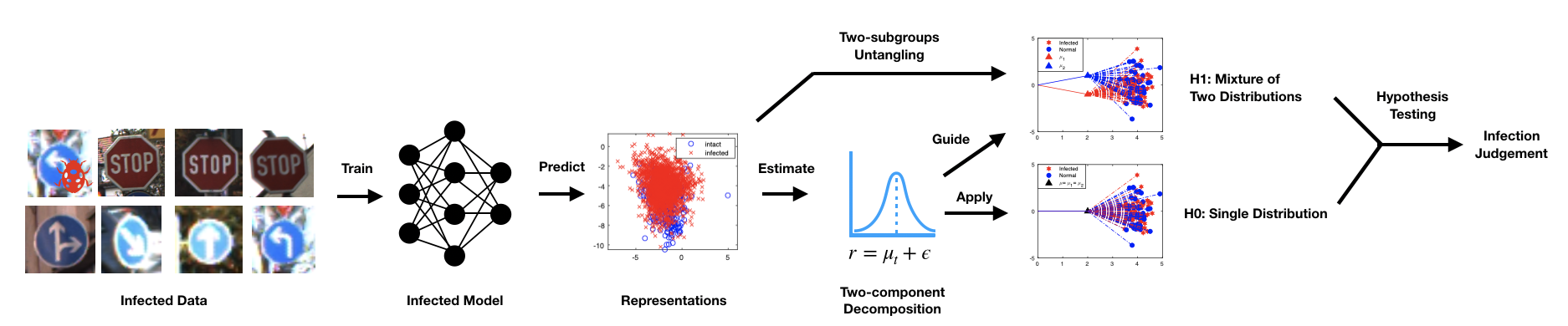}
  \end{subfigure}

	\caption{An illustration of Statistical Contamination Analyzer.}
	\label{fig:defense}

\end{figure*}




\vspace{5pt}\noindent\textbf{Algorithm}.
Our defence adopts a maximum likelihood approach, and utilizes an Expectation-Maximization (EM) algorithm to estimate the most likely parameters of the mixture model. Precisely, it consists of four steps, as demonstrated in Fig.~\ref{fig:defense}

\noindent\textit{Step 1:} Leverage the target model to generate representations for all input data (images) from training set that contains both the infected and normal images, and test the overall performance of the model. In general, the overall performance should be good. 

\noindent\textit{Step 2:} Estimate the parameters in model (\ref{decompose}) and decompose every representation.

\noindent\textit{Step 3:} For samples in each class, fit the representations into a mixture model (\ref{mixturemodel}) containing two subgroups. 

\noindent\textit{Step 4:} For the samples in each class, conduct the likelihood ratio test on their representations using the mixture model (\ref{mixturemodel}) (from step 3) against the the null hypothesis model (\ref{decompose}) (from step 2); if the null hypothesis is rejected, the corresponding class is reported to be infected (i.e., the training data labelled as this class contain contaminated samples).


\subsection{Key Techniques}
\label{subsec:approach}

Next, we present the technical details of our Statistical Contamination Analyzer (SCAn) approach. Some mathematical details are presented in the the Appendix. Notably, in this section, the representation vectors are centralized prior to further analyses.

\vspace{5pt}\noindent\textbf{Two-components decomposition}.
Based on the assumptions above, a representation vector can expressed as the superposition of two latent vectors: $r = \mu + \varepsilon$, and $\mu$ and $\varepsilon$ each follows a normal distribution

\ignore{
explain the representation $r$ as a random variable equaling to the sum of other two independent random variables $\mu$ and $\varepsilon$.

\begin{equation}
\begin{array}{c@{\quad}l}
r = \mu + \varepsilon
\end{array}
\label{eqn:r}
\end{equation}

We assume that the latent variable $\mu$ and $\varepsilon$ follow two normal distributions:
}

\begin{equation}
\begin{array}{c@{\quad}l}
\mu \sim N(0,S_{\mu}) \\
\varepsilon \sim N(0,S_{\varepsilon}) \\
\end{array}
\end{equation}

\noindent where $S_{\mu}$ and $S_{\varepsilon}$ are two unknown covariance matrices, which needs to be estimated from the representations of input data.
Notably, $S_{\mu}$ may be approximated by the between-class covariance matrix and $S_{\varepsilon}$ by the within-class covariance matrix; however, these approximations were previously shown to be less effective than an EM algorithm~\cite{chen2012bayesian}. Therefore, we adopt an EM algorithm to estimate most likely model parameters.

\noindent{\em E-step:} Following the decomposition model (\ref{decompose}), we express the relationship between
the observation $\mathbf{r} = [r_1;...;r_m]$ (e.g., for $m$ images) and the latent vectors $\mathbf{h} = [\mu;\varepsilon_1;...;\varepsilon_m]$ in the matrix form as:

\begin{equation}
\begin{array}{c@{\quad}l}
\mathbf{r} = \mathbf{T} \mathbf{h}, & \text{where  } \mathbf{T} =
\begin{bmatrix}
\mathbf{I} & \mathbf{I} & \mathbf{0} & \cdots & \mathbf{0} \\
\mathbf{I} & \mathbf{0} & \mathbf{I} & \cdots & \mathbf{0} \\
\vdots & \vdots & \vdots & \ddots & \vdots \\
\mathbf{I} & \mathbf{0} & \mathbf{0} & ... & \mathbf{I} \\
\end{bmatrix}
\end{array}
\end{equation}
Thus, $\mathbf{h} \sim N(0,\Sigma_h)$, where
\begin{equation}
\begin{array}{c@{\quad}l}
\Sigma_h =
\begin{bmatrix}
S_{\mu} & \mathbf{0} & \mathbf{0} & \cdots & \mathbf{0} \\
\mathbf{0} & S_{\varepsilon} & \mathbf{0} & \cdots & \mathbf{0} \\
\mathbf{0} & \mathbf{0} & S_{\varepsilon} & \cdots & \mathbf{0} \\
\vdots & \vdots & \vdots & \ddots & \vdots \\
\mathbf{0} & \mathbf{0} & \mathbf{0} & ... & S_{\varepsilon} \\
\end{bmatrix} \\
\end{array}
\end{equation}
and $\mathbf{r} \sim N(0,\Sigma_r)$, where
\begin{equation}
\begin{array}{c@{\quad}l}
\Sigma_r =
\begin{bmatrix}
S_{\mu}+S_{\varepsilon} & S_{\mu} & \cdots & S_{\mu}\\
S_{\mu} & S_{\mu}+S_{\varepsilon} & \cdots & S_{\mu} \\
\vdots & \vdots & \ddots & \vdots \\
S_{\mu} & S_{\mu} & ... & S_{\mu}+S_{\varepsilon} \\
\end{bmatrix} \\
\end{array}
\end{equation}
Hence, given the observation $\mathbf{r}$ and model parameters $S_{\mu}$ and $S_{\varepsilon}$, the expectation of $\mathbf{h}$ can be computed ~\cite{Multivar47:online} by:

\begin{equation}
\begin{array}{c@{\quad}l}
E(\mathbf{h} | \mathbf{r}) = \Sigma_h \mathbf{T}^T \Sigma_{r}^{-1} \mathbf{r}
\end{array}
\label{eqn:global_expectation}
\end{equation}

\noindent{\em M-step:} In this step, we try to obtain the most likely parameters of $S_{\mu}$ and $S_{\varepsilon}$ leading to the maximum expectation of $\mathbf{h}$ (i.e., $\mu$ and $\varepsilon$) in Eqn.~\ref{eqn:global_expectation}. A straightforward update rule of $S_{\mu}$ and $S_{\varepsilon}$ is:
\begin{equation}
\begin{array}{c@{\quad}l}
S_{\mu} &= \mathbf{cov}(\mu) \\
S_{\varepsilon} &= \mathbf{cov}(\varepsilon) \\
\end{array}
\end{equation}
where $\mu$ and $\varepsilon$ are the expected vectors computed in the previous E-step. Even though we do not have a formal proof for the convergence of the Expectation-Maximization algorithm, it is shown to be effective in our experiments.

\vspace{5pt}\noindent\textbf{Two-subgroups untangling}.
We assume the representations of the infected samples follow a mixture model of two Gaussian distributions, one for the group of cover samples ($N(\mu_1, S_1)$) and one for the group of attack samples ($N(\mu_2, S_2)$. If the class labels are known for each sample, a hyperplane that maximizes between-class variation versus within-class variation among the projected of representation vectors can be determined by using Linear Discriminant Analysis (LDA) ~\cite{}, which
maximizes the Fisher's Linear Discriminant (FLD):

\begin{equation}
\begin{array}{c@{\quad}l}
\text{FLD}(v) = v^T \Sigma_B v / v^T \Sigma_W v
\end{array}
\end{equation}
where
\begin{equation}
\notag
\begin{array}{c@{\quad}l}
\Sigma_B &= (\mu_1-\mu_2)(\mu_1-\mu_2)^T \\
\Sigma_W &= S_1 + S_2
\end{array}
\end{equation}
Empirically, greater FLD corresponds to distant projected means and concentrated projected vectors for each of the two subgroups of samples (Appendix~\ref{app:local_model}).
However, in our case, the labels ({\em cover} or {\em attack}) of the representations are unknown, and thus we can not estimate the mean and covariance matrix for every subgroup of samples. To address this challenge, we first assume $S_1 = S_2 = S_{\varepsilon}$ according to the {\em universal variation assumption},
and use an iterative algorithm to simultaneously estimate the model parameters ($\mu_1$, $\mu_2$ and $S_{\varepsilon}$) and the subgroup label for each sample in the class.

\noindent{\em Step-1:} Suppose we know the class labels for the $m$ input samples, Then we can estimate the model parameters ($\mu_1$, $\mu_2$ and $S_{\varepsilon}$) on the representations of cover samples and attack samples, respectively, using the similar method as presented above in the {\em E-step}.

\noindent{Step-2:} After obtaining the model parameters, we compute the optimal discriminating hyperplane (denoted by its normal vector $v$) by maximizing the FLD,

\begin{equation}
\begin{array}{c@{\quad}l}
v &= S_{\varepsilon}^{-1} (\mu_1-\mu_2)
\end{array}
\end{equation}
from which we can re-compute the subgroup label $c_i$ for each sample $i$
($c_i=1$ if the $i$-th sample is a cover sample, and $c_i=2$ if it is an attack sample),
\begin{equation}
\begin{array}{c@{\quad}l}
c_i =
\begin{cases}
1, v^T r < t\\
2, v^T r \ge t\\
\end{cases}
\end{array}
\label{eq:z}
\end{equation}
where
\begin{equation}
\notag
\begin{array}{c@{\quad}l}
t &= \frac{1}{2} (\mu_1^T S_{\varepsilon}^{-1} \mu_1 - \mu_2^T S_{\varepsilon}^{-1} \mu_2)
\end{array}
\end{equation}

\noindent{Step 3:} Iteratively execute Step-1 and Step-2 until convergence. Finally, we will simultaneously obtain the model parameters and the subgroup labels of all samples in the class of interest.

\noindent\textbf{Hypothesis testing.} For each class $t$, we aim to determine if the samples in the class is contaminated by using a likelihood ratio test over the samples ($\mathcal{R}_t$) in the class based on two hypotheses:

\noindent (null hypothesis) $\mathbf{H_0}:$ $\mathcal{R}_t$ is drawn from a single normal distribution.

\noindent (alternative hypothesis) $\mathbf{H_1}:$ $\mathcal{R}_t$ is drawn from a mixture of two normal distributions.

\noindent and the test statistic is defined as:
\begin{equation}
\begin{array}{c@{\quad}l}
J_t = -2 \log{\frac{P(\mathcal{R}_t | \mathbf{H_0})}{P(\mathcal{R}_t | \mathbf{H_1})}}
\end{array}
\label{eqn:criteria}
\end{equation}
where
\begin{equation}
\notag
\begin{array}{c@{\quad}l}

P(\mathcal{R}_t | \mathbf{H_0}) &= \Pi_{r \in \mathcal{R}_t} N(r | \mu_t,S_{\varepsilon})\\
P(\mathcal{R}_t | \mathbf{H_1}) &= \Pi_{i: c_i=1} N(r_i| \mu_1, S_{\varepsilon}) \Pi_{i: c_i=0} N(r_i | \mu_2, S_{\varepsilon})\\
\end{array}
\end{equation}
Applying Eqn.~\ref{eqn:criteria}, we can simplify the likelihood ratio,
\begin{equation}
\notag
\begin{array}{c@{\quad}l}
J_t & = 2 \log(P(\mathcal{R}_t | \mathbf{H_1}) / {P(\mathcal{R}_t | \mathbf{H_0}))} \\
& = \sum_{r \in \mathcal{R}_t}[(r-\mu_t)^T S_{\varepsilon}^{-1} (r-\mu_t) - (r-\mu_j)^T S_{\varepsilon}^{-1} (r-\mu_j)]
\end{array}
\end{equation}
where $j \in \{1,2\}$ is the subgroup label of the representation $r$.

According to Wilks' theorem~\cite{wilks1938large}, our test statistic $J_t$ follows a $\chi^2$ distribution with the degree of freedom equal to the difference between the numbers of free parameters in the null and alternative hypotheses, respectively. In our case, however, the degree of freedom may be as large as tens of thousands, and thus it is difficult to compute the p-value using the $\chi^2$ distribution
Fortunately, according to the central limit theorem~\cite{wiki:chi-squared},
\begin{equation}
\notag
\begin{array}{c@{\quad}l}
\bar{J}_t = (J_t-k) / \sqrt{2k}
\end{array}
\end{equation}
approaches the standard normal distribution, where $k$ is the degree of freedom of the underlying $\chi^2$ distribution. Therefore, we leverage the normal distribution of the Median Absolute Deviation (MAD)~\cite{leys2013detecting} to detect the class(es) with abnormally great values of $J$. Specifically, we use
\begin{equation}
\notag
\begin{array}{c@{\quad}l}
J_t^* = |\bar{J}_t-\tilde{J}| / (\text{MAD}(\bar{J}) * 1.4826)
\end{array}
\end{equation}
where
\begin{equation}
\notag
\begin{array}{c@{\quad}l}
\tilde{J} &= median(\{\bar{J}_t: t \in \mathcal{L} \}) \\
\text{MAD}(\bar{J}) &= median(\{|\bar{J}_t-\tilde{J}|: t \in \mathcal{L}\}) \\
\end{array}
\end{equation}
Here, the constant (1.4826) is a normalization value for the standard normal distribution followed by $\bar{J}_t$. Therefore, when $J_t^* > 7.3891 = exp(2)$, the null hypothesis $\mathbf{H_0}$ can be rejected with $> (1-1e^{-9})$ confidence, and thus the class $t$ is reported to be contaminated.



\subsection{Effectiveness}

In this section, we demonstrate the effectiveness of SCAn by defeating the TaCT, the attack has beaten all of existing defenses, on three fields: various tasks and triggers, less known clean data and absence of the target class.

\begin{figure}[ht]
	\centering
  \begin{subfigure}{0.1\textwidth}
		\includegraphics[width=\textwidth]{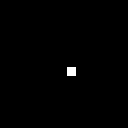}
		\caption{Square}
  \end{subfigure}
  \begin{subfigure}{0.1\textwidth}
		\includegraphics[width=\textwidth]{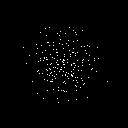}
		\caption{Normal}
  \end{subfigure}
  \begin{subfigure}{0.1\textwidth}
		\includegraphics[width=\textwidth]{figure/uniform.png}
		\caption{Uniform}
  \end{subfigure}
  \begin{subfigure}{0.1\textwidth}
		\includegraphics[width=\textwidth]{figure/watermark.png}
		\caption{Watermark}
  \end{subfigure}
	\caption{Four triggers}
	\label{fig:triggers}
\end{figure}

\vspace{3pt}\noindent\textbf{Various tasks and triggers}. We respectively launched TaCT on our three datasets (Table~\ref{tb:datasets_models}) and used four different triggers (Fig.~\ref{fig:triggers}). These three datasets not only cover different tasks but also distributions of data: GTSRB has small number of classes and images, ILSVRC2012 has many classes and each class has large number of images, MegaFace has tremendous classes but each class may has only a few images. Those four triggers also represents four different distributions of the trigger: square trigger is concentrated and with small norm, normal trigger is sparse and with small norm, uniform trigger is sparse and with large norm and  watermark trigger is concentrated and with large norm.

Specifically, to the attacker's advantage, we injected $2\%$ attack images and $1\%$ cover images into the training set of the target model. As illustrated in Table~\ref{tb:scan_top1} and \ref{tb:scan_mis}, all these infected models achieved compatible performance with the uninfected models and high targeted misclassification rate. As a defender, we assume 50\% of the whole dataset is clean and known previously. and estimate the parameters of model (\ref{decompose}) on these clean data. From our results, we discover that SCAn is very effective to defeat TaCT. The test statistic $J^*$ for the target class is greater by several orders of magnitude from those of non-target classes. We illustrated $ln(J^*)$ on Fig.~\ref{fig:effect_4triggers_3tasks}, which demonstrates, no matter on which task nor the trigger, SCAn supports a strong security guarantee and has achieved no false positive detected.

\begin{table}[tbh]
\centering
\caption{Top-1 accuracy of infected models.}
\begin{tabular}{|c|c|c|c|}
\hline
     & GTSRB & ILSVRC2012 & MegaFace\\
    \hline
Square & 96.6\% & 76.3\% & 71.1\% \\
\hline
Normal & 96.1\% & 76.1\% & 71.2\%\\
\hline
Uniform & 95.9\% & 75.9\% & 71.2\%\\
\hline
Watermark & 96.5\% & 75.5\% & 70.9\%\\
\hline
Uninfected & 96.4\% & 76.0\% & 71.4\%\\
\hline
\end{tabular}
\label{tb:scan_top1}
\end{table}

\begin{table}[tbh]
\centering
\caption{Targeted misclassification rate of infected models.}
\begin{tabular}{|c|c|c|c|}
\hline
   & GTSRB & ILSVRC2012 & MegaFace\\
    \hline
Square & 98.5\% & 98.2\% & 98.1\% \\
\hline
Normal & 82.4\% & 83.8\% & 81.4\%\\
\hline
Uniform & 94.9\% & 90.5\% & 88.2\%\\
\hline
Watermark & 99.3\% & 98.4\% & 97.1\%\\
\hline
\end{tabular}
\label{tb:scan_mis}
\end{table}

\begin{figure}[ht]
	\centering
  \begin{subfigure}{0.23\textwidth}
		\includegraphics[width=\textwidth]{figure/scan_square.eps}
		\caption{Square}
  \end{subfigure}
  \begin{subfigure}{0.23\textwidth}
		\includegraphics[width=\textwidth]{figure/scan_normal.eps}
		\caption{Normal}
  \end{subfigure}
  \begin{subfigure}{0.23\textwidth}
		\includegraphics[width=\textwidth]{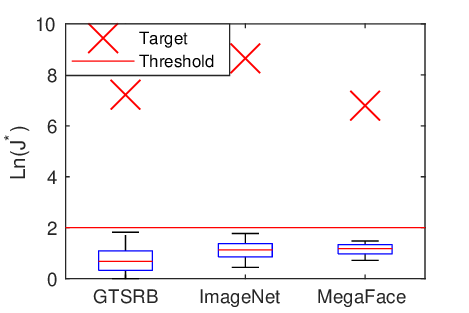}
		\caption{Uniform}
  \end{subfigure}
  \begin{subfigure}{0.23\textwidth}
		\includegraphics[width=\textwidth]{figure/scan_watermark.eps}
		\caption{Watermark}
  \end{subfigure}
	\caption{Detection results of SCAn on different datasets and triggers.}
	\label{fig:effect_4triggers_3tasks}
\end{figure}

\vspace{3pt}\noindent\textbf{Less known clean data}
SCAn relays on correct building the decomposition model (Eqn.~\ref{decompose}), which requires previous knowing a set of clean data. Our experiments demonstrates that even a small set of data is known before, SCAn still maintains the high effectiveness.

Specifically, using the similar manner as above, we injected 2\% attack images an 1\% cover images into the training set, but we change the knowing ratio of the clean data.

\subsection{Robustness}
\label{subsec:robustbess}

\vspace{3pt}\noindent\textbf{Extension to no clean data}

\vspace{3pt}\noindent\textbf{Customized attacks}

Guided by~\cite{wangneural2019} where authors lists five types of advanced backdoor attacks, we discuss two of them in this section.
The rest three types attacks actually have been covered in Section~\ref{subsec:effectivness} where our defense works well on them.

\vspace{3pt}\noindent$\bullet$\textit{Multiple Infected Labels with Separate Triggers}

In this experiment, for $n$ triggers case, we set $(i-1)$ as the target label for the $i^{th}$ trigger. Fig.~\ref{fig:mvm} demonstrates results of different $n$. We observe that when the $n$ increase, our defence performance drop. We discuss that this is because the global model we built is affected by the infected data.

There is a very easy mitigation method. We build the global model just on the predictions of hold-on data. Fig.~\ref{fig:mit_mvm} shows that this mitigation works perfect. While this mitigation requires big enough hold-on dataset which is one limitation of our defence.

\vspace{3pt}\noindent$\bullet$\textit{Single Infected Label with Multiple Triggers}

As usual, we set 0 as the target label, and try to inject multiple triggers for label 0. As shown in Fig.~\ref{fig:ovm}, when the number of triggers increase, $\tilde{J}_0$ varies and the classification performance decrease. And once more than 8 triggers have been injected the classification performance drop significantly. This will break the stealthiness of attack.

Actually, in the view of our defence, this more-in-one attack is no different between normal cases, compact multiple classes into one class. Thus our defence works well as usual.

}

\vspace{-0.1in}
\section{Discussion}
\label{sec:discussion}
\vspace{-0.1in}
\noindent\textbf{Limitations of SCAn}. As mentioned earlier, SCAn utilizes a set of clean data for the contamination analysis. We believe that this requirement is reasonable, since a small clean dataset is often provided by the model provider for testing the model's performance, as also assumed in the prior studies~\cite{strip,sentinet}. Note that the size of this clean dataset can be just 1\% of the training set for defeating the attack involving up to 8 triggers (Section~\ref{subsec:robustbess}). Also, our approach relies on the presence of attack images (carrying triggers) to identify an infected class. 
Further, we only evaluated SCAn on image classification tasks. However, we believe that there is a potential to extend our approach to mitigate the threat posed by the backdoor using a non-image trigger. Behind SCAn is our insight that the globally statistical information about a model's representations can help untangle a specific class. Such information is described in our research using the covariance matrix ($S_{\varepsilon}$) of multivariate normal distribution, which helps effectively untangle different classes. This finding indicates that the multivariate normal distribution provides a good description of high-dimensional representations generated from a large amount of data (tens of thousands images). Since such representations also characterize some non-vision tasks, such as code analysis, it is likely that our modeling can also apply to identify Trojaned inputs in these tasks. Further exploration on this direction is left to the future research.

\vspace{2pt}\noindent\textbf{Future research}. Down the road, we will seek more efficient techniques to untangle mixed representations, e.g., using deep learning with GPU acceleration, and more precise approximation for a specific task. As an interesting observation, our experiments 
on MegaFace show that the classes containing both baby and adult images have a higher $J^*$ than other normal classes, even though this anomaly is still well below those of infected classes. This may indicate that our method could help mine hard-negative examples, for evaluating a DNN model's classification quality.

\vspace{-0.1in}
\section{Related Works}
\label{sec:relatedWork}
\vspace{-0.1in}

We present a new protection SCAn that can defeat our attack TaCT designed for injecting source-specific triggers into the target model. Such a trigger has been briefly mentioned in NC~\cite{wangneural2019} and STRIP~\cite{strip}, without details about how to launch the attack. The NC paper discussed the potential to detect source-specific triggers when running NC $O(N log_2(N))$ rounds for $N$ classes. We argue that the computational complexity will increase to $O(N^2)$ in the presence of TaCT, given that NC's recall is just $6.5\%$ on TaCT,  as demonstrated in Section~\ref{subsec:current_defences}. As a result, the divided-and-conquer algorithm cannot be used to reduce the complexity, which makes the approach less practical when $N$ is huge (Table~\ref{tb:datasets_models}). 
By comparison, SCAn defeats TaCT with a complexity $O(N)$, by testing every class once. 
Liu et al.~\cite{liu2018fine} proposed the fine-pruning method. They first prune neurons that are dormant when processing clean data until the accuracy tested on a hold-on dataset below a threshold, and then fine-tune the pruned model to recover the accuracy. Their defense relies on extensive interactions with the training process. In contrast, our approach only needs to go through dataset in two rounds and is independent of the training of the target model.
Other related approaches, as discussed in Section~\ref{subsec:current_defences}, are all defeated by TaCT, with SCAn being the only solution working on the attack. 
Nelson et al.~\cite{nelson2009misleading} and Baracaldo et al.~\cite{baracaldo2017mitigating} proposed two general protections against backdoor attack. Both methods require extensive retraining of the model on the datasets with the similar size as the original one, which is often infeasible for DNNs. Additionally, they detect infected data by evaluating the overall performance of the model. However, the overall performance of the infected model often remains good under current advanced attacks (like TaCT), and thus these methods will become ineffective against these attacks.
In the traditional statistical analysis domain, a review written by Victoria et al.~\cite{hodge2004survey} summarizes several effective outlier detection methods, including k-nearest neighbors (k-nn)~\cite{knox1998knn}, k-means~\cite{nairac1999kmeans} and principal components analysis (pca)~\cite{korn1998pca}. To find out whether directly applying them to sample representations can detect infected classes, we ran these methods on the representations produced by a TaCT infected model for the images in the target class. The results on Fig.~\ref{fig:trad_stati} show that these methods cause many false positives. 

\vspace{-0.2in}
\section{Conclusion}
\label{sec:conclusion} 
\vspace{-0.1in}
Our work demonstrated that backdoors created by conventional data poisoning attacks are source-agnostic and characterized by unique representations generated for attack images, which are mostly determined by the trigger, regardless of other image content, and clearly distinguishable from those for normal images. Those four existing detection techniques rely on these proprieties and all fail to raise the bar to black-box attacks injecting source-specific backdoors like TaCT. 
Based on leveraging the distribution of the sample representations through a two-component model, we designed a statistical method SCAn to untangle representations of each class into a mixture model, and utilized a likelihood ratio test to detect an infected class. The effectiveness and robustness of SCAn were demonstrated through extensive experiments. 
Our study takes a step forward to understand the mechanism of implanting a backdoor within a DNN model and how a backdoor looks like from the perspective of model's representations. It may lead to deeper understanding of neural networks.


\section*{Acknowledgment}
\vspace{-0.1in}
We thank our anonymous reviewers for their comprehensive feedback.
This work was supported in part by the General Research Funds (Project No. 14208019) established under the University Grant Committee of the Hong Kong SAR., the Chinese University of Hong Kong research contract agreement (Contract No. TS1711490), and the IARPA (Grant No. W91NF-20-C-0034) the TrojAI project.

\bibliographystyle{plain}
\bibliography{main}

\begin{thebibliography}{10}

\bibitem{how_to_backdoor}
Eugene Bagdasaryan, Andreas Veit, Yiqing Hua, Deborah Estrin, and Vitaly
  Shmatikov.
\newblock How to backdoor federated learning.
\newblock {\em CoRR}, abs/1807.00459, 2018.

\bibitem{baracaldo2017mitigating}
Nathalie Baracaldo, Bryant Chen, Heiko Ludwig, and Jaehoon~Amir Safavi.
\newblock Mitigating poisoning attacks on machine learning models: A data
  provenance based approach.
\newblock In {\em Proceedings of the 10th ACM Workshop on Artificial
  Intelligence and Security}, pages 103--110. ACM, 2017.

\bibitem{box1978statistics}
George~EP Box, William~Gordon Hunter, J~Stuart Hunter, et~al.
\newblock {\em Statistics for experimenters}, volume 664.
\newblock John Wiley and sons New York, 1978.

\bibitem{ac_attack}
Bryant Chen, Wilka Carvalho, Nathalie Baracaldo, Heiko Ludwig, Benjamin
  Edwards, Taesung Lee, Ian Molloy, and Biplav Srivastava.
\newblock Detecting backdoor attacks on deep neural networks by activation
  clustering.
\newblock In {\em Workshop on Artificial Intelligence Safety 2019 co-located
  with the Thirty-Third {AAAI} Conference on Artificial Intelligence 2019
  (AAAI-19), Honolulu, Hawaii, January 27, 2019.}, 2019.

\bibitem{chen2012bayesian}
Dong Chen, Xudong Cao, Liwei Wang, Fang Wen, and Jian Sun.
\newblock Bayesian face revisited: {A} joint formulation.
\newblock In {\em Computer Vision - {ECCV} 2012 - 12th European Conference on
  Computer Vision, Florence, Italy, October 7-13, 2012, Proceedings, Part
  {III}}, pages 566--579, 2012.

\bibitem{zoo}
Pin{-}Yu Chen, Huan Zhang, Yash Sharma, Jinfeng Yi, and Cho{-}Jui Hsieh.
\newblock {ZOO:} zeroth order optimization based black-box attacks to deep
  neural networks without training substitute models.
\newblock In Bhavani~M. Thuraisingham, Battista Biggio, David~Mandell Freeman,
  Brad Miller, and Arunesh Sinha, editors, {\em Proceedings of the 10th {ACM}
  Workshop on Artificial Intelligence and Security, AISec@CCS 2017, Dallas, TX,
  USA, November 3, 2017}, pages 15--26. {ACM}, 2017.

\bibitem{chen2017targeted}
Xinyun Chen, Chang Liu, Bo~Li, Kimberly Lu, and Dawn Song.
\newblock Targeted backdoor attacks on deep learning systems using data
  poisoning.
\newblock {\em CoRR}, abs/1712.05526, 2017.

\bibitem{sentinet}
Edward Chou, Florian Tram{\`{e}}r, Giancarlo Pellegrino, and Dan Boneh.
\newblock Sentinet: Detecting physical attacks against deep learning systems.
\newblock {\em CoRR}, abs/1812.00292, 2018.

\bibitem{strip}
Yansong Gao, Change Xu, Derui Wang, Shiping Chen, Damith~Chinthana Ranasinghe,
  and Surya Nepal.
\newblock {STRIP:} a defence against trojan attacks on deep neural networks.
\newblock In David Balenson, editor, {\em Proceedings of the 35th Annual
  Computer Security Applications Conference, {ACSAC} 2019, San Juan, PR, USA,
  December 09-13, 2019}, pages 113--125. {ACM}, 2019.

\bibitem{gu2017badnets}
Tianyu Gu, Brendan Dolan{-}Gavitt, and Siddharth Garg.
\newblock Badnets: Identifying vulnerabilities in the machine learning model
  supply chain.
\newblock {\em CoRR}, abs/1708.06733, 2017.

\bibitem{he2016deep}
Kaiming He, Xiangyu Zhang, Shaoqing Ren, and Jian Sun.
\newblock Deep residual learning for image recognition.
\newblock In {\em Proceedings of the IEEE conference on computer vision and
  pattern recognition}, pages 770--778, 2016.

\bibitem{hodge2004survey}
Victoria~J. Hodge and Jim Austin.
\newblock A survey of outlier detection methodologies.
\newblock {\em Artif. Intell. Rev.}, 22(2):85--126, 2004.

\bibitem{blackbox_attack}
Andrew Ilyas, Logan Engstrom, Anish Athalye, and Jessy Lin.
\newblock Black-box adversarial attacks with limited queries and information.
\newblock In Jennifer~G. Dy and Andreas Krause, editors, {\em Proceedings of
  the 35th International Conference on Machine Learning, {ICML} 2018,
  Stockholmsm{\"{a}}ssan, Stockholm, Sweden, July 10-15, 2018}, volume~80 of
  {\em Proceedings of Machine Learning Research}, pages 2142--2151. {PMLR},
  2018.

\bibitem{knox1998knn}
Edwin~M Knox and Raymond~T Ng.
\newblock Algorithms for mining distancebased outliers in large datasets.
\newblock In {\em Proceedings of the international conference on very large
  data bases}, pages 392--403. Citeseer, 1998.

\bibitem{likelihood_ratio}
Karl{-}Rudolf Koch.
\newblock {\em Parameter estimation and hypothesis testing in linear models}.
\newblock Springer, 1988.

\bibitem{koh2017understanding}
Pang~Wei Koh and Percy Liang.
\newblock Understanding black-box predictions via influence functions.
\newblock In {\em Proceedings of the 34th International Conference on Machine
  Learning-Volume 70}, pages 1885--1894. JMLR. org, 2017.

\bibitem{korn1998pca}
Flip Korn, Alexandros Labrinidis, Yannis Kotidis, Christos Faloutsos, Alex
  Kaplunovich, and Dejan Perkovic.
\newblock Quantifiable data mining using principal component analysis.
\newblock Technical report, 1998.

\bibitem{krizhevsky2009learning}
Alex Krizhevsky and Geoffrey Hinton.
\newblock Learning multiple layers of features from tiny images.
\newblock Technical report, Citeseer, 2009.

\bibitem{lecun1998gradient}
Yann LeCun, L{\'e}on Bottou, Yoshua Bengio, and Patrick Haffner.
\newblock Gradient-based learning applied to document recognition.
\newblock {\em Proceedings of the IEEE}, 86(11):2278--2324, 1998.

\bibitem{leys2013detecting}
Christophe Leys, Christophe Ley, Olivier Klein, Philippe Bernard, and Laurent
  Licata.
\newblock Detecting outliers: Do not use standard deviation around the mean,
  use absolute deviation around the median.
\newblock {\em Journal of Experimental Social Psychology}, 49(4):764--766,
  2013.

\bibitem{li2018printracker}
Zhengxiong Li, Aditya~Singh Rathore, Chen Song, Sheng Wei, Yanzhi Wang, and
  Wenyao Xu.
\newblock Printracker: Fingerprinting 3d printers using commodity scanners.
\newblock In {\em Proceedings of the 2018 ACM SIGSAC Conference on Computer and
  Communications Security}, pages 1306--1323. ACM, 2018.

\bibitem{liu2018fine}
Kang Liu, Brendan Dolan{-}Gavitt, and Siddharth Garg.
\newblock Fine-pruning: Defending against backdooring attacks on deep neural
  networks.
\newblock In Michael Bailey, Thorsten Holz, Manolis Stamatogiannakis, and
  Sotiris Ioannidis, editors, {\em Research in Attacks, Intrusions, and
  Defenses - 21st International Symposium, {RAID} 2018, Heraklion, Crete,
  Greece, September 10-12, 2018, Proceedings}, volume 11050 of {\em Lecture
  Notes in Computer Science}, pages 273--294. Springer, 2018.

\bibitem{abs}
Yingqi Liu, Wen{-}Chuan Lee, Guanhong Tao, Shiqing Ma, Yousra Aafer, and
  Xiangyu Zhang.
\newblock {ABS:} scanning neural networks for back-doors by artificial brain
  stimulation.
\newblock In Lorenzo Cavallaro, Johannes Kinder, XiaoFeng Wang, and Jonathan
  Katz, editors, {\em Proceedings of the 2019 {ACM} {SIGSAC} Conference on
  Computer and Communications Security, {CCS} 2019, London, UK, November 11-15,
  2019}, pages 1265--1282. {ACM}, 2019.

\bibitem{liu2017trojaning}
Yingqi Liu, Shiqing Ma, Yousra Aafer, Wen{-}Chuan Lee, Juan Zhai, Weihang Wang,
  and Xiangyu Zhang.
\newblock Trojaning attack on neural networks.
\newblock In {\em 25th Annual Network and Distributed System Security
  Symposium, {NDSS} 2018, San Diego, California, USA, February 18-21, 2018},
  2018.

\bibitem{mika1999fisher}
Sebastian Mika, Gunnar Ratsch, Jason Weston, Bernhard Scholkopf, and
  Klaus-Robert Mullers.
\newblock Fisher discriminant analysis with kernels.
\newblock In {\em Neural networks for signal processing IX: Proceedings of the
  1999 IEEE signal processing society workshop (cat. no. 98th8468)}, pages
  41--48. Ieee, 1999.

\bibitem{nairac1999kmeans}
Alexandre Nairac, Neil Townsend, Roy Carr, Steve King, Peter Cowley, and Lionel
  Tarassenko.
\newblock A system for the analysis of jet engine vibration data.
\newblock {\em Integrated Computer-Aided Engineering}, 6(1):53--66, 1999.

\bibitem{nech2017level}
Aaron Nech and Ira Kemelmacher-Shlizerman.
\newblock Level playing field for million scale face recognition.
\newblock In {\em Proceedings of the IEEE Conference on Computer Vision and
  Pattern Recognition}, 2017.

\bibitem{nelson2009misleading}
Blaine Nelson, Marco Barreno, Fuching~Jack Chi, Anthony~D Joseph, Benjamin~IP
  Rubinstein, Udam Saini, Charles Sutton, JD~Tygar, and Kai Xia.
\newblock Misleading learners: Co-opting your spam filter.
\newblock In {\em Machine learning in cyber trust}, pages 17--51. Springer,
  2009.

\bibitem{ng2014data}
Hong-Wei Ng and Stefan Winkler.
\newblock A data-driven approach to cleaning large face datasets.
\newblock In {\em 2014 IEEE International Conference on Image Processing
  (ICIP)}, pages 343--347. IEEE, 2014.

\bibitem{GAN_defense}
Ximing Qiao, Yukun Yang, and Hai Li.
\newblock Defending neural backdoors via generative distribution modeling.
\newblock In Hanna~M. Wallach, Hugo Larochelle, Alina Beygelzimer, Florence
  d'Alch{\'{e}}{-}Buc, Emily~B. Fox, and Roman Garnett, editors, {\em Advances
  in Neural Information Processing Systems 32: Annual Conference on Neural
  Information Processing Systems 2019, NeurIPS 2019, 8-14 December 2019,
  Vancouver, BC, Canada}, pages 14004--14013, 2019.

\bibitem{ILSVRC15}
Olga Russakovsky, Jia Deng, Hao Su, Jonathan Krause, Sanjeev Satheesh, Sean Ma,
  Zhiheng Huang, Andrej Karpathy, Aditya Khosla, Michael Bernstein,
  Alexander~C. Berg, and Li~Fei-Fei.
\newblock {ImageNet Large Scale Visual Recognition Challenge}.
\newblock {\em International Journal of Computer Vision (IJCV)},
  115(3):211--252, 2015.

\bibitem{shafahi2018poison}
Ali Shafahi, W~Ronny Huang, Mahyar Najibi, Octavian Suciu, Christoph Studer,
  Tudor Dumitras, and Tom Goldstein.
\newblock Poison frogs! targeted clean-label poisoning attacks on neural
  networks.
\newblock In {\em Advances in Neural Information Processing Systems}, pages
  6103--6113, 2018.

\bibitem{simonyan2014very}
Karen Simonyan and Andrew Zisserman.
\newblock Very deep convolutional networks for large-scale image recognition.
\newblock In {\em 3rd International Conference on Learning Representations,
  {ICLR} 2015, San Diego, CA, USA, May 7-9, 2015, Conference Track
  Proceedings}, 2015.

\bibitem{sitawarin2018darts}
Chawin Sitawarin, Arjun~Nitin Bhagoji, Arsalan Mosenia, Mung Chiang, and
  Prateek Mittal.
\newblock {DARTS:} deceiving autonomous cars with toxic signs.
\newblock {\em CoRR}, abs/1802.06430, 2018.

\bibitem{Stallkamp2012}
Johannes Stallkamp, Marc Schlipsing, Jan Salmen, and Christian Igel.
\newblock Man vs. computer: Benchmarking machine learning algorithms for
  traffic sign recognition.
\newblock {\em Neural Networks}, 32:323--332, 2012.

\bibitem{szegedy2017inception}
Christian Szegedy, Sergey Ioffe, Vincent Vanhoucke, and Alexander~A Alemi.
\newblock Inception-v4, inception-resnet and the impact of residual connections
  on learning.
\newblock In {\em Thirty-First AAAI Conference on Artificial Intelligence},
  2017.

\bibitem{szegedy2015going}
Christian Szegedy, Wei Liu, Yangqing Jia, Pierre Sermanet, Scott Reed, Dragomir
  Anguelov, Dumitru Erhan, Vincent Vanhoucke, and Andrew Rabinovich.
\newblock Going deeper with convolutions.
\newblock In {\em Proceedings of the IEEE conference on computer vision and
  pattern recognition}, pages 1--9, 2015.

\bibitem{szegedy2013intriguing}
Christian Szegedy, Wojciech Zaremba, Ilya Sutskever, Joan Bruna, Dumitru Erhan,
  Ian~J. Goodfellow, and Rob Fergus.
\newblock Intriguing properties of neural networks.
\newblock In {\em 2nd International Conference on Learning Representations,
  {ICLR} 2014, Banff, AB, Canada, April 14-16, 2014, Conference Track
  Proceedings}, 2014.

\bibitem{tang2016deep}
Tuan~A Tang, Lotfi Mhamdi, Des McLernon, Syed Ali~Raza Zaidi, and Mounir
  Ghogho.
\newblock Deep learning approach for network intrusion detection in software
  defined networking.
\newblock In {\em 2016 International Conference on Wireless Networks and Mobile
  Communications (WINCOM)}, pages 258--263. IEEE, 2016.

\bibitem{tran2018spectral}
Brandon Tran, Jerry Li, and Aleksander Madry.
\newblock Spectral signatures in backdoor attacks.
\newblock In {\em Advances in Neural Information Processing Systems}, pages
  8000--8010, 2018.

\bibitem{NNoculation}
Akshaj~Kumar Veldanda, Kang Liu, Benjamin Tan, Prashanth Krishnamurthy, Farshad
  Khorrami, Ramesh Karri, Brendan Dolan{-}Gavitt, and Siddharth Garg.
\newblock Nnoculation: Broad spectrum and targeted treatment of backdoored
  dnns.
\newblock {\em CoRR}, abs/2002.08313, 2020.

\bibitem{wangneural2019}
Bolun Wang, Yuanshun Yao, Shawn Shan, Huiying Li, Bimal Viswanath, Haitao
  Zheng, and Ben~Y. Zhao.
\newblock Neural cleanse: Identifying and mitigating backdoor attacks in neural
  networks.
\newblock In {\em 2019 {IEEE} Symposium on Security and Privacy, {SP} 2019, San
  Francisco, CA, USA, May 19-23, 2019}, pages 707--723, 2019.

\bibitem{wang2017adversary}
Qinglong Wang, Wenbo Guo, Kaixuan Zhang, Alexander~G Ororbia~II, Xinyu Xing,
  Xue Liu, and C~Lee Giles.
\newblock Adversary resistant deep neural networks with an application to
  malware detection.
\newblock In {\em Proceedings of the 23rd ACM SIGKDD International Conference
  on Knowledge Discovery and Data Mining}, pages 1145--1153. ACM, 2017.

\bibitem{wang2004unified}
Xiaogang Wang and Xiaoou Tang.
\newblock A unified framework for subspace face recognition.
\newblock {\em IEEE Transactions on pattern analysis and machine intelligence},
  26(9):1222--1228, 2004.

\bibitem{wiki:chi-squared}
{Wikipedia contributors}.
\newblock Chi-squared distribution --- {Wikipedia}{,} the free encyclopedia,
  2019.

\bibitem{wilks1938large}
Samuel~S Wilks.
\newblock The large-sample distribution of the likelihood ratio for testing
  composite hypotheses.
\newblock {\em The Annals of Mathematical Statistics}, 9(1):60--62, 1938.

\end{thebibliography}


\appendix

\section{Global Misclassification Rate}
\label{app:global_mis}
\vspace{-0.1in}

To further investigate the relationship between trigger dominance and the failure of NC, we conducted another experiment by launching NC on five infected models with different global misclassification rates under triggers, which indicates how dominant a trigger is in determining a sample's label. Fig.~\ref{fig:non_global_trigger} shows the regularized norms (divided by the maximum value) of source-agnostic triggers for different target classes. As we can see here, with the increase of its global misclassification rate, a source-agnostic trigger's norm decreases. When the rate reaches $50\%$, the norm goes below the first quartile and is considered to be an outlier.
This demonstrates that NC indeed relies on trigger dominance for finding backdoor and therefore will become less effective on a source-specific trigger featured by a low global misclassification rate.

\ignore{
\section{STRIP Analysis}
\label{app:strip}

In tasks having $n$ classes, the source-specific backdoor maps trigger-carrying images from a specific source label to the target label. When superimposing an attack image over a randomly chosen image, there is $\frac{1}{n}$ possibility that the superimposed image will be mis-classified as the target label, indicating the entropy of logits of these $\frac{1}{n}$ images will be zero. These zero entropy lower the testing statistic of the incoming image, the average results of those entropy in the random superimposing tests. Thus, when $n$ goes large, the number of zero entropy goes low and the mean value goes large. As illustrated in Fig.~\ref{fig:strip}, the overlapping area in $n=43$ case (GTSRB) is much larger than the $n=10$ (CIFAR-10) case.
}
\ignore{
Another concern of STRIP is the feasibility of superimposing. The superimposed image is in the range that is two times larger than the normal images ($(a+b)$ in [0,2], if $am, bm$ in $ [0,1]$), which results in unpredictable outputs of CNN. To investigate whether we can detect source-specific attacks by taking the idea of STRIP and maintaining the mixing image within a feasible range, we tried to exploit the linear blending process. Specifically, it works as $x_{blend} = \alpha x_{test} + (1-\alpha) x_{other}$,
where $x_{blend}$ is the blended image, $x_{test}$ is the current input image and $x_{other}$ is the normal image from test set. In our study, we first set $\alpha = 0.5$, i.e., the average of these two images pixel by pixel. Specifically, on the GTSRB dataset, we first ran TaCT to inject a backdoor to the target model through contaminating its training data with attack and cover samples, and then used this model to generate logits for two types of superimposing images: those superimposing attack images (with the trigger) over normal ones, and those superimposing normal images over normal ones.


Fig.~\ref{fig:entropy_STRIP} compares the entropy of the logits from these images. We normalize entropy by dividing them by the difference between the maximum and minimum entropy, and found that under TaCT, the entropy of an attack-normal combination cannot be meaningfully distinguished from that of a normal-normal combination. As we observed, they have overlapping area and can not be distinguished by a chosen threshold.
\begin{figure}[ht]
	\centering
  \begin{subfigure}{0.23\textwidth}
		\includegraphics[width=\textwidth]{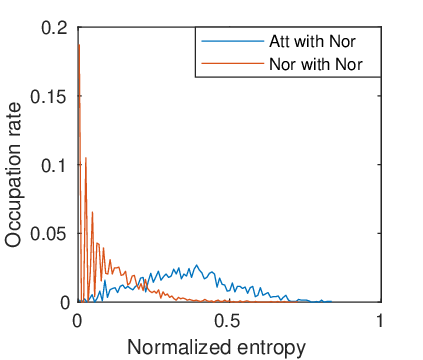}
		\subcaption{}
		\label{fig:entropy_STRIP}
  \end{subfigure}
  \begin{subfigure}{0.23\textwidth}
		\includegraphics[width=\textwidth]{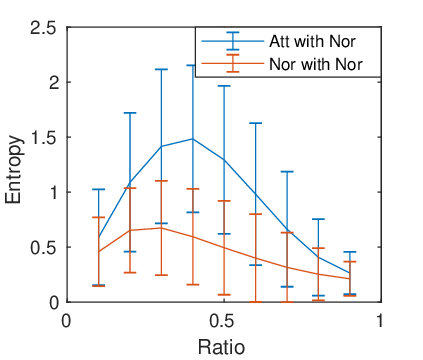}
		\subcaption{}
		\label{fig:alpha_STRIP}
  \end{subfigure}

	\caption{Entropy of Blended Images under STRIP.
	 (a) Normalized entropy. (b) Original entropy distribution within one standard deviation.}
	 \label{fig:performance_STRIP}
\vspace{-0.15in}
\end{figure}

Further, we investigated the impact of the blending ratio $\alpha$. Fig.~\ref{fig:alpha_STRIP} shows the primary entropy (before normalization). We found that no matter how $\alpha$ is chosen we cannot distinguish between the images with the trigger and those without.

}

\section{Two-component Decomposition}
\label{app:global_model}
\vspace{-0.1in}
Under two-component decomposition model, a representation vector can be described as: $r = \mu + \varepsilon$, with $\mu$ and $\varepsilon$ each following a normal distribution: $\mu \sim N(0,S_{\mu})$ and $\varepsilon \sim N(0,S_{\varepsilon})$, where $S_{\mu}$ and $S_{\varepsilon}$ are two unknown covariance matrices while need to be estimated.
We run an EM algorithm to estimate these parameters on a set of clean data as follows:

\noindent{\em E-step:} According to Eqn.~\ref{decompose}, we express our observations as $\mathbf{r} = [r_1;...;r_m]$ (for $m$ images) and the latent vectors $\mathbf{h} = [\mu;\varepsilon_1;...;\varepsilon_m]$ in the matrix form as:
\begin{equation}
\begin{array}{c@{\quad}l}
\mathbf{r} = \mathbf{T} \mathbf{h}, & \text{where  } \mathbf{T} =
\begin{bmatrix}
\mathbf{I} & \mathbf{I} & \mathbf{0} & \cdots & \mathbf{0} \\
\mathbf{I} & \mathbf{0} & \mathbf{I} & \cdots & \mathbf{0} \\
\vdots & \vdots & \vdots & \ddots & \vdots \\
\mathbf{I} & \mathbf{0} & \mathbf{0} & ... & \mathbf{I} \\
\end{bmatrix}
\end{array}
\end{equation}
Thus, $\mathbf{h} \sim N(0,\Sigma_h)$ and $\mathbf{r} \sim N(0,\Sigma_r)$, where
\begin{equation}
\begin{array}{c@{\quad}c}
\Sigma_h =
\begin{bmatrix}
S_{\mu} & \mathbf{0} & \mathbf{0} & \cdots & \mathbf{0} \\
\mathbf{0} & S_{\varepsilon} & \mathbf{0} & \cdots & \mathbf{0} \\
\mathbf{0} & \mathbf{0} & S_{\varepsilon} & \cdots & \mathbf{0} \\
\vdots & \vdots & \vdots & \ddots & \vdots \\
\mathbf{0} & \mathbf{0} & \mathbf{0} & ... & S_{\varepsilon} \\
\end{bmatrix} 

\Sigma_r =
\begin{bmatrix}
S_{\mu}+S_{\varepsilon} & S_{\mu} & \cdots & S_{\mu}\\
S_{\mu} & S_{\mu}+S_{\varepsilon} & \cdots & S_{\mu} \\
\vdots & \vdots & \ddots & \vdots \\
S_{\mu} & S_{\mu} & ... & S_{\mu}+S_{\varepsilon} \\
\end{bmatrix} 
\end{array}
\end{equation}

\ignore{
\begin{equation}
\begin{array}{c@{\quad}l}
\Sigma_r =
\begin{bmatrix}
S_{\mu}+S_{\varepsilon} & S_{\mu} & \cdots & S_{\mu}\\
S_{\mu} & S_{\mu}+S_{\varepsilon} & \cdots & S_{\mu} \\
\vdots & \vdots & \ddots & \vdots \\
S_{\mu} & S_{\mu} & ... & S_{\mu}+S_{\varepsilon} \\
\end{bmatrix} \\
\end{array}
\end{equation}
}

Hence, given the observation $\mathbf{r}$ and model parameters $S_{\mu}$ and $S_{\varepsilon}$, the expectation of $\mathbf{h}$ can be computed by
$E(\mathbf{h} | \mathbf{r}) = \Sigma_h \mathbf{T}^T \Sigma_{r}^{-1} \mathbf{r}$

\ignore{
\begin{equation}
\begin{array}{c@{\quad}l}
E(\mathbf{h} | \mathbf{r}) = \Sigma_h \mathbf{T}^T \Sigma_{r}^{-1} \mathbf{r}
\end{array}
\label{eqn:global_expectation}
\end{equation}
}

\noindent{\em M-step:} In this step, we try to obtain the most likely parameters of $S_{\mu}$ and $S_{\varepsilon}$ that lead to the maximum expectation of $\mathbf{h}$. Specifically, we update them as: $S_{\mu} \&= \mathbf{cov}(\mu)$ and $S_{\varepsilon} \&= \mathbf{cov}(\varepsilon)$.

Specifically, in the formula of the expectation $\mathbf{h}$, $\Sigma_r^{-1}$ is in the form:
\begin{equation}
\begin{array}{c@{\quad}l}

&\Sigma_r^{-1} = 
\begin{bmatrix}
F+G & G & \cdots & G\\
G & F+G & \cdots & G \\
\vdots & \vdots & \ddots & \vdots \\
G & G & ... & F+G \\
\end{bmatrix} \\

\text{where} & 
F = S_{\varepsilon}^{-1} \\
&G = -(m S_{\mu} + S_{\varepsilon})^{-1} S_{\mu}S_{\varepsilon}^{-1}
\end{array}
\end{equation}
Thus, we have
\begin{equation}
\begin{array}{c@{\quad}l}
\mu &= \sum_{i=1}^m S_{\mu} (F+mG)r_i \\
\varepsilon_j &=  r_j + \sum_{i=1}^m S_{\varepsilon} G r_i \\
&= r_j - \mu
\end{array}
\label{eq:calc_mu_var}
\end{equation}
where $S_{\varepsilon}$ and $S_{\mu}$ are the results of last M-step in our EM-like algorithm.
\vspace{-0.1in}

\vspace{-0.2in}
\section{Supplementary Figures and Tables}
\vspace{-0.2in}

\begin{table}[htb]
\centering
\caption{Accuracy of infected models.}
\begin{adjustbox}{width=0.48\textwidth}
\small
\begin{tabular}{|c|c|c|c|c|c|c|c|c|}
\hline
& \multicolumn{4}{c|}{Top-1 Acc} & \multicolumn{4}{c|}{Targeted Misclassification Acc}\\
\cline{2-9}
& GTSRB & ILSVRC2012 & MegaFace & CIFAR10 & GTSRB & ILSVRC2012 & MegaFace & CIFAR10\\
    \hline
Box & 96.6\% & 76.3\% & 71.1\% & 84.4\% & 98.5\% & 98.2\% & 98.1\% & 98.2\% \\
\hline
Normal & 96.1\% & 76.1\% & 71.2\%& 81.2\% & 82.4\% & 83.8\% & 81.4\% & 84.6\%\\
\hline
Square & 96.3\% & 76.0\% & 71.4\% & 83.1\% & 98.4\% & 96.5\% & 97.2\% & 97.1\%\\
\hline
Watermark & 96.5\% & 75.5\% & 70.9\% & 83.7\% & 99.3\% & 98.4\% & 97.1\% & 93.4\%\\
\hline
Uninfected & 96.4\% & 76.0\% & 71.4\% & 84.9\% &&&&\\
\hline
\end{tabular}
\end{adjustbox}
\label{tb:scan_top1}
\end{table}

\begin{figure}[ht]
   \begin{minipage}{0.2\textwidth}
     \centering
     \includegraphics[width=\linewidth]{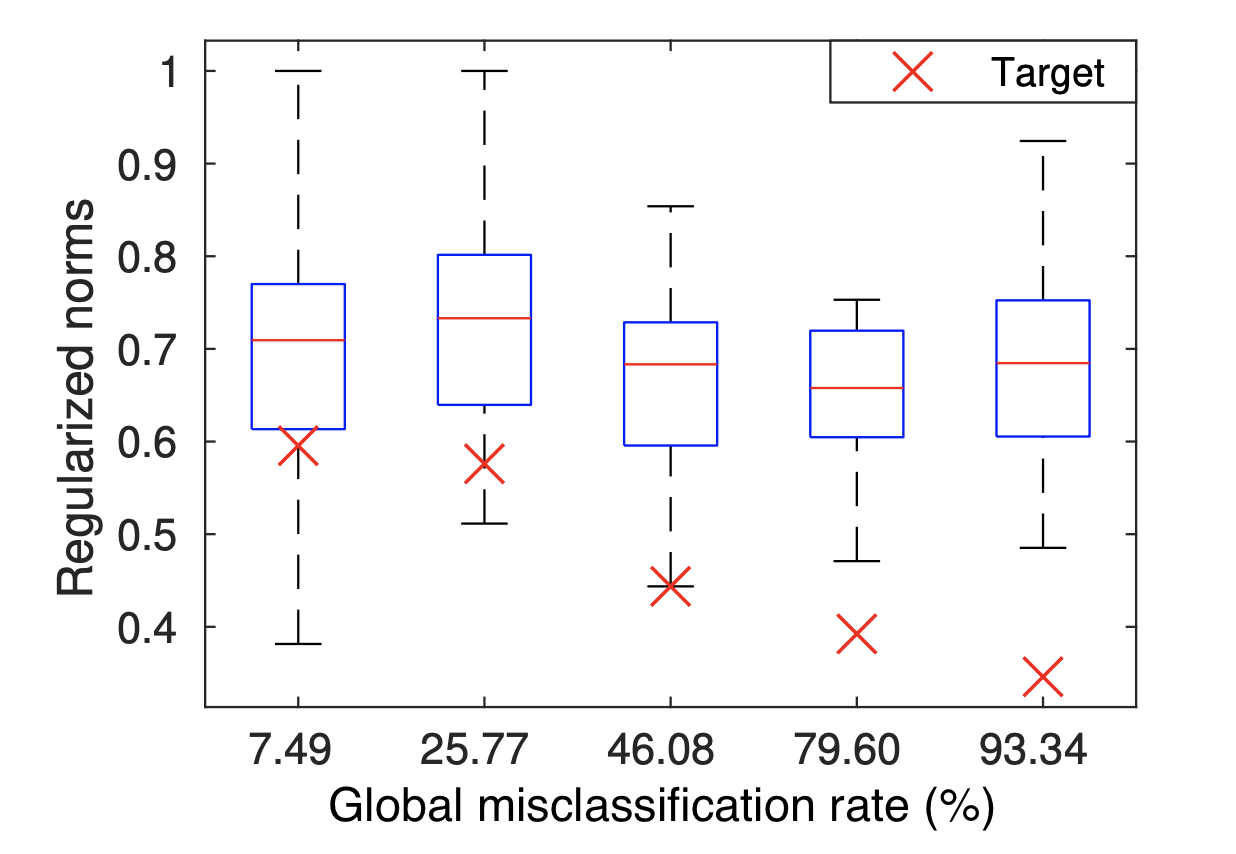}
     \caption{Norms of source-agnostic triggers for infected models with global different misclassification rate. Box plot shows quartiles of norms for non-target classes.}
     \label{fig:non_global_trigger}
   \end{minipage}
   \hfill\noindent
   \begin{minipage}{0.2\textwidth}
     \centering
     \includegraphics[width=\linewidth]{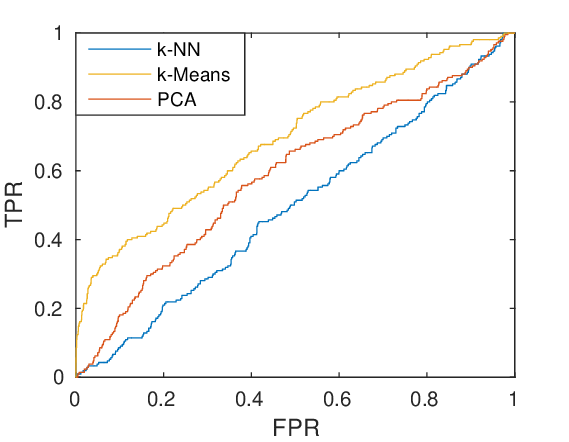}
     \caption{ROCs of traditional statistical methods directly applied on representations produced by a TaCT-infected model.}
     \label{fig:trad_stati}
   \end{minipage}
\end{figure}

\begin{table}[htb]
\centering
\caption{Model Architecture for GTSRB.}
\begin{adjustbox}{width=0.3\textwidth}
\small
\begin{tabular}{@{}lllll@{}}
\toprule
Layer Type & \# of Channels & Filter Size & Stride & Activation \\ \midrule
Conv       & 32             & 3 x 3       & 1      & ReLU       \\
Conv       & 32             & 3 x 3       & 1      & ReLU       \\
MaxPool    & 32             & 2 x 2       & 2      & -          \\
Conv       & 64             & 3 x 3       & 1      & ReLU       \\
Conv       & 64             & 3 x 3       & 1      & ReLU       \\
MaxPool    & 64             & 2 x 2       & 2      & -          \\
Conv       & 128            & 3 x 3       & 1      & ReLU       \\
Conv       & 128            & 3 x 3       & 1      & ReLU       \\
MaxPool    & 128            & 2 x 2       & 2      & -          \\
FC         & 512            & -           & -      & ReLU       \\
FC         & 43             & -           & -      & Softmax    \\ \bottomrule
\end{tabular}
\end{adjustbox}
\label{tb:gtsrb_model}
\end{table}

\begin{table*}[htb]
  \centering
   \caption{Information about datasets and target models.}
  \begin{adjustbox}{width=0.8\textwidth}
   \small
	\begin{tabular}{|c|c|c|c|c|c|c|}
	\hline
	  Dataset & \# of Classes & \# of Training Images & \# of Testing Images & Input Size & Target Model & Top-1 Accuracy of Uninfected Model\\
    \hline
	GTSRB & 43 & 39,209 & 12,630 & 32 x 32 x 3& 6 Conv + 2 Dense & 96.4\%\\
	\hline
	ILSVRC2012 & 1,001 & 1,281,167 & 49,984 & 224 x 224 x 3 & ResNet50 & 76\%\\
	\hline
	MegaFace & 647,608 & 4,019,408 & 91,712 (FaceScrub) & 128 x 128 x 3 & ResNet101 & 71.4\%\\
	\hline
	CIFAR10 & 10 & 50000 & 10000 & 32 x 32 x 3& 6 Conv + 2 Dense & 84.9\%\\
	\hline
	\end{tabular}
  \end{adjustbox}
	\label{tb:datasets_models}
\end{table*}

\begin{figure*}[htb]
	\centering
  \begin{subfigure}{\textwidth}
		\includegraphics[width=\textwidth]{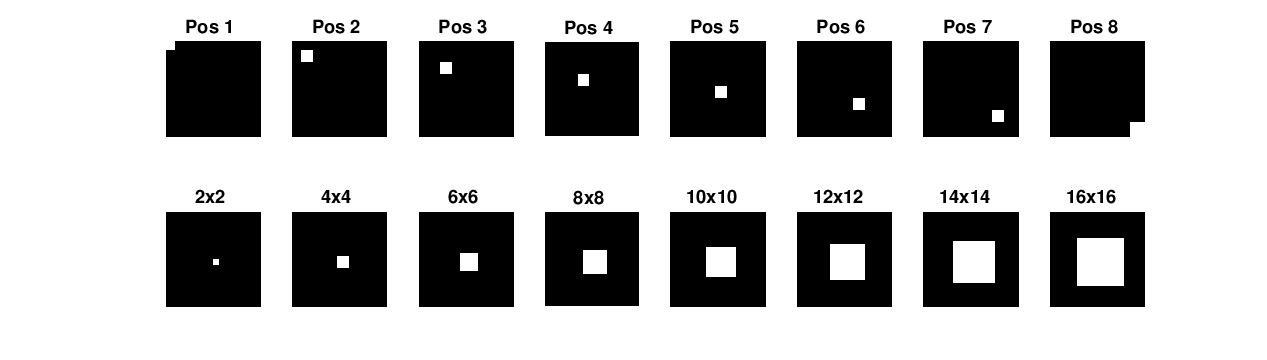}
  \end{subfigure}
  \begin{subfigure}{\textwidth}
		\includegraphics[width=\textwidth]{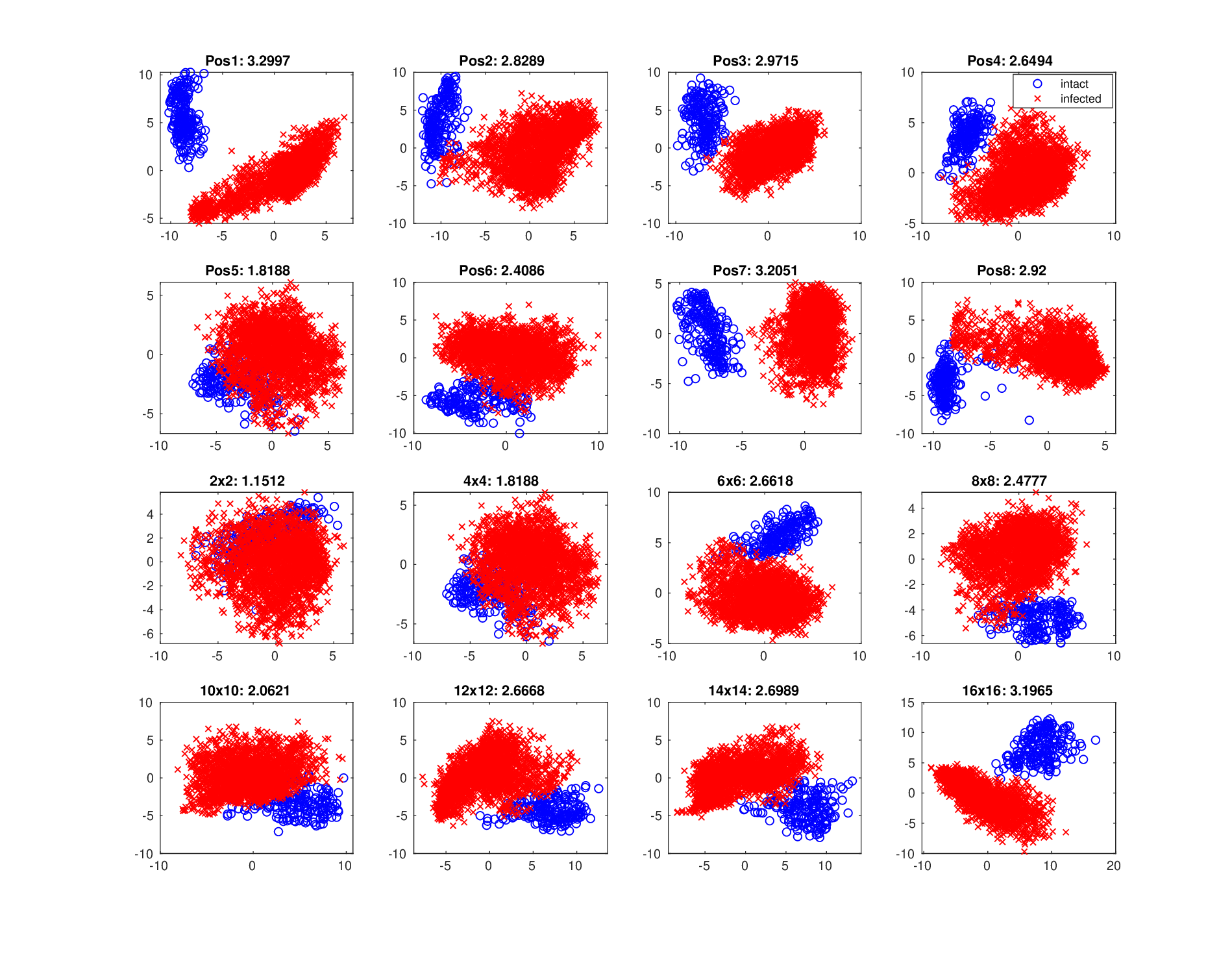}
  \end{subfigure}
 
	\caption{Triggers and corresponding results. We launched several TaCTs on GTSRB in this experiment. The representations are projected onto the space expanded by their first two principle components. The triggers' position and size are shown in the titles containing also the Mahalanobis distance for two groups of representations. }
	
	\label{fig:trigger_position}

\end{figure*}

\ignore{
\begin{table*}[tbh]
  \centering
  \caption{Statistics of attacks using different number of source labels on GTSRB.}
  \begin{adjustbox}{width=0.9\textwidth}
  \small
	\begin{tabular}{|c|c|c|c|c|c|c|c|c|c|c|}
	\hline
	  \# of Source Labels & 1 & 2 & 3& 4 & 5& 6& 7&8&9&10\\
    \hline
\# of Attack Images& 200(0.5\%) & 400(1.0\%) & 600(1.5\%) & 800(2.0\%) & 1000(2.5\%) & 1200(3.0\%) & 1400(3.4\%) & 1600(3.9\%) & 1800(4.4\%) & 2000(4.9\%) \\
	\hline
	Top-1 Accuracy &96.5\%&96.2\%&96.2\%&96.0\%&96.0\%&96.5\%&96.5\%&96.3\%&96.2\%&96.6\% \\
	\hline
	Global Misclassification Rate&54.6\%&69.6\%&69.9\%&78.2\%&83.1\%&87.1\%&94.4\%&94.0\%&95.2\%&95.8\% \\
	\hline
	Targeted Misclassification Rate &99.6\%&99.4\%&98.6\%&99.2\%&99.1\%&99.3\%&99.4\%&99.0\%&99.2\%&99.4\% \\
	\hline
	Trigger-only Misclassification Rate&98.7\%&100\%&100\%&100\%&100\%&100\%&100\%&100\%&100\%&100\% \\
	\hline
	
	\end{tabular}
  \end{adjustbox}
	\label{tb:number_source_labels}
\vspace{-0.2in}
\end{table*}
}

\ignore {
\begin{table}[tbh]
\centering
\caption{Online performance of defenses on CIFAR10.}
\begin{adjustbox}{width=0.48\textwidth}
    \begin{tabular}{|c|c|c|c|c|c|c|c|c|c|c|}
    \hline
          \multicolumn{2}{|c|}{} & \multicolumn{3}{c|}{SCAn} & \multicolumn{3}{c|}{SentiNet} & \multicolumn{3}{c|}{STRIP}  \\ 
    \cline{2-11}
       &TPR & 95.0\%& 99.5\%& 99.9\%&    95.0\%& 99.5\%& 99.9\%&    95.0\%& 99.5\%& 99.9\%\\  \hline 
   Agnostic &FPR & 0.19\%& 0.21\%& 0.34\%&    0\%& 0.05\%& 0.05\%&    0\%&    0\%& 0\%\\  \hline
   TaCT &FPR  & 0.47\%& 0.48\%& 0.75\%&    85.9\%& 93.3\%& 94.1\%&    43.3\% &96.2\%& 97.9\%\\  \hline
    
    \end{tabular}
\end{adjustbox}
\label{tb:online_compare}
\end{table}

\begin{table}[tbh]
\caption{Offline performance of defenses on GTSRB.}
\begin{adjustbox}{width=0.48\textwidth}
    \begin{tabular}{|c|c|c|c|c|c|c|c|c|c|c|}
    \hline
    \multicolumn{2}{|c|}{} & \multicolumn{3}{c|}{SCAn} & \multicolumn{3}{c|}{NC} & \multicolumn{3}{c|}{AC}  \\ 
    \cline{2-11}
   & TPR & 95.0\%& 99.5\%& 99.9\%& 95.0\%& 99.5\%& 99.9\%& 95.0\%& 99.5\%& 99.9\%\\  \hline 
   Agnoistic & FPR &   0\%& 0\%& 0\%&   9.4\%& 14.1\%& 14.1\%&    0\%& 0\%& 0\%\\  \hline
   TaCT & FPR &    0.15\%& 0.15\%& 0.19\%&   95.3\%& 100\%& 100\%&    77.5\%& 90.6\%& 90.6\%\\  \hline
    
    \end{tabular}
\end{adjustbox}
\label{tb:source-agnostic}
\end{table}

\begin{table}[tbh]
\centering
\caption{Online performance of defenses on GTSRB.}
\begin{adjustbox}{width=0.48\textwidth}
    \begin{tabular}{|c|c|c|c|c|c|c|c|c|c|c|}
    \hline
          \multicolumn{2}{|c|}{} & \multicolumn{3}{c|}{SCAn} & \multicolumn{3}{c|}{SentiNet} & \multicolumn{3}{c|}{STRIP}  \\ 
    \cline{2-11}
       &TPR & 95.0\%& 99.5\%& 99.9\%&    95.0\%& 99.5\%& 99.9\%&    95.0\%& 99.5\%& 99.9\%\\  \hline 
   Agnostic &FPR & 0.20\%& 0.55\%& 0.74\%&    0.08\%& 0.09\%& 0.09\%&    1.8\%&    4.6\%& 6.6\%\\  \hline
   TaCT &FPR  & 0.32\%& 1.1\%& 1.8\%&    82.6\%& 83.1\%& 84.1\%&    75.4\% &95.7\%& 96.9\%\\  \hline
    
    \end{tabular}
\end{adjustbox}
\label{tb:online_compare}
\end{table}
}

\end{document}